\documentclass[11pt,cite,epsfig,psfrag]{article}
\usepackage[utf8]{inputenc}
\usepackage{placeins}
\usepackage{geometry}
 \usepackage{graphicx}
 \usepackage{subfig}
  \usepackage{amsmath} 
  \graphicspath{{images_pss_massive/}}
  \usepackage[usenames, dvipsnames]{color}
\usepackage{wrapfig}
\usepackage{boxedminipage}
\usepackage{multirow}

\usepackage{array}
\usepackage{multirow}

\usepackage{fullpage}
\usepackage{amsmath}
\usepackage{amssymb}
\usepackage{graphicx}
\usepackage{mathrsfs}
\usepackage{wrapfig}
\usepackage{boxedminipage}
\usepackage{epsfig}
\usepackage{setspace}
\usepackage{float}
\usepackage{hyperref}
\usepackage{enumerate}
\usepackage{cite}
\usepackage[font=footnotesize,labelfont=rm]{caption}

\usepackage[numbers,sort&compress]{natbib}

\def\be{\begin{equation}}
\def\ee{\end{equation}}
\def\bea{\begin{eqnarray}}
\def\eea{\end{eqnarray}}

\numberwithin{equation}{section}

 \newcommand{\RN}[1]{%
   \textup{\uppercase\expandafter{\romannumeral#1}}%
 }
\begin{document}

\thispagestyle{empty}

\vskip 2cm

\begin{center}
{\Large \bf Topological charge and black hole photon spheres  in massive gravity}
\end{center}

\vskip .2cm

\vskip 1.2cm

\centerline{ \bf   Pavan Kumar Yerra~\footnote{pk11@iitbbs.ac.in} and Chandrasekhar Bhamidipati~\footnote{chandrasekhar@iitbbs.ac.in}  }

\vskip 7mm 

\begin{center}{ $^{1}$ 	Institute of Fundamental Physics and Quantum Technology, \\ Department of Physics, School of Physical Science and Technology, \\ Ningbo University, Ningbo, Zhejiang 315211, China}
\end{center}

\begin{center}{ $^{2}$ Department of Physics\\ 
		Indian Institute of Technology Bhubaneswar \\ Bhubaneswar, Odisha, 752050, India}
\end{center}

\vskip 1.2cm
\centerline{\bf Abstract}
\vskip 0.5cm
\noindent

In this paper, we investigate the existence and nature of the photon spheres (PS) in the background of four dimensional static and spherically symmetric black holes in the de Rham-Gabadadze-Tolley (dRGT) massive gravity theory. Apart from the known case of one PS, there are regions in the parameter space of massive gravity where either two or no PS's exist outside the event horizon. Topological arguments show that, the case of one PS falls in the category of Einstein gravity (with topological charge $-1$)~\cite{Wei:2020rbh},  whereas, the cases with two or zero PS's belong to a different topological class with total charge zero.  PS's of horizonless compact objects, also belong to the same class with total topological charge $0$~\cite{Cunha:2017qtt,Cunha:2020azh,Wei:2020rbh}, though, one distinction can be made with the black holes in massive gravity. While in the former case, the inner PS is stable, in the later case, it is the outer PS which is stable (the inner PS is unstable). We also study the landscape of possible regions of existence of standard and exotic photon spheres in the massive gravity parameter space, and correlate it with the horizon structure of the black holes.

\newpage
\setcounter{footnote}{0}
\noindent

\baselineskip 15pt

\section{Introduction}

The detection of gravitational waves \cite{Abbott} and the images of black hole shadow\cite{EventHorizonTelescope:2019dse,EventHorizonTelescope:2019pgp,EventHorizonTelescope:2019ggy} ushered in a new aeon in astrophysical observations. These experimental observations have paved the way for deep investigations into the geometrical features of the event horizon of black holes.  The study of orbits of particles is important for testing novel phenomena associated to compact astrophysical objects, in both Einstein and various versions of modified theories of gravity. For instance, interesting physical properties of spacetime when gravity effects are strong or weak, lead to novel understanding of ring down in a binary system of black holes\cite{LIGOScientific:2016aoc}, shadows of black holes\cite{EventHorizonTelescope:2022wkp,EventHorizonTelescope:2019dse}, and other observational signatures \cite{Grandclement:2014msa,Grould:2017rzz,Teodoro:2020kok,Teodoro:2021ezj,Gibbons:1999uv,Herdeiro:2000ap,Diemer:2013fza,Delgado:2021jxd,Bambi:2019tjh,Vagnozzi:2022moj,Khodadi:2024ubi}. It is well appreciated that in the black hole mergers, especially, during the ring down stage,  the knowledge of quasinormal modes gives crucial information on the radiation carried away by the gravitational waves. Stable orbits of massive probes around black holes give information on accretion discs, while the unstable ones shine light on the shadows~\cite{Stuchlik:1999qk,Claudel:2000yi,Virbhadra:1999nm,Claudel:2000yi,Virbhadra:2002ju,Adler:2022qtb,Levin:2008mq,Pugliese:2010ps,Hackmann:2010zz,Villanueva:2013zta,Wei:2017mwc,Chandrasekhar:2018sjg,Lehebel:2022yyz,Bozza:2002zj,Cardoso:2008bp,Hioki:2009na,Collodel:2017end,Ye:2023gmk,Cunha:2020azh,Cunha:2022gde,Wei:2022mzv,Ghosh:2021txu,Ghosh:2023kge}. In the eikonal limit, there are interesting connections between the formation of black hole shadows, quasinormal modes and the photon spheres or the light-rings of general rotating black holes \cite{Cardoso:2008bp,Glampedakis:2017dvb,Berti:2005eb,Cardoso:2016rao}. The nature and stability aspects of photon spheres may be helpful in differentiating objects with or without horizon, e.g.,  black holes, ultracompact objects and naked singularities, and this leads to the study of the images of light sources around the compact objects in general\cite{Decanini:2009mu,Stefanov:2010xz,Cunha:2018acu,Claudel:2000yi}. These are thus important physical reasons to investigate the geometrical and topological properties of photon spheres or the light-rings around gravitational objects, which may unravel crucial details of the space-time. \\

\noindent
Different methods have been employed recently, with emphasis on geometrical and topological techniques, to study orbits around spherically symmetric as well as general rotating compact objects\cite{Hod:2018kql,Lehebel:2022yyz,Bozza:2002zj,Cardoso:2008bp,Hioki:2009na,Collodel:2017end,Ye:2023gmk,Cunha:2020azh,Cunha:2022gde,Wei:2022mzv,Guo:2022ghl,Ye:2023gmk,Wu:2023eml,Sadeghi:2024itx,Afshar:2024bgi,Liu:2024soc,Song:2025vhw,Sadeghi:2023dxy,NooriGashti:2025rwl}.
In line with the above objectives, Cunha, Berti, and Herdeiro put forward a novel proposal on inferring the light-ring stability around ultra compact objects \cite{Cunha:2017qtt}. In particular, they showed that, without recourse to field equations the computation of certain topological quantities can prove important results, such as, establishing that non-degenerate light rings of  horizonless ultra compact objects should come in pairs, with one of them being stable (inner PS) \cite{Cunha:2017qtt,Cunha:2020azh}.  
There are of course certain exceptions when the light-rings become degenerate\cite{Hod:2017zpi,Hod:2013jhd}. There are also further important implications for four dimensional stationary and axi-symmetric black holes, which are either asymptotically flat, anti de Sitter or de Sitter, namely, the existence of at least one light ring for each rotation sense. The computations required for reaching these conclusions involved calculating the winding number of a vector field, following from the effective potential.  Relevant to the investigations of this work are the results obtained in\cite{Wei:2020rbh}, showing that  the computation of topological charges points towards the existence of at least one photon sphere in general spherically symmetric black holes which are asymptotically flat, anti de Sitter or de Sitter. In this paper, our aim is to study photon spheres and their stability around black holes in massive theories of gravity, using topological techniques. In particular, we emphasize that apart from the existence of a single unstable photon sphere (in agreement with~\cite{Wei:2020rbh}), there is a certain range of the massive gravity parameters, where it is possible to have a pair of photon spheres (stable and unstable) or actually no photon spheres at all, outside the event horizon of both neutral and charged black holes. We also investigate the reasons for the existence of novel topological classes of photon spheres in the massive gravity framework, in a certain range of parameters. Especially, we bring out the contrasting asymptotic behaviour of the lapse function
of the black hole geometry, and a certain everywhere regular effective potential function, which decides the topological classification of photon spheres. \\

\noindent
Before proceeding it is helpful to recollect broad motivations for interest in various massive theories of gravity.
Einstein's general theory of relativity is an extremely successful theory which has passed several experimental investigations, in addition to recent observations from the LIGO collaboration~\cite{LIGO2017,deRham2014Review} with regard to gravitational wave signals as well. Nonetheless, it is well appreciated that there do exist a host of other physical phenomena which needs modified theories of gravity, such as the cosmological constant issue, the accelerated expansion of our universe, the hierarchy problem, among others. Keeping these issues in mind, certain massive theories of gravity have been proposed ~\cite{MassiveIb,MassiveIc}, where the various limits set on the gravity mass could be met by recent observations~\cite{Abbott}. Of course, the idea of massive gravity is not new and models were constructed starting from 1939 by Firez and Pauli~\cite{Fierz1939}, with several modifications, leading to new massive gravity theories\cite{BDghost,Newmasssive,dRGTI,dRGTII}, being studied actively\cite{NewM1,NewM2,NewM3,NewM4,NewM5,HassanI,HassanII}. There are interesting classes of black holes in these theories with novel thermodynamic behaviour~\cite{BHMassiveI,BHMassiveII,BHMassiveIII,BHMassiveIV} apart from intriguing cosmological as well as astrophysics applications~\cite{Katsuragawa,Saridakis,YFCai,Leon,Hinterbichler,Fasiello,Bamba}, where one can investigate the deviations from Einstein's theory of gravity. Certain versions of massive gravity theories were constructed in~\cite{Vegh} (see also~\cite{Hinterbichler:2011tt, deRham:2014zqa}), which have the right set of features required to address some of the shortcomings of Einstein's theory, as emphasised in~\cite{Gumrukcuoglu,Gratia,Kobayash,DeffayetI,DeffayetII,DvaliI,
DvaliII,Will,Mohseni,GumrukcuogluII,NeutronMass,Ruffini,EslamPanah:2018evk}. For instance, there is a possibility of comprehending  
the current observations of dark matter
\cite{Schmidt-May2016DarkMatter} which in correlation with the accelerating
expansion of the universe may relax the requirement of  dark energy 
\cite{MassiveCosmology2013,MassiveCosmology2015}. It is important to mention that many of the massive theories of gravity naturally fit in the string theory and holographic framework~\cite{MGinString2018,Geng:2020qvw,Geng:2020fxl}. On the thermodynamic side of development, when the cosmological constant is considered to a variable, leading to the introduction of pressure in the black hole chemistry program (see\cite{Ahmed:2023snm,Mann:2025xrb} and references therein), the study of massive gravity theories~\cite{Cai2015,PVMassV,PVMassIV,Alberte,Zhou,Dehyadegari,Magmass,Dehghani:2019thq,Hendi:2015eca,Akbarieh:2021vhv,Chen:2023ddv} and the black hole microstructures has yielded interesting results~\cite{Yerra:2020tzg,Yerra:2020oph,Yerra:2021hnh,Hogas:2021saw,Caravano:2021aum,Chabab:2019mlu,Wu:2020fij}.\\

\noindent
For the purpose of studying topological properties of photon spheres in this work, we concentrate on the dRGT massive gravity theories studied in~\cite{deRham:2010kj,Ghosh:2015cva,Hendi:2022qgi}, which contain black hole solutions, where the action is that of  four dimensional Einstein-Maxwell gravity with a coupling to certain nonlinear interaction terms.  There are interesting features of these models, including stability of the solutions and the possibility of not having ghosts, which are all quite important while discussing various issues of black holes\cite{HZhang,PVMassI,PVMassII,PVMassIII,PVMassIV,PVMassV}. A key feature of this set of theories is the presence of a  non dynamical reference metric $f_{\mu\nu}$, breaking the diffeomorphism invariance, thus leading to some singular nature~\cite{Vegh}. This by itself is not a problem, as in certain holographic models, a theory of gravity with a massive graviton in AdS spacetime, together with a singular (degenerate) reference metric has been used to model certain classes of strongly interacting quantum field theories which have  broken translational symmetry (that is, momentum dissipation). This is due to the Lorentz breaking mass term of the graviton. Further, the massive graviton can be helpful in modelling a lattice in holographic models of conductors. For example, the Drude peak possessed by conductivity becomes a delta function in the massless gravity limit. The auxiliary reference metric referred to above is used in yielding Boulware-Deser ghost-free theories~\cite{deRham2014Review,Hinterbichler:2011tt,HassanII}. In fact, every choice of the auxiliary metric can lead to a different non-degenerate massive gravity theory~\cite{HassanII}. Some of these features are helpful in holographic applications, e.g., while building a model for conductors with momentum dissipation and non-zero (finite) DC conductivity~\cite{Vegh,Davison:2013jba}, which may not be possible in the massless gravity, such as, in higher derivative gravity where the DC conductivity is infinite~\cite{Hartnoll:2008vx,Hartnoll:2008kx,Gregory:2009fj,Barclay:2010up}. Moreover, other well known phases of condensed matter theories where the translational invariance is broken, such as solids/liquids/perfect fluids ~\cite{Alberte,Alberte:2015isw,Alberte:2017oqx}, or cosmological~\cite{Zhang:2019oes} and general holographic situations~\cite{Vegh,Davison:2013jba,Blake:2013bqa,Cao:2015cza,Baggioli:2014roa} require the current set up. We should also mention that there have also been efforts to make the auxiliary reference metric invertible and dynamical for a more viable description of massive gravity theories, in the so called bimetric gravity program~\cite{Schmidt-May2016DarkMatter,MassiveCosmology2013,MassiveCosmology2015,Hogas:2021saw,Caravano:2021aum} (see also~\cite{Gialamas:2023aim,Gialamas:2023lxj,Gialamas:2023fly}).
  In addition, there may be other alternatives to the use of a reference metric $f_{\mu\nu}$, where one introduces different sets of St\"{u}ckelberg fields, varying on the dimension of the reference metric~\cite{dRGTII}. Holographic conductivity has been studied in this St\"{u}ckelberg set up by making the fields dynamical\cite{Alberte}, while also obtaining certain coordinate independent expressions. The methods and techniques developed in this work though are very general, and can be carried over to study these interesting variations of  massive gravity theories as well.\\

\noindent
Plan of the rest of the paper is as follows. Section-(\ref{2}), contains a brief overview of the static and spherically symmetric solutions of black holes in a theory of massive gravity in four dimensions. We identify certain range of parameters where there is a possibility of multiple horizons with rich geometric structure.  Sections-(\ref{3}) and (\ref{4}) contain our main results. First, in section-(\ref{3}) we set up the basic equations required to see the presence of photon spheres. To do this, we fix some of the parameters of the model, and vary the others in a way that gives us qualitatively different types of stable and unstable photon spheres. Here, we show the existence of a single unstable photon sphere (in agreement with earlier results in~\cite{Wei:2020rbh}), and confirm this by computing the topological charges. We then go on to identify a novel range of the massive gravity parameters, where (unlike in~\cite{Wei:2020rbh}) it is possible to have a pair of photon spheres (stable and unstable) or actually no photon spheres at all outside the event horizon of both neutral and charged black holes. In subsection-(\ref{3.1}), these new photon spheres are classified using their topological charges, and the reason for their presence is attributed to a subtle behaviour of the potential function (which aids in the topological classification) in massive gravity theories, as compared to standard Einstein gravity.  In section-(\ref{4}), we present the landscape of parameter space of massive gravity theory, where the nature of black hole photon spheres is summarised based on topological arguments.  Conclusions are presented in section-(\ref{5}).

\section{Static, spherically symmetric black holes in massive gravity} \label{2}

Let us start with the dRGT massive theory of gravity. Here, the four dimensional action of the Einstein-Maxwell gravity with certain nonlinear interaction terms giving the graviton a mass $m_g$, is taken as~\footnote{For convenience, henceforth, we set the Newtons gravitational constant ($G$) and the speed of light ($c$) to be $G=c=1$.} as~\cite{deRham:2010kj,Ghosh:2015cva,Hendi:2022qgi}:
	\begin{equation}\label{Act:dRGT}
	I=~\frac{1}{16\pi}\int d^{4}x \sqrt{-g}\Big(\mathcal{R} - \mathcal{F} +m_{g}^{2}\, \mathcal{U}(g,\Phi^{a})\Big)\,,
	\end{equation}
where $\mathcal{R}$ stands  for the  Ricci scalar.  The Maxwell invariant is denoted as $\mathcal{F} = F_{\mu\nu}F^{\mu\nu}$  where $F_{\mu\nu}= \partial_\mu A_\nu - \partial_\nu A_\mu$ is expressed through gauge potential $A_\mu$.  The quantity $\mathcal{U}$ denotes the effective potential for the graviton, which in four dimensions is: 
\begin{equation*}
	\mathcal{U}(g,\Phi^{a})=~\mathcal{U}_{2}+\alpha_{3}\mathcal{U}_{3}+\alpha_{4}\mathcal{U}_{4}\, .
\end{equation*}	
In the above expression,  $ \alpha_{3}$ and $\alpha_{4} $ are dimensionless parameters which can be conveniently expressed in terms of new parameters, $ \alpha$ and $\beta $, as $ \alpha_{3}=\frac{\alpha-1}{3} $ and $ \alpha_{4}=\frac{\beta}{4}+\frac{1-\alpha}{12} $. 
The expressions for $\mathcal{U}_i$'s are taken to be:
\begin{eqnarray}
\mathcal{U}_{2} &\equiv &[\mathcal{K}]^{2}-[\mathcal{K}^{2}],  \notag \\
\mathcal{U}_{3} &\equiv &[\mathcal{K}]^{3}-3[\mathcal{K}][\mathcal{K}^{2}]+2[%
\mathcal{K}^{3}],  \notag \\
\mathcal{U}_{4} &\equiv &[\mathcal{K}]^{4}-6[\mathcal{K}]^{2}[\mathcal{K}%
^{2}]+8[\mathcal{K}][\mathcal{K}^{3}]+3[\mathcal{K}^{2}]^{2}-6[\mathcal{K}%
^{4}],
\end{eqnarray}%
where
\begin{equation*}
\mathcal{K}_{\,\,\,\nu }^{\mu }=\delta _{\nu }^{\mu }-\sqrt{g^{\mu \sigma
	}f_{ab}\partial _{\sigma }\Phi ^{a}\partial _{\nu }\Phi ^{b}},
\end{equation*}%
where  $[\mathcal{K}]=\mathcal{K}%
_{\,\,\,\mu }^{\mu }$ and $[\mathcal{K}^{n}]=(\mathcal{K}^{n})_{\,\,\,\mu
}^{\mu }$ denote the traces, and $f_{ab}$ is a reference metric.  Further, $ \Phi^{a} $ represent the St\"{u}ckelberg scalars, brought in to keep the action in a covariant form.
\vskip 0.2cm
\noindent The above action is known to possess static and spherically symmetric black hole solutions, where the corresponding line element and reference metric are
\begin{eqnarray}\label{eq:bh_metric}
ds^2 &=& -f(r)dt^2 +\frac{dr^2}{f(r)} +r^2(d\theta^2 + \sin^2\theta d\varphi^2)\, , \\
f_{\mu \nu } &=& diag\left( 0,0,h^{2},h^{2}\sin ^{2}\theta \right) \, .
\end{eqnarray}
$h$ appearing above is a positive constant.With an ansatz for the gauge
potential chosen as  $A_\mu = \big(A(r),0,0,0 \big)$, the lapse function $f(r)$ turns out to be
	\begin{equation}\label{eq:f_gen}
	f(r)=~1-\frac{2M}{r}+\frac{Q^{2}}{r^{2}}+\frac{\Lambda}{3}r^{2}+\gamma r +\zeta \, ,
	\end{equation}
	where $ M $ and $Q$ are the mass  and charge of the hole, respectively, with the other parameters coming out to be
		\begin{eqnarray}
	\Lambda &= & 3m_{g}^{2}(1+\alpha+\beta)\, , \nonumber	\\
	\gamma &=& -h m_{g}^{2}(1+2\alpha+3\beta)\, , \nonumber	\\
		\zeta &=& h^{2}m_{g}^{2}(\alpha+3\beta) \,.
	\end{eqnarray}
In~\cite{Ghosh:2015cva}, it was found that there exist various regions in the parameter space of $(\alpha,\beta)$, where the charged (neutral) black holes can have up to four (three) event horizons as well. $\Lambda$	and $\gamma$ play a crucial role in determining the asymptotic structure of the solution, with $\Lambda$ symbolizing the cosmological constant. In the limit when the graviton mass  is set to zero, i.e., $m_g=0$, the standard Reissner-Nordstrom black hole solution is recovered. Moreover, the parameters $(\alpha, \beta)$, as well as reference metric can be chosen in way that theory admits, asymptotically de-Sitter/anti-de Sitter solutions, and even global monopole solutions (more details on various possibilities can be found in~\cite{Ghosh:2015cva}).

\section{ Photon Spheres and Topological Charge } \label{3}
\begin{figure}[t!]
	
	{\centering
		
		\subfloat[]{\includegraphics[width=2.7in]{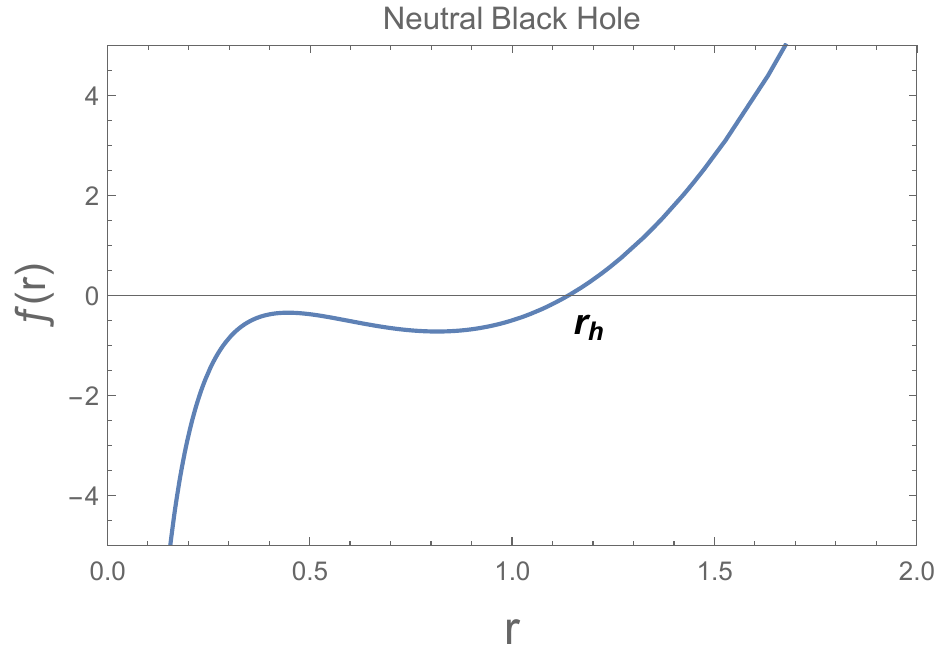}}\hspace{1.4cm}	
		\subfloat[]{\includegraphics[width=2.7in]{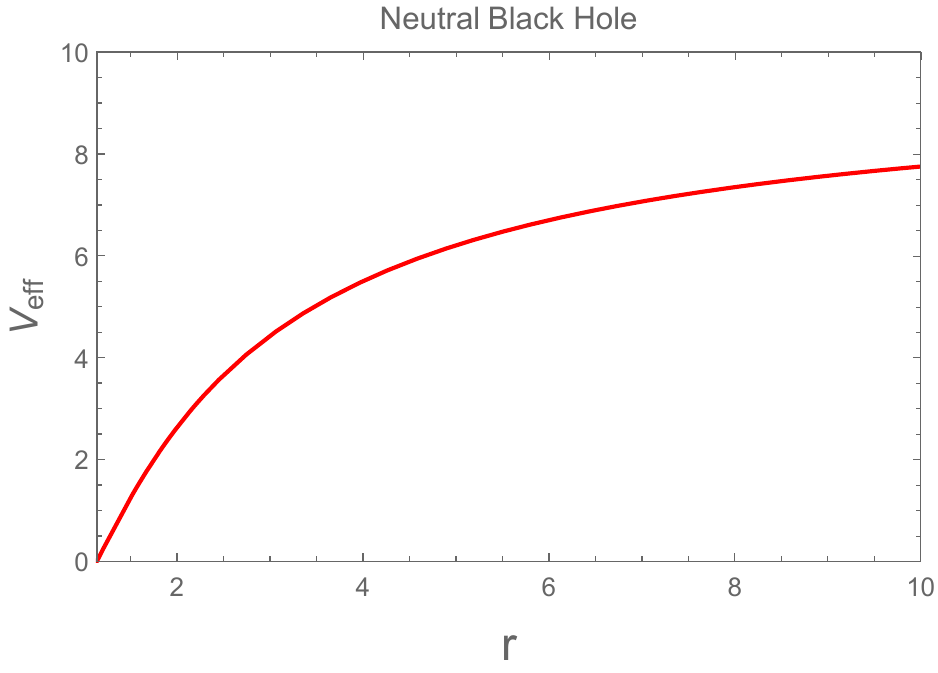}}\hspace{1.4cm}	
		\subfloat[]{\includegraphics[width=2.7in]{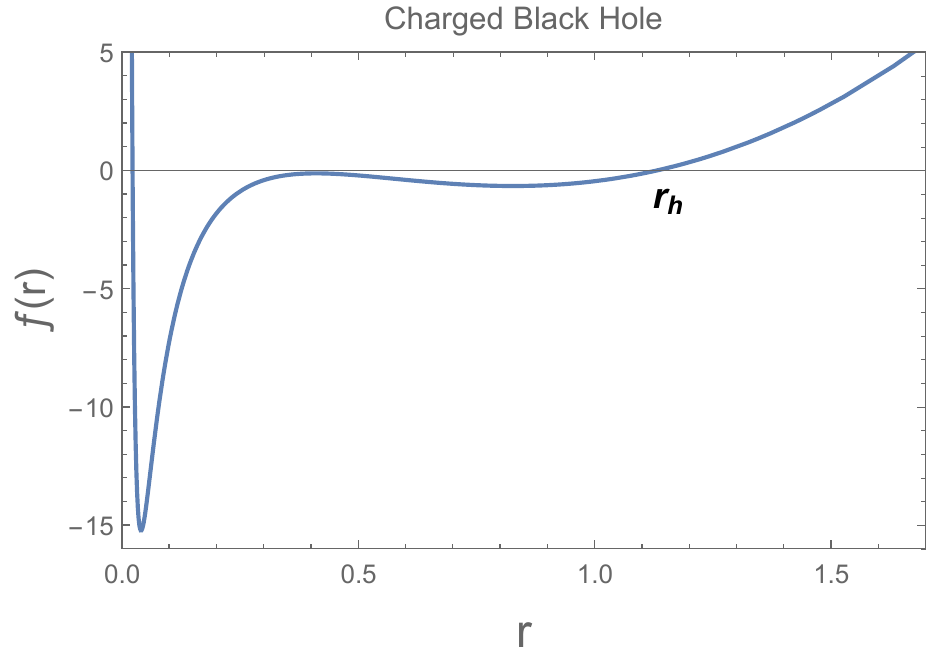}}\hspace{1.4cm}	
		\subfloat[]{\includegraphics[width=2.7in]{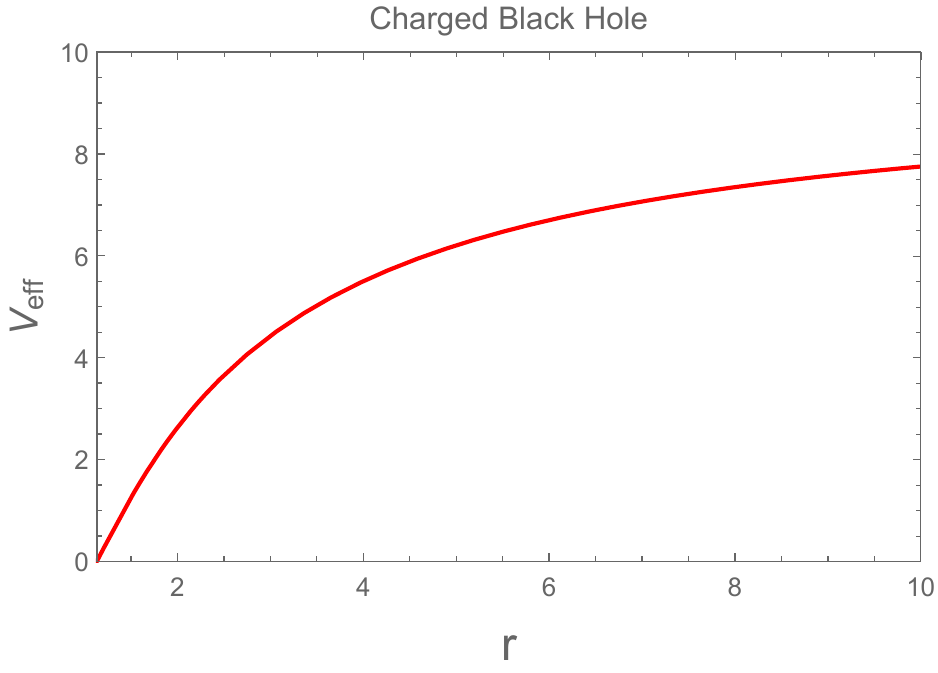}}	
		
		\caption{\footnotesize Left: behavior of the lapse function $f(r)$. Right:  effective potential~\eqref{eq:veff_photons}  showing the absent  of  photon spheres outside the event horizon. Top row:  for a neutral black hole with  single horizon at $r_h = 1.1396$ (Here, we set $\alpha =8, \beta =0.5$). Bottom row:  for a charged black hole with two horizons (inner and outer) where the outer horizon is located at $r_h = 1.1329$   (Here, we set $ Q=0.2, \alpha =8, \beta =0.5$).}
		\label{fig:fv_neucha_0ps}	}
	
\end{figure}

In this section, our first aim is to study the circular null geodesics in static and spherically symmetric black  holes in massive gravity. We then point out the possibility of existence of different number of photon spheres as well as their stability, which depend on the choice of the parameters $(\alpha, \beta)$. We recall the null geodesic equation corresponding to the line element in eqn. ~\eqref{eq:bh_metric} to be:
\begin{equation}\label{eq_geodesic}
g_{\mu\nu}\dot{x}^\mu \dot{x}^\nu = 0,
\end{equation}
where  the dot  indicates a derivative with respect to the affine parameter.
\begin{figure}[t!]
	
	{\centering
		
		\subfloat[]{\includegraphics[width=2.7in]{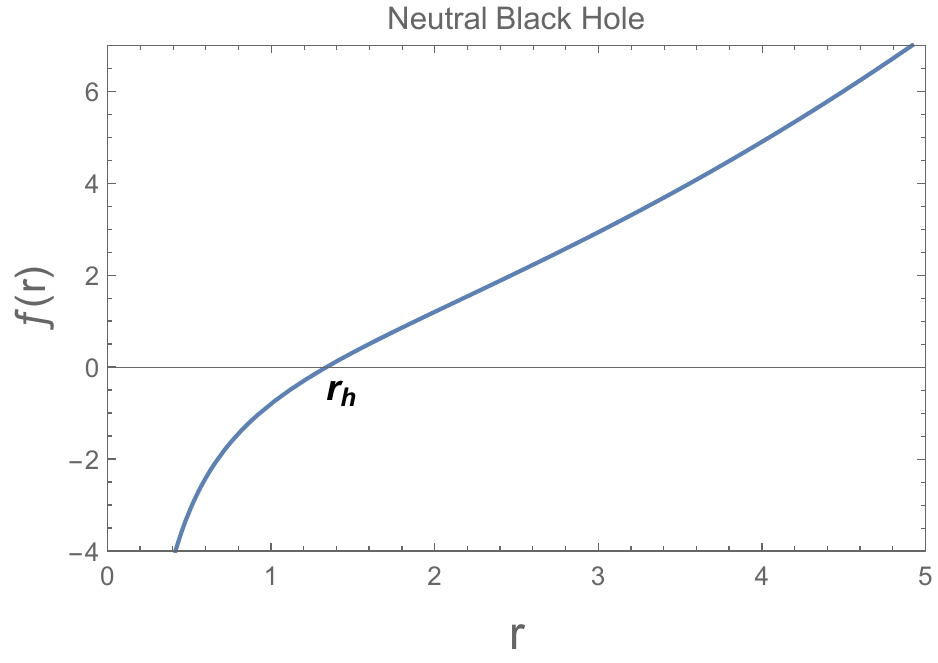}}\hspace{1.4cm}	
		\subfloat[]{\includegraphics[width=2.7in]{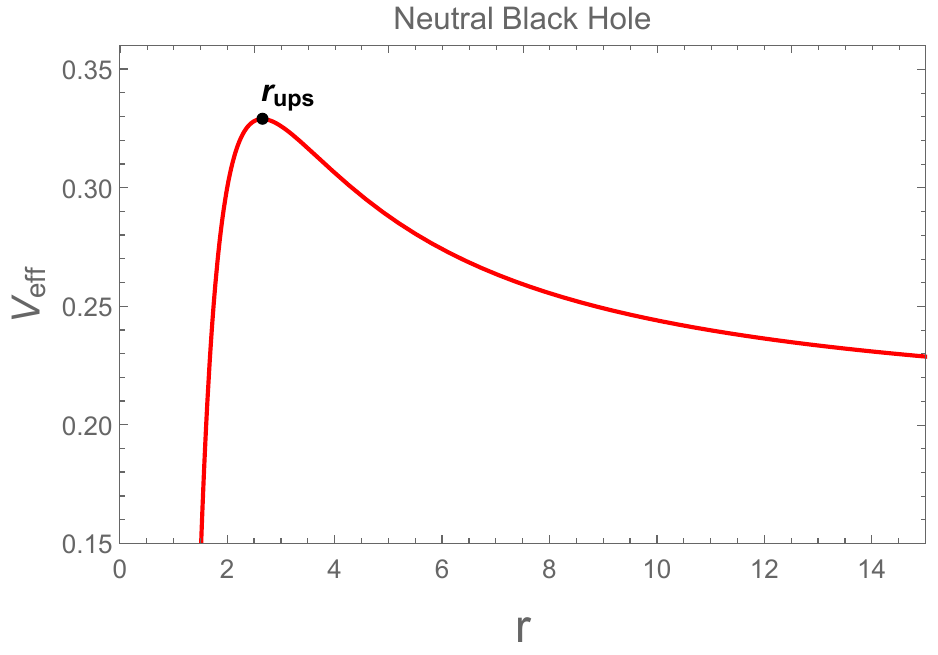}}\hspace{1.4cm}	
		\subfloat[]{\includegraphics[width=2.7in]{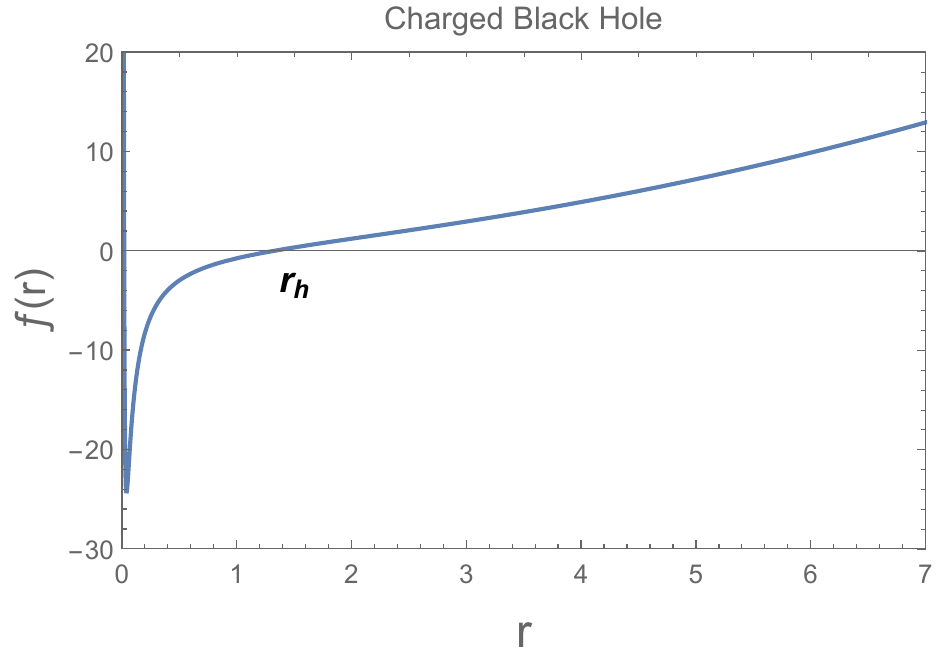}}\hspace{1.4cm}	
		\subfloat[]{\includegraphics[width=2.7in]{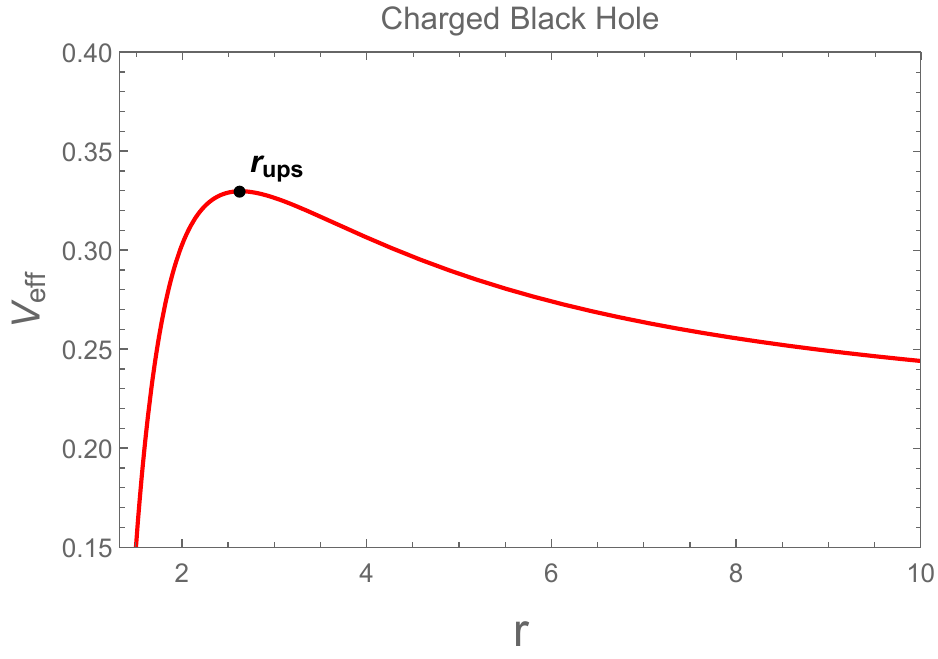}}	
		
		\caption{\footnotesize Left: behavior of the lapse function $f(r)$. Right:  effective potential~\eqref{eq:veff_photons}  showing the existence of single unstable photon sphere (ups) at $r_{\text{ups}}$ outside the event horizon. Top row:  for a neutral black hole with  single horizon at $r_h = 1.3387$ and $r_{\text{ups}} = 2.6533$ (here, we set $\alpha =-1, \beta =0.2$). Bottom row:  for a charged black hole with two horizons (inner and outer) where the outer horizon is located at $r_h = 1.3277$  and $r_{\text{ups}} = 2.635$  (here, we set $ Q=0.2, \alpha =-1, \beta =0.2$).}
		\label{fig:fv_neucha_1ps}	}
	
\end{figure}
\vskip 0.2cm \noindent
Considering further, the equatorial null geodesic equations with $\theta = \pi/2$, and using the existence of conserved quantities (such as the energy $ E = - g_{tt}\dot{t}$, and the orbital angular momentum $L = g_{\varphi \varphi} \dot{\varphi}$ of the photon)  corresponding  to the symmetries of the spacetime, the  equation~\eqref{eq_geodesic} yields:
\begin{equation}
\dot{r}^2 + V_{\text{eff}} \, =  E^2,
\end{equation}
with the effective potential $V_{\text{eff}}$ turning out to be,
\begin{equation}\label{eq:veff_photons}
 V_{\text{eff}} = f(r)\bigg(\frac{L^2}{r^2} \bigg).
\end{equation}
On solving $V_{\text{eff}} = E^2 \quad \text{and} \quad  V^{'}_{\text{eff}}=0$, where the prime indicates the derivative with respect to $r$, we can obtain the photon sphere (PS) at $r_{\text{ps}}$. 
Further, a PS is deemed to be stable (unstable) if the corresponding effective potential obeys $V^{''}_{\text{eff}}(r_{\text{ps}}) \, > (<) \, 0 $. 
\begin{figure}[t!]
	
	{\centering
		
		\subfloat[]{\includegraphics[width=2.7in]{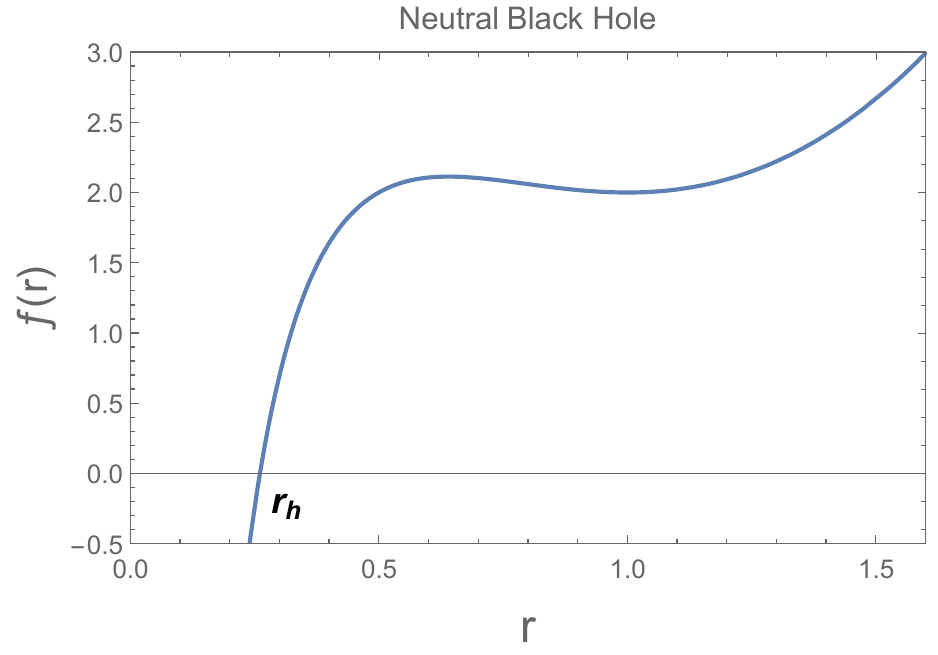}}\hspace{1.4cm}	
		\subfloat[]{\includegraphics[width=2.7in]{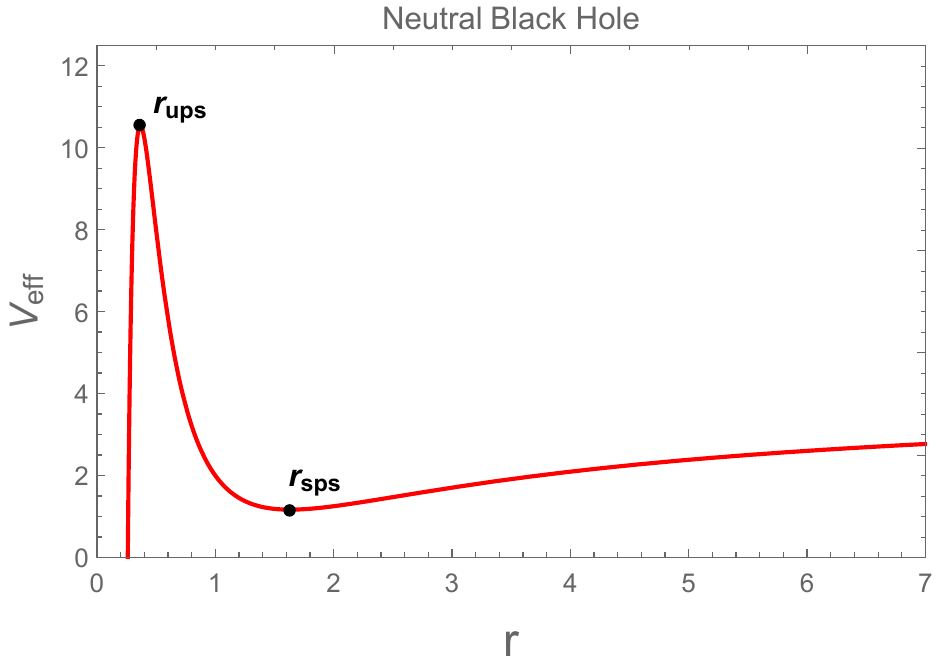}}\hspace{1.4cm}	
		\subfloat[]{\includegraphics[width=2.7in]{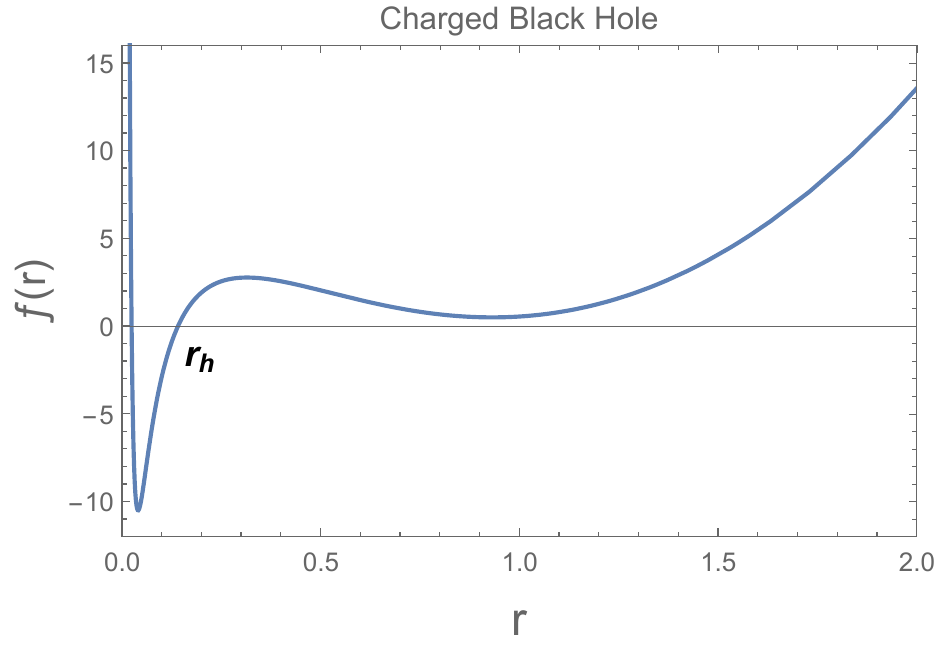}}\hspace{1.4cm}	
		\subfloat[]{\includegraphics[width=2.7in]{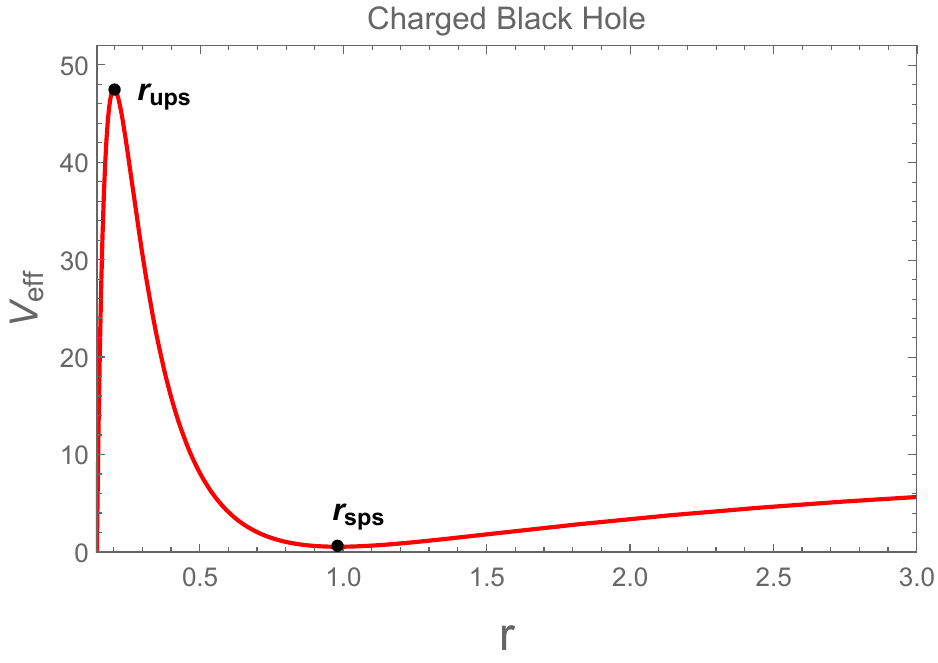}}	
		
		\caption{\footnotesize Left: behavior of the lapse function $f(r)$. Right:  effective potential~\eqref{eq:veff_photons}  showing the existence of a pair of unstable photon sphere (ups) and stable photon sphere (sps) at $r_{\text{ups}}$ and  $r_{\text{sps}}$, respectively, outside the event horizon. Top row:  for a neutral black hole with  single horizon at $r_h = 0.2610$, $r_{\text{ups}} = 0.3675$ and $r_{\text{sps}} = 1.6325$ (here, we set $\alpha =0, \beta =3$). Bottom row:  for a charged black hole with two horizons (inner and outer) where the outer horizon is located at $r_h = 0.1414$, $r_{\text{ups}} = 0.2012$  and $r_{\text{sps}} = 0.9828$  (here, we set $ Q=0.2, \alpha =10, \beta =1.5$).}
		\label{fig:fv_neucha_2ps}	}
	
\end{figure}
\vskip 0.2cm
\noindent
Plotting the effective potential~\footnote{In this manuscript, we fix  the free parameters as $L=M=h=m_g =1.$ } in eqn.~\eqref{eq:veff_photons} for chosen values of parameters $(\alpha, \beta)$ in fig.~\ref{fig:fv_neucha_1ps}, we observe the existence of a single unstable PS outside the event horizon, in both the neutral and charged black hole case. This result is in agreement  with the observation in ref.~\cite{Wei:2020rbh},  which is that, there exists an odd number of PS's outside the static and spherically symmetric black holes.
\vskip 0.2cm
\noindent
However, in contrast to ref.~\cite{Wei:2020rbh}, we had earlier observed in ~\cite{Yerra:2024stj}, that there exist certain regions in the parameter space of  $(\alpha, \beta)$, for black holes in massive gravity (both the neutral and charged cases), where one can also have even number of photon spheres outside the event horizon~\footnote{We note here that, although ref.~\cite{Hendi:2022qgi} indicated the existence of two PSs for the charged black hole case, their analysis however is not sufficient to conclude on the presence of these PS's,outside the black hole.}. In fact, as we show in fig.~\ref{fig:fv_neucha_2ps}, in massive gravity theories considered here, there can be two PS's in general. Interestingly, there also exist other choices of parameters $(\alpha,\beta)$ for the black holes in massive gravity, where there may not be any photon spheres outside the event horizon at all, as shown in fig.~\ref{fig:fv_neucha_0ps}.
\noindent
Although, a comprehensive analysis of the photon spheres in the full range of the parameter space of $(\alpha,\beta)$ is important, this is not straightforward due to the presence of several free parameters, such as $M,Q,m_g,h,\alpha,\beta$. We report some progress using numerical techniques in the sequel. For the moment, it is helpful to proceed forthwith by studying the topological properties of the PS's, and bringing out the novel features in massive gravity theories, as compared to the situation in standard Einstein gravity case. 

\subsection{Topological Charge} \label{3.1}

To set up the study of the topological properties associated with the photon spheres, we follow~\cite{Wei:2020rbh,Cunha:2020azh,Cunha:2017qtt}, and start by analysing the properties of the everywhere regular potential function,
\begin{equation}\label{eq:new_eff_potential_H}
H(r, \theta) = \sqrt{\frac{-g_{tt}}{g_{\varphi \varphi}}} = \frac{1}{\text{sin}\theta} \bigg( \frac{f(r)}{r^2} \bigg)^{\frac{1}{2}}\, .
\end{equation} 
The purpose of this function is to define the vector field $\phi (\phi^r, \phi^\theta)$, in the following way:
\begin{equation}
\phi^r = \frac{\partial_r H}{\sqrt{g_{rr}}} = \sqrt{f(r)} \partial_r H , \quad \phi^\theta = \frac{\partial_\theta H}{\sqrt{g_{\theta \theta}}} = \frac{\partial_\theta H}{r}.
\end{equation}
\begin{figure}[t!]	
 	{\centering

 		\subfloat[]{\includegraphics[width=2.5in]{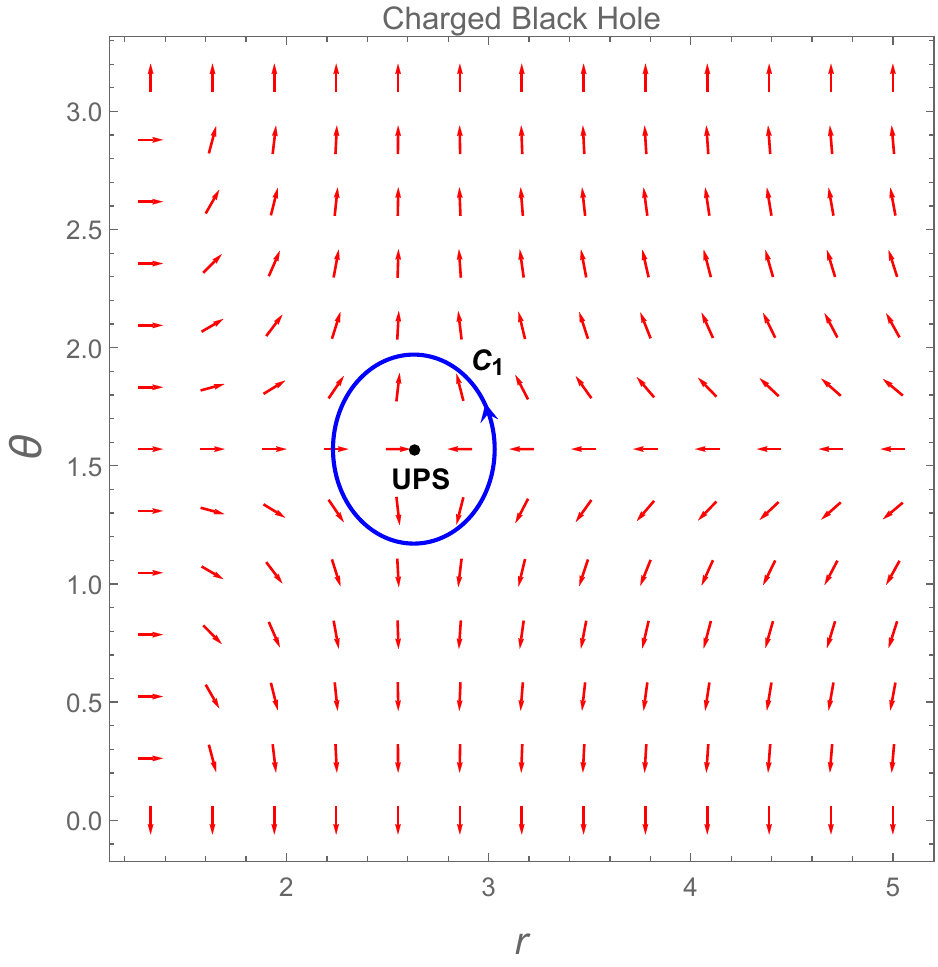}}\hspace{1.4cm}
 		\subfloat[]{\includegraphics[width=2.5in]{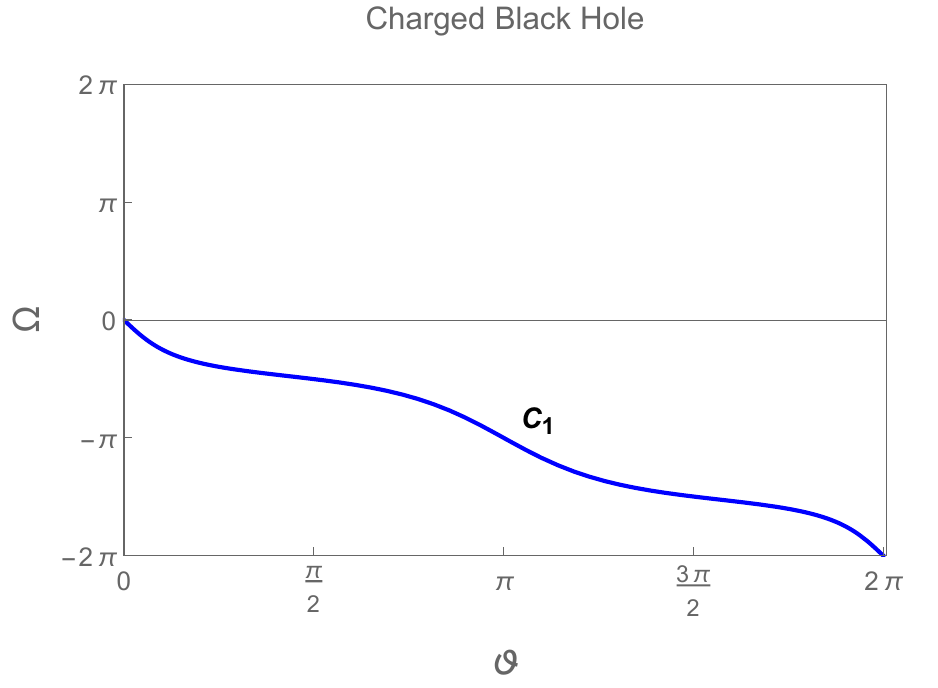}}\hspace{1.4cm}	
 		\subfloat[]{\includegraphics[width=2.5in]{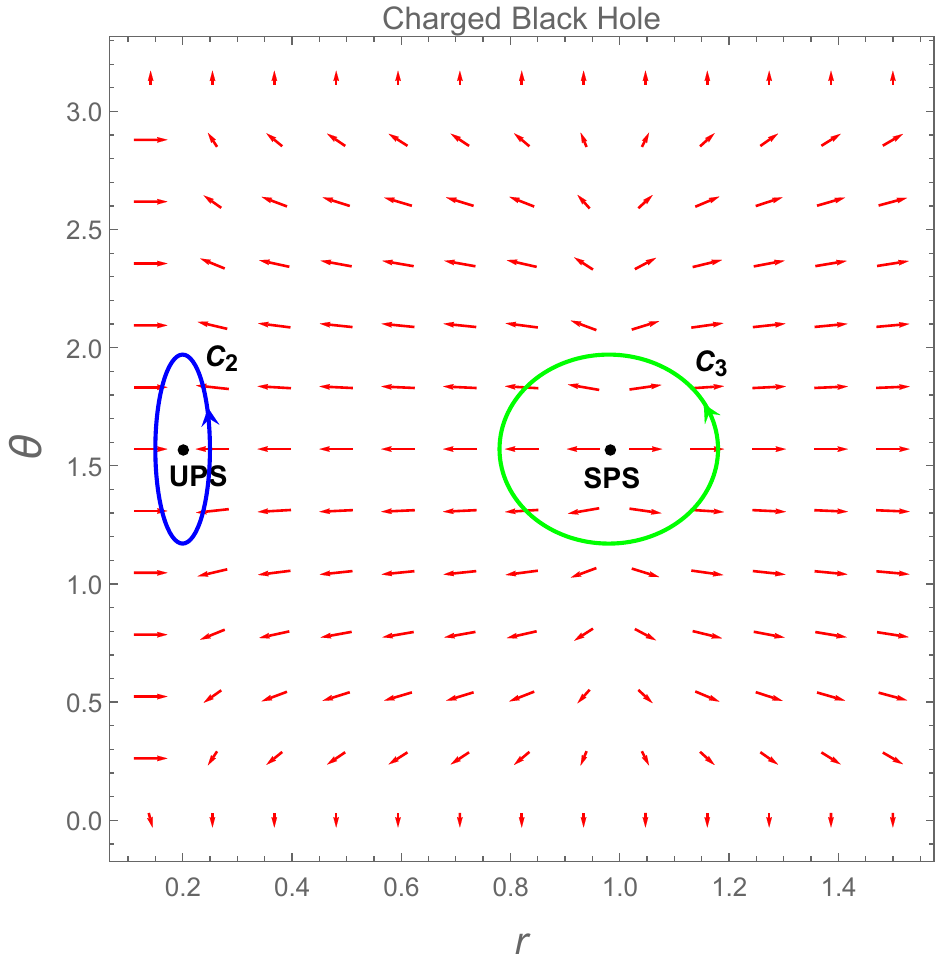}}\hspace{1.4cm}
 		\subfloat[]{\includegraphics[width=2.5in]{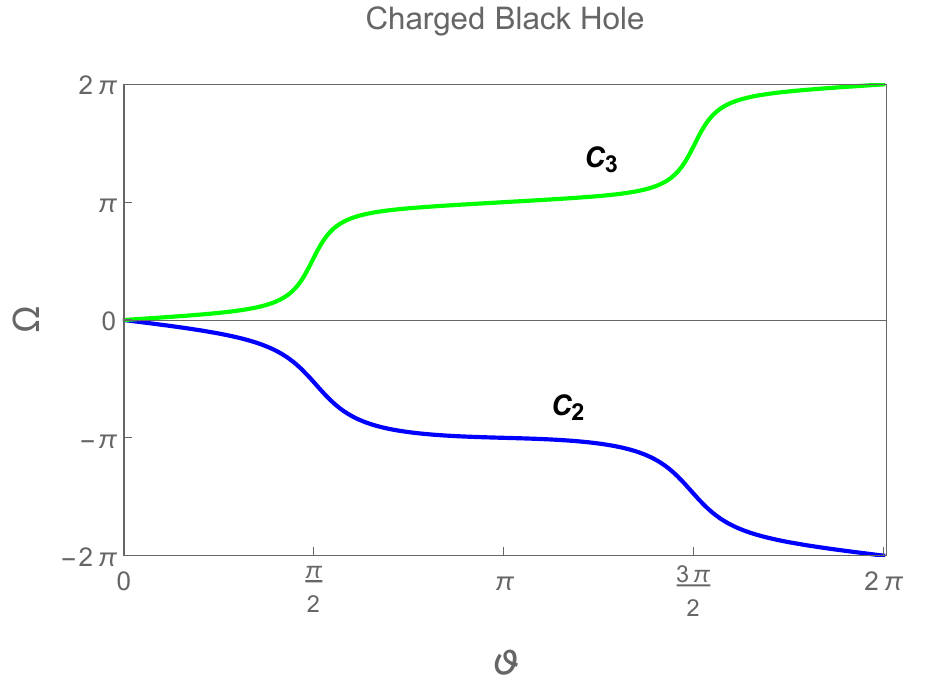}}
 		\caption{\footnotesize Left:  the normalised vector field $n$, in $\theta -r$ plane, showing that the points where it vanishes correspond exactly to the location of photon spheres.   Right:  the deflection angle $\Omega (\vartheta)$  of the vector field, giving the winding number -1 for the unstable photon sphere (UPS), and +1 for the stable photon sphere (SPS). Top row: for charged black hole with one photon sphere. 	Bottom row: for charged black hole with two photon spheres. (here,  the circular contour $C_i$ is  parameterised by the angle $ \vartheta \in (0, 2\pi) $, with  $(r, \theta) = (a\cos\vartheta+r_0, b\sin\vartheta+\frac{\pi}{2} )$. The chosen values are $(r_0, a, b) = (2.63, 0.4 ,0.4)$ for $C_1, (0.2, 0.05, 0.4)$ for $C_2, (0.98, 0.2, 0.4)$ for $C_3$). Similar results can be obtained for neutral black hole as well (not shown).  \label{fig:topology12ps}}
 	}	
 \end{figure}
\vskip 0.2cm 
\noindent
 The location of the photon spheres can now be obtained through the zero points of this vector field $\phi$. Following Duan's $\phi$-mapping procedure~\cite{Duan:1984ws,Duan:2018rbd}, one can assign a  topological charge $Q_t$ (called the winding number $w$) for each photon sphere, by analysis of a conserved topological current $j^\mu$ (satisfying $\partial_{\mu}j^{\mu}=0$), which is given as~\cite{Wei:2020rbh,Cunha:2020azh,Cunha:2017qtt}:
 \begin{eqnarray}
 j^{\mu}=\frac{1}{2\pi}\epsilon^{\mu\nu\rho}\epsilon_{ab}\partial_{\nu}n^{a}\partial_{\rho}n^{b},
 \quad \mu,\nu,\rho=0,1,2, \, \, \text{and} \, \, a, b = 1,2,
 \end{eqnarray}
 where  $n=(\frac{\phi^r}{||\phi||}, \frac{\phi^\theta}{||\phi||})$ denotes the normalized vector field.
  \noindent 
The definition of topological charge  ensues from the above construction as~\cite{Wei:2020rbh,Cunha:2020azh,Cunha:2017qtt} 
\begin{equation}
Q_t   =\int_\Sigma j^0 d^2x  = \Sigma_{i=1} w_i, 
\end{equation}
contained in a region $\Sigma$. Here, $w_i$ denotes the winding number of the $i^{\rm th}$ point corresponding to zero of $\phi$. 
In order to compute the winding number $w_i$, we consider a piecewise smooth and positively oriented closed curve $C_i$ that encloses the $i^{\rm th}$ zero point of the vector field $\phi$. 
Then, one can obtain  the topological charge (i.e.,winding number $w_i$) by measuring the deflection $\Omega$ of the vector field $\phi$ along the  contour $C_i$ as
\begin{equation}
 w_i = \frac{1}{2\pi} \oint_{C_{i}} d\Omega, \quad \text{where}\, \, \Omega= \text{arctan}(\phi^2 / \phi^1).
\end{equation}
 \noindent
 We plot the normalized vector field $n$ in fig.~\ref{fig:topology12ps}, showing that the points where it vanishes correspond exactly to the location of the photon spheres. The computation of the deflection angle $\Omega$ gives the winding number $-1$ for the unstable photon sphere, and $+1$ for the stable photon sphere. Thus, the total topological charge (TTC) would be $-1$ for black holes with a single photon sphere, and zero for those with two photon spheres.

\vskip 0.2cm  \noindent
Further,  the  vector field $n$ for both neutral and charged black holes with no  photon spheres, plotted in fig.~\ref{fig:topology0ps}, shows qualitatively its non-vanishing nature. We go ahead and compute the total topological charge associated with the black holes with no PS's, by considering a rectangular contour $C$ (as shown in fig.~\ref{fig:topology0ps}) that encloses the entire parameter space outside the black hole, as follows:  

\begin{figure}[t!]	
	{\centering

		\subfloat[]{\includegraphics[width=2.4in]{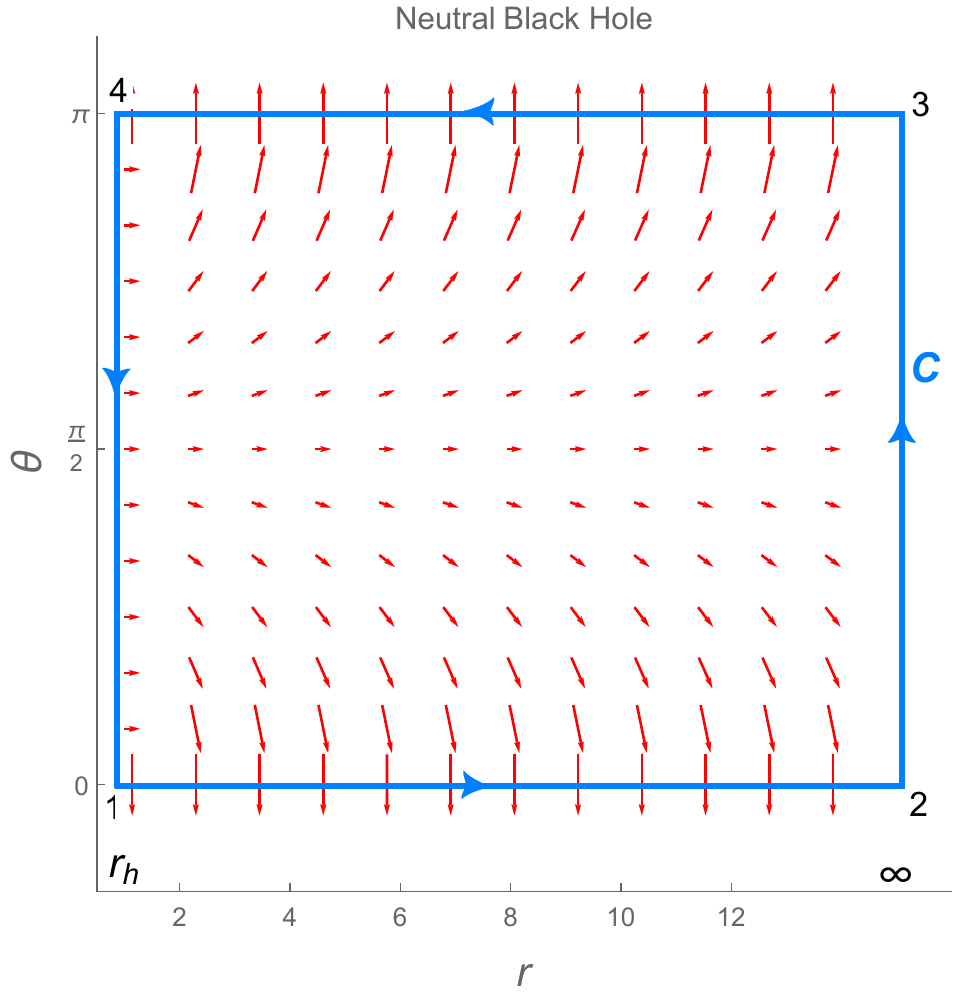}}\hspace{1.4cm}
		\subfloat[]{\includegraphics[width=2.4in]{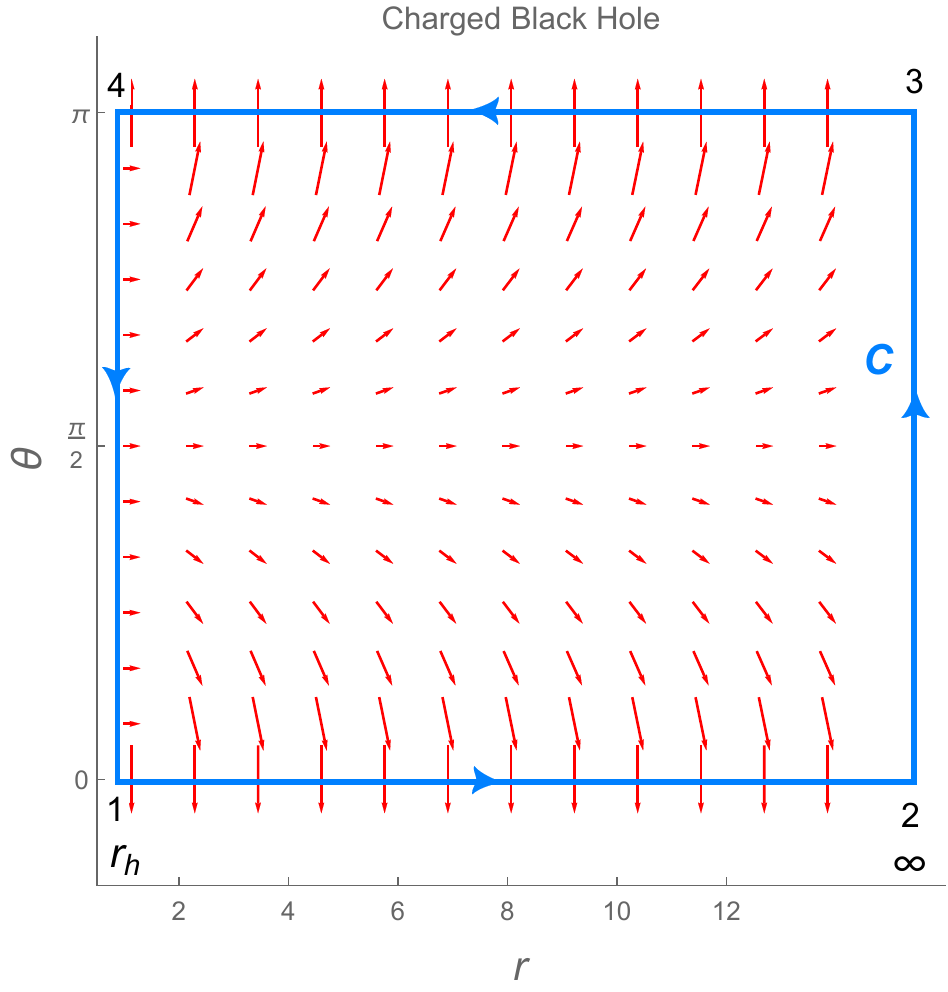}}	
		\caption{\footnotesize   The normalised vector field $n$, in $\theta -r$ plane, showing its non-vanishing behavior.   The  rectangular contour $C$,  denotes the entire boundary of the	parameter space outside the black hole.  (a) for neutral black hole with no photon spheres. (b) for charged black hole with no photon spheres. \label{fig:topology0ps}}
	}	
\end{figure}
\vskip 0.2cm 
\noindent
 \textbf{At $\theta = 0, \pi$:} Along the line segment $3 \rightarrow 4$ of the contour $C$,  the direction of the vector field $\phi$ is upwards (i.e., $\Omega =+\frac{\pi}{2}$) at $\theta= \pi$, while along the line segment $1 \rightarrow 2$, it is downwards (i.e., $\Omega = -\frac{\pi}{2}$) at $\theta= 0$. Thus,  we get $\int\limits_{r_{\rm h}}^{\infty} d\Omega \Big\vert_{ \theta =0} = \int\limits_{\infty}^{r_{\rm h}} d\Omega \Big\vert_{ \theta =\pi}= 0 $ (as $d\Omega \big\vert_{\theta =0, \pi} = 0$).
 \noindent
\par \textbf{At $r = r_{\rm h}$:} We can see that along the line segment $ 4 \rightarrow 1$ of the contour $C$,  the direction of the vector field $\phi$ at $r=r_{\rm h}$, is rightward.

\vskip 0.2cm 
\noindent
Since, along the line segment $4 \rightarrow 1$, the vector field  is changing the direction from upwards (i.e., $\Omega = +\frac{\pi}{2}$) at $\theta = \pi$ to downwards  (i.e., $\Omega = -\frac{\pi}{2}$) at $\theta = 0$, in a clockwise direction, we get 
\begin{equation}
\int\limits_{\pi}^{0} d\Omega \Big\vert_{ r =r_{\rm h}} = \Omega \Big\vert_{\theta =0}-\Omega \Big\vert_{\theta =\pi} = -\frac{\pi}{2}-\frac{\pi}{2}=-\pi. 
\end{equation}
\noindent
\par \textbf{At $r =\infty$:} Along the line segment $2 \rightarrow  3$ of the contour $C$,  the direction of the vector field $\phi$ at $r \rightarrow \infty$, is rightward.

\noindent
Since, along the line segment $2 \rightarrow 3$, the vector field  is changing the direction from downwards (i.e., $\Omega = -\frac{\pi}{2}$) at $\theta = 0$ to upwards  (i.e., $\Omega = +\frac{\pi}{2}$) at $\theta = \pi$, in a counterclock wise direction, we get
\begin{equation}
\int\limits_{0}^{\pi} d\Omega \Big\vert_{ r =\infty} = \Omega \Big\vert_{\theta =\pi}-\Omega \Big\vert_{\theta =0} = \frac{\pi}{2}+\frac{\pi}{2}= +\pi. 
\end{equation}
Therefore,  the total topological charge associated with the black holes with no photon spheres, as expected, would be   
\begin{equation}
Q_t=\frac{1}{2\pi} \oint_C d\Omega = \frac{1}{2\pi} \bigg(\int\limits_{r_{\rm h}}^{\infty} d\Omega \Big\vert_{ \theta =0} + \int\limits_{0}^{\pi} d\Omega \Big\vert_{ r =\infty} +\int\limits_{\infty}^{r_{\rm h}} d\Omega \Big\vert_{ \theta =\pi} + \int\limits_{\pi}^{0} d\Omega \Big\vert_{r =r_{\rm h}}\bigg) = 0.
\end{equation}
\noindent
Thus, both neutral and charged black holes in massive gravity, possessing two PSs (UPS and SPS) and no PSs, are endowed with the total topological charge $Q_t = 0$, and hence they belong to same topological class.
On the otherhand, both neutral and charged black holes in massive gravity, possessing one PS (UPS), have the  total topological charge $Q_t = -1$, and hence they belong to a different topological class. 
	\begin{table}[t!]\caption{Notations for PS based on Topological charge}\label{table:definitions} 
		\centering{		
			\begin{tabular}{|m{2.5 cm} |m{4.5 cm} |m{2.5 cm}| m{2.2 cm}|} 
					\hline
			\vskip 0.3cm	Conditions for PS existence & \quad Stability criterion for PS &  Topological  charge & Naming convention \\ \hline
			\multirow{4}{2cm \vskip 0.15cm }  {$V_{\text{eff}} = E^2,$   \text{\quad and } \quad $ V^{'}_{\text{eff}}=0.$}  & \vskip 0.3cm Unstable, if  $V^{''}_{\text{eff}}(r_{\text{ps}})  <  \, 0 $. & \vskip 0.3cm \quad $-1$ & \vskip 0.3cm Standard PS\\ 
			 & &  &  \\ \cline{2-4} 
			  & \vskip 0.3cm  Stable, if $V^{''}_{\text{eff}}(r_{\text{ps}})  > \, 0 $.  & \vskip 0.3cm \quad $+1$ & \vskip 0.3cm Exotic PS
			  \\
			   & &  &  \\ \hline

			\end{tabular} }
		\end{table}
\vskip 0.2cm
\noindent
The notations which are now standard in literature on the classification of topological charges and the naming conventions for PS's are given in the table-\ref{table:definitions}.   The PS carrying the topological charge $+1$ is named exotic, as it is generally associated with the horizonless ultra-compact objects (UCOs) which are assumed to form with some sort of exotic matter with unknown mechanism~\cite{Cunha:2017qtt,Junior:2021svb,Cunha:2020azh,Cunha:2022gde,Wei:2020rbh}.

\begin{figure}[t!]	
	{\centering

		\subfloat[]{\includegraphics[width=2.7in]{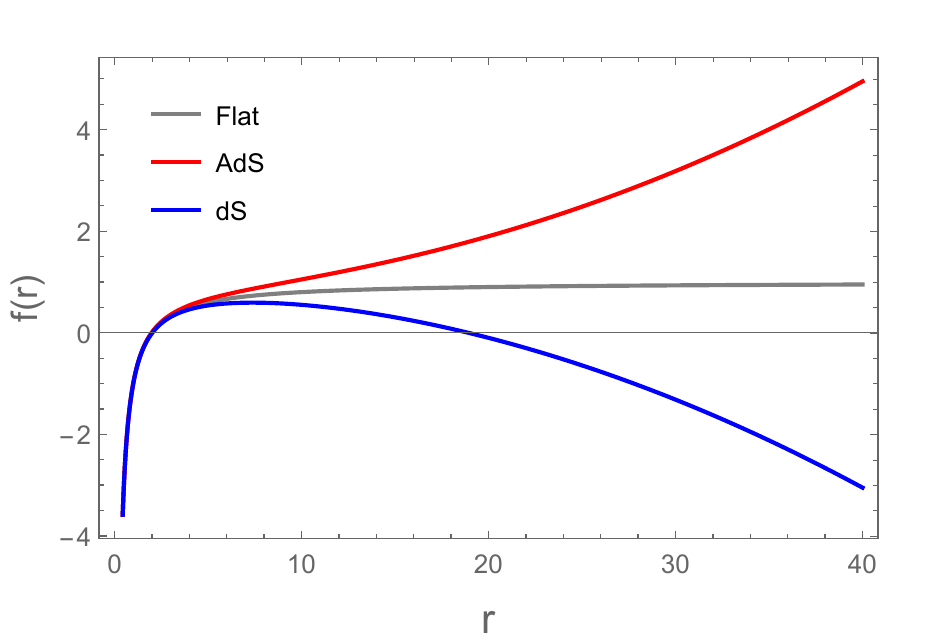}}\hspace{1.4cm}
		\subfloat[]{\includegraphics[width=2.7in]{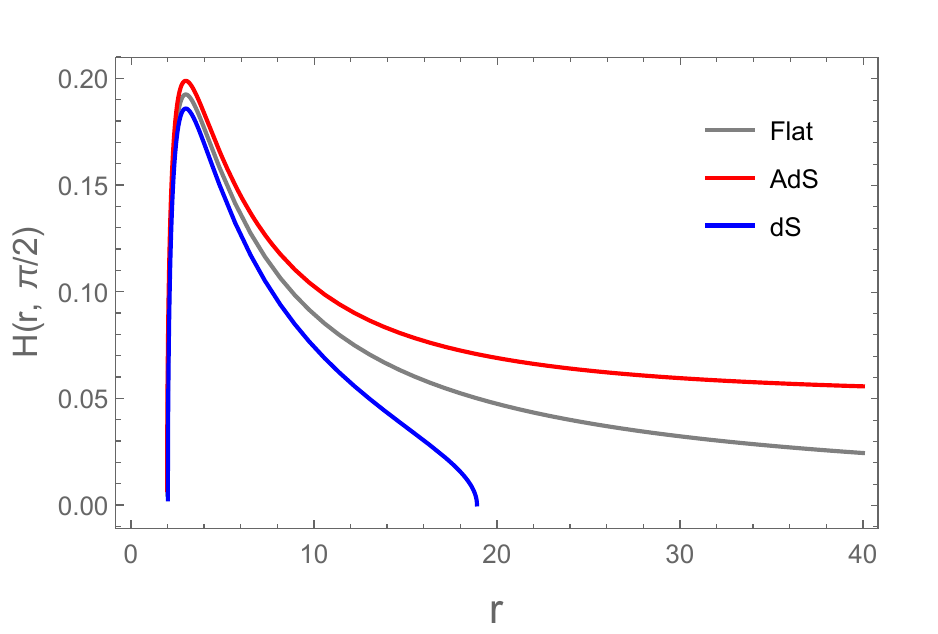}}	
			\caption{\footnotesize For black holes in asymptotically flat, AdS and dS spacetimes (a)  The lapse function $f(r)$ shows different asymptotic behaviors. (b) The effective potential function $H (r, \theta)$  shows same asymptotic behavior. (here, plots  are drawn for Schwarzschild black hole in flat, AdS, dS spacetimes in standard Einstein gravity as an example). \label{fig:analysis_einstein}}
	}	
\end{figure}

\begin{figure}[t!]	
	{\centering

		\subfloat[]{\includegraphics[width=2in]{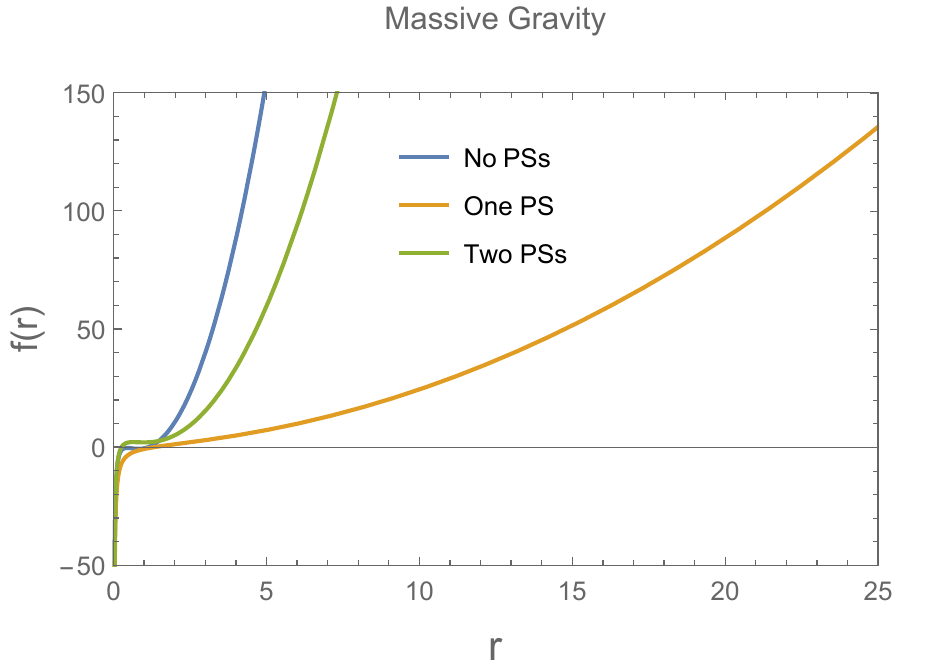}}\hspace{0.2cm}
		\subfloat[]{\includegraphics[width=2in]{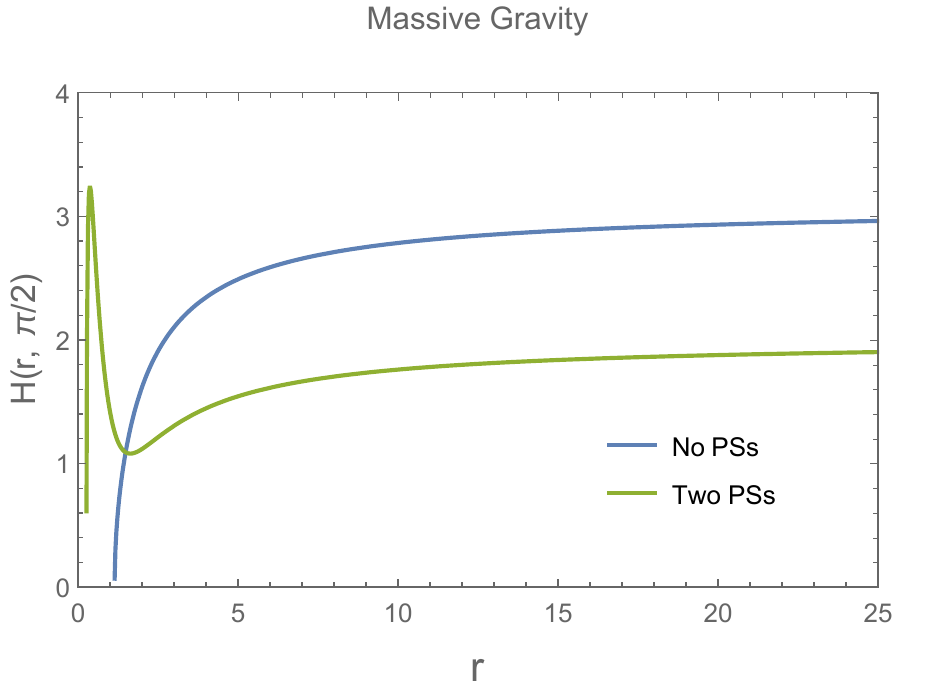}}\hspace{0.2cm}
		\subfloat[]{\includegraphics[width=2in]{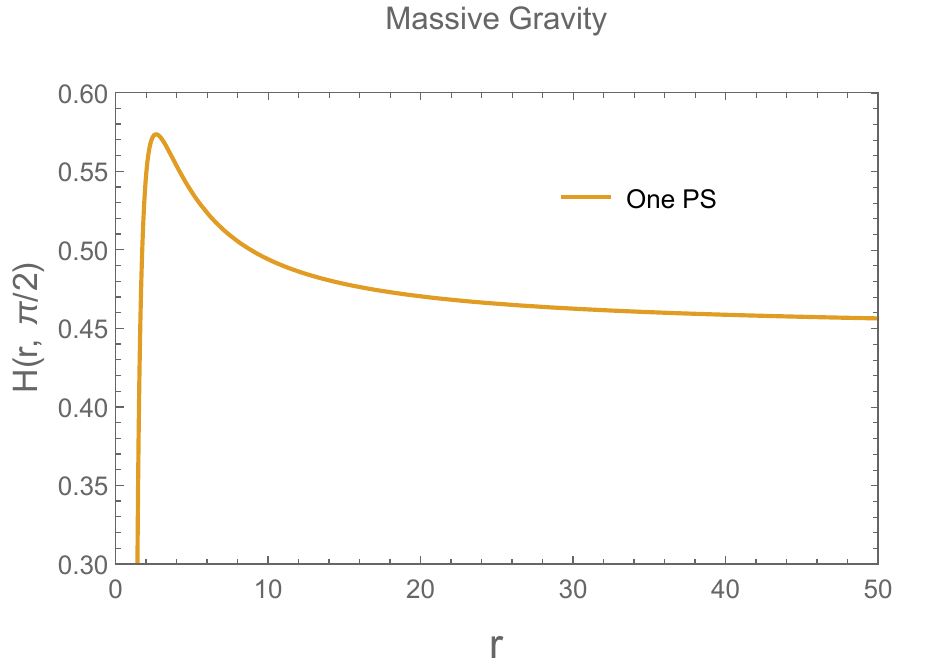}}
		\caption{\footnotesize for the  black holes in massive gravity with one PS, two PSs, and zero PSs. (a)  The lapse function $f(r)$   shows same asymptotic behavior.  The effective potential function $H (r, \theta)$ shows different asymptotic behaviors. (b) increasing behavior for two and zero PSs. and (c) decreasing behavior for one PS. (Here, plots  are shown for  neutral case. Similar behavior exists for charged case as well).
		 \label{fig:analysis_massive}}
	}	
\end{figure}

\vskip 0.2 cm \noindent 
Having classified the topological classes of stable and unstable PS's, we can now try to interpret the reason for the black holes in massive gravity possessing total topological charge $Q_t = 0$, in contrast to the value $Q_t = -1$ found for the PS's for black holes in standard Einstein gravity in ref.~\cite{Wei:2020rbh}. We first note that, for the various classes of black holes considered in ref.~\cite{Wei:2020rbh}, the asymptotic behaviour of the effective potential potential $H (r,\theta)$ (computed from the definition in equation~\eqref{eq:new_eff_potential_H}) is pretty similar. In other words, $H (r,\theta)$ for general black holes (irrespective of the asymptotic behaviour of the background geometry) of ref.~\cite{Wei:2020rbh}, plotted in
fig.~\ref{fig:analysis_einstein}(b) decreases as $r \rightarrow \infty$, although the  behaviour of the lapse function is quite different for asymptotically flat, AdS, or dS cases (see
fig.~\ref{fig:analysis_einstein}(a)).  Since, the topological classification relies on the behaviour of $H (r,\theta)$, one has the result that, the PS's for the black holes considered in ref.~\cite{Wei:2020rbh} always belong to the same topological class, with the total charge $Q_t = -1$. Contrast this with the situation in the massive gravity theories considered here. From fig.~\ref{fig:analysis_massive}(a), we conclude that for the massive gravity theories considered in this study, particularly, for black holes having zero, one or two PS's, the asymptotic behaviour of the lapse function is pretty similar. Yet, the asymptotic behaviour of the potential function $H (r,\theta)$ is quite different.
\vskip 0.2 cm \noindent 
 For the case of even number of PS's (i.e., zero or two), $H (r,\theta)$ shows an increasing behaviour from fig.~\ref{fig:analysis_massive}(b), whereas, for the case of odd number of PS (one, in this case), it shows a decreasing behaviour as seen in fig.~\ref{fig:analysis_massive}(c) (similar to the case in fig.~\ref{fig:analysis_einstein}(b)).     
Therefore, the asymptotic behavior of the $H (r, \theta) $ plays a key role  (rather than the asymptotic geometry of the background ) in deciding the total topological charge of the black holes PS's.  The result is that, the cases of black holes with even and odd number of PS's belong to distinct topological classes, as dictated by the computation of their total topological charge. 

\section{Photon sphere landscape}\label{4}
\begin{figure}[t!]	
	{\centering

		\subfloat[]{\includegraphics[width=2in]{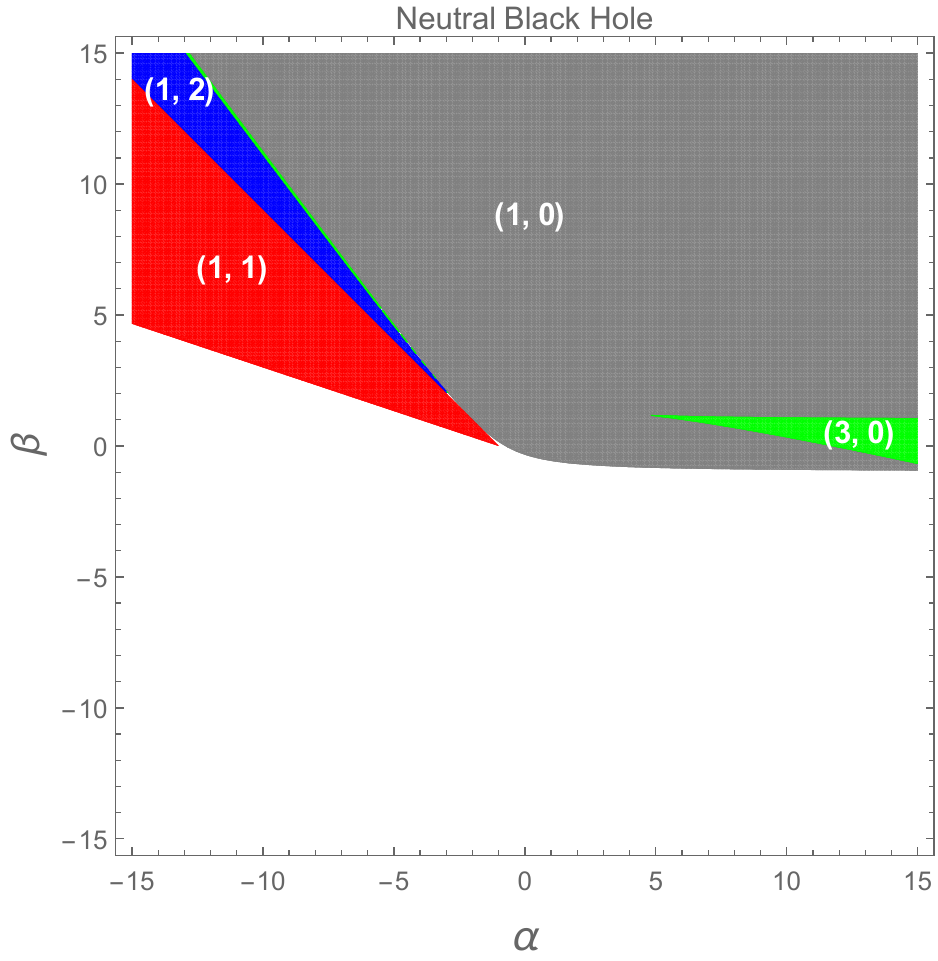}}\hspace{0.2cm}
		\subfloat[]{\includegraphics[width=2in]{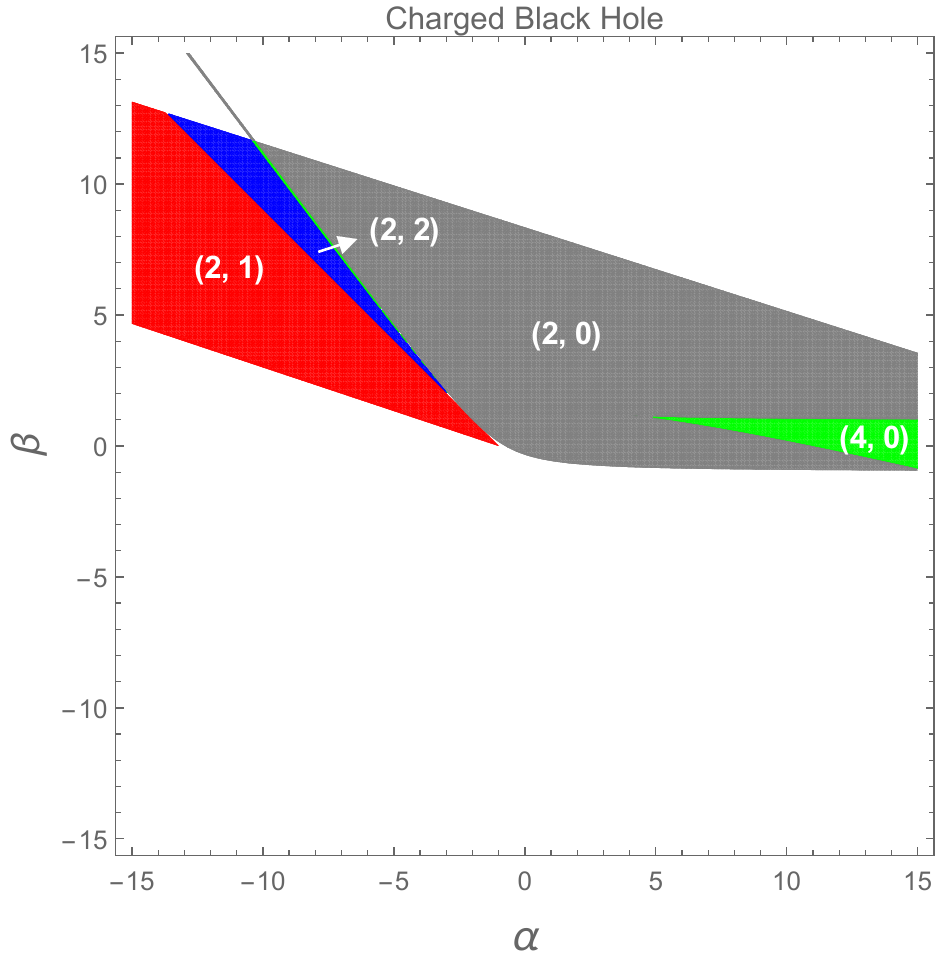}}\hspace{0.2cm}
		\subfloat[]{\includegraphics[width=2in]{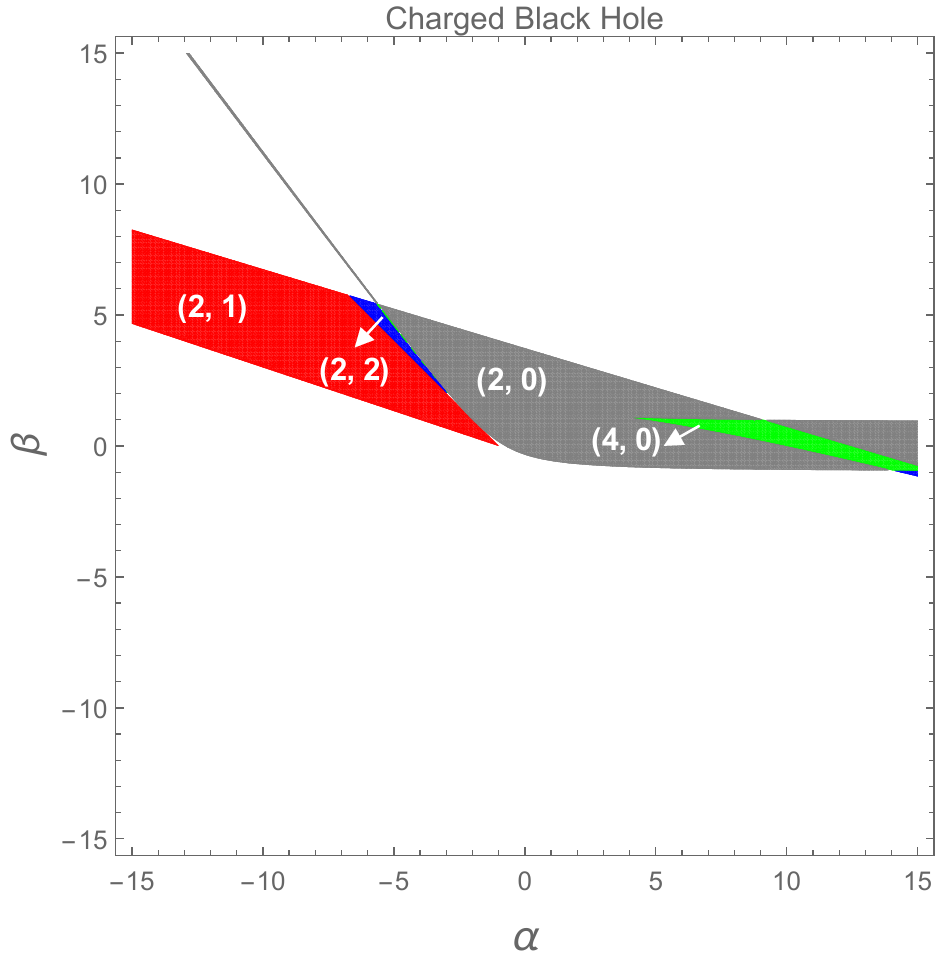}}
		\caption{\footnotesize The sample parameterspace of $(\alpha, \beta)$ showing the  black holes satisfying $f^{'}(r_h)>0$,  with rich horizon structure. Each region is denoted with a parenthesis indicating the black holes, possessing number of black hole horizons and number of cosmological horizons, i.e., as (Number of black hole horizons, Number of cosmological horizons). (a)  Neutral black hole. (b)  Charged black hole with $Q=0.2$. (c)  Charged black hole with $Q=0.3$. The uncolored regions contain other objects such as extremal black holes, naked singularities and Nariai limits. \label{fig:bh_area}}
	}	
\end{figure}

\begin{figure}[t!]	
	{\centering

		\subfloat[]{\includegraphics[width=2in]{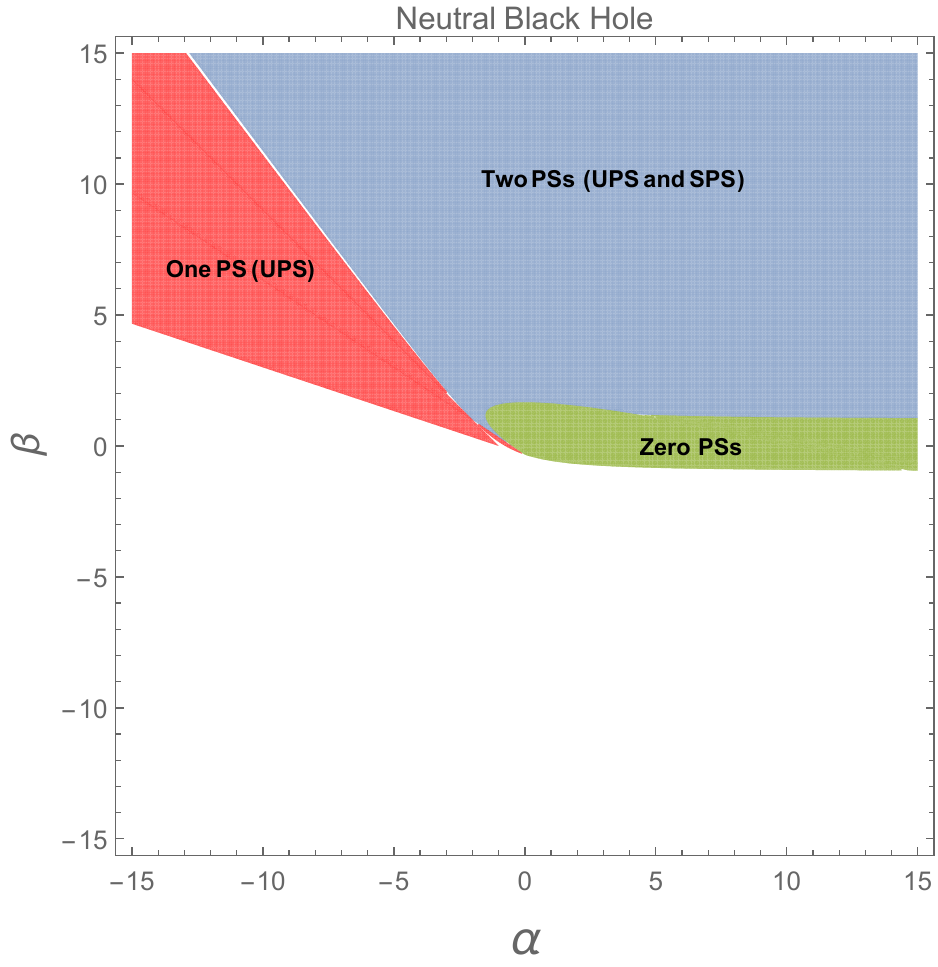}}\hspace{0.15cm}
		\subfloat[]{\includegraphics[width=2in]{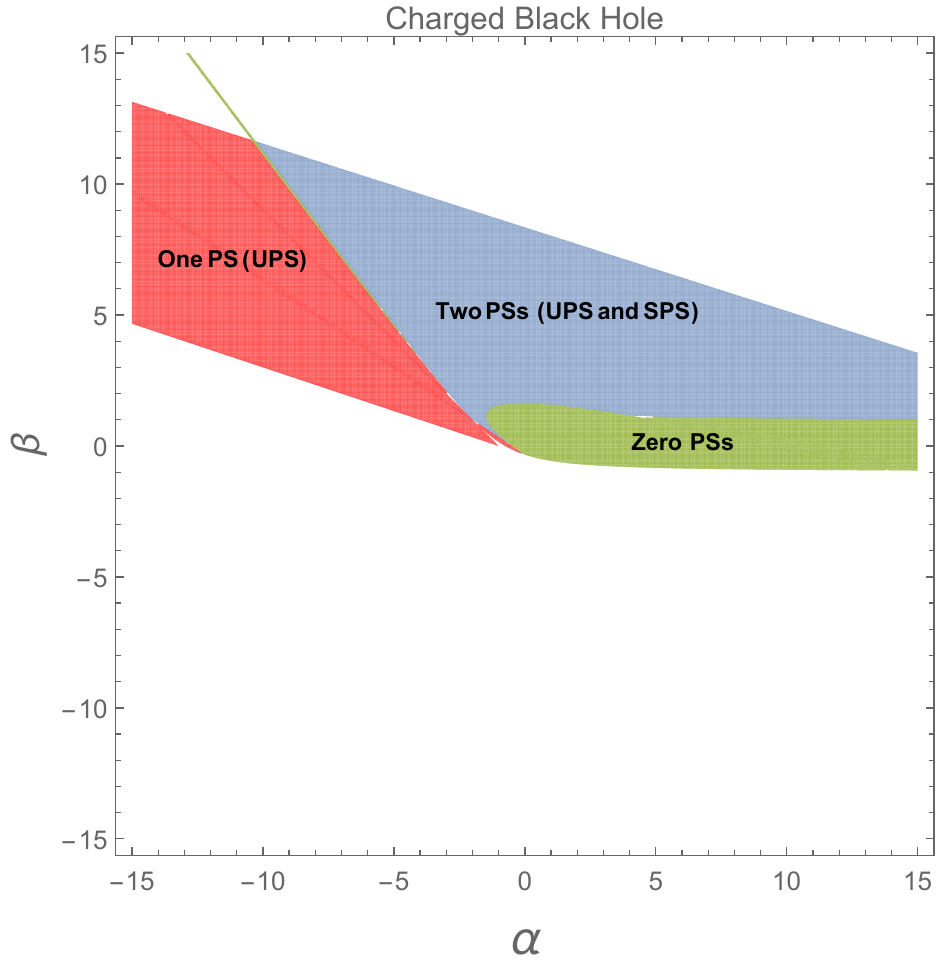}}\hspace{0.15cm}	
		\subfloat[]{\includegraphics[width=2in]{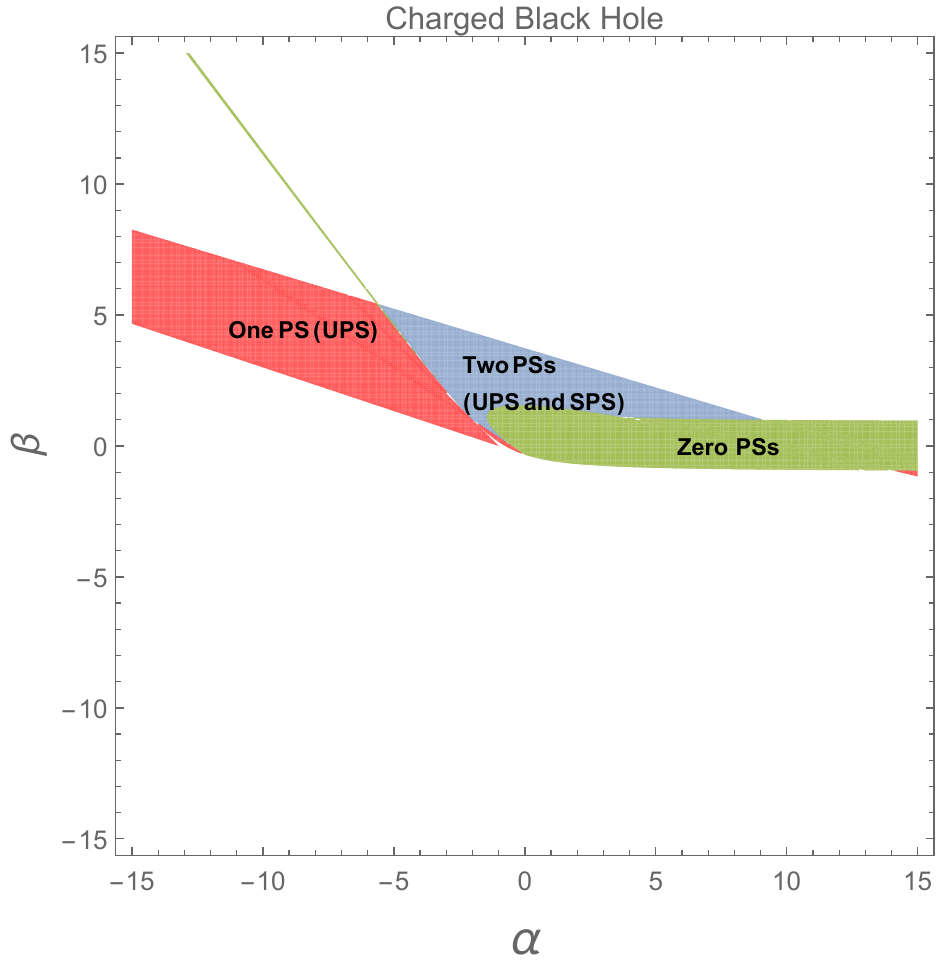}}
		\caption{\footnotesize The sample parameterspace of $(\alpha, \beta)$ showing the existence of two (unstable photon sphere (UPS) and stable photon sphere (SPS)), one (UPS), and zero photon spheres (PSs) outside the black hole. (a) Neutral black hole. (b) Charged black hole with $Q=0.2$. (c)  Charged black hole with $Q=0.3$. \label{fig:ps_area}}
	}	
\end{figure}

From the analysis thus far, we found that both neutral and charged black holes in massive gravity theories have rich horizon structure as well as different number of photon spheres outside the event horizon,  depending on how the values of $(\alpha, \beta)$ are chosen. In this section, our aim is make an attempt at identifying the possible parameter space of $(\alpha, \beta)$, where different number horizon exist, and within this range, how many number and type of PS's are present. Considering that these questions are not tractable analytically, we now fix the parameters of the black hole as before (i.e., $ M=h=m_g=1$) and proceed numerically.  Thus, within the parameter space of $(\alpha, \beta)$, we first find out the various possible horizons of black holes (fig.~\ref{fig:bh_area}) and then check the number of PS's possible (fig.~~\ref{fig:ps_area}). Collating the information from these two inputs, will allow us to present our results. 
\vskip 0.2cm 
\noindent
First, from the lapse function~\eqref{eq:f_gen} of the black holes, one can see the possibility of having at most four (three) horizons for the charged (neutral) case. Depending on the asymptotic geometry (flat, de Sitter or anti de Sitter), in addition to the black hole horizons, cosmological horizons could exist as well.
As the lapse function~\eqref{eq:f_gen} is quadratic in $r$ when $M=Q=0$, we can have at most two cosmological horizons for both neutral and charged black holes.   
 As we consider only the non-extremal~\footnote{For extremal black holes, PS coincides with the black hole horizon~\cite{Wei:2020rbh}.} black holes where $f^{'}(r_h) > 0 $, our landscape of the parameter space should exclude the other possible objects such as, naked singularities (NSs),  Nariai limit, and extremal black holes.
 \vskip 0.2cm 
\noindent
The neutral and charged black holes satisfying the above criteria can thus be identified with in  the parameter space of $(\alpha, \beta)$. We show this in fig.~\ref{fig:bh_area} and use the notation where the numbers in the round parenthesis are taken to be:  (the number of black hole horizons, number of cosmological horizons). \\

\noindent
For the case of neutral black holes, we find two possible cases as shown in fig.~\ref{fig:bh_area}(a) :

\begin{itemize}
\item 
One or three horizons with zero cosmological horizons (shown as $(1, 0)$ and $(3, 0)$ regions respectively).

\item One horizon in the presence of one or two cosmological horizons (shown as $(1, 1)$ and $(1, 2)$ regions respectively) as well.

\end{itemize}

\noindent
For charged black holes, there are again two possible cases, shown in figs.~\ref{fig:bh_area}(b) and  ~\ref{fig:bh_area}(c):

\begin{itemize}
\item 
Two or four horizons with zero cosmological horizons (shown as $(2, 0)$ and $(4, 0)$ regions respectively).

\item  Two horizons in the presence of one or two cosmological horizons (shown as $(2, 1)$ and $(2, 2)$ regions respectively) as well.

\end{itemize}

Next, there exist three (two) PSs for the charged (neutral) black holes, which can be seen by solving the equation $V^{'}_{\text{eff}}=0$. However, there is no guarantee that all these PS's exist outside the black hole~\footnote{If there are cosmological horizons in addition, the PS's should lie between the black hole and the inner cosmological horizon.}.
\vskip 0.2cm 
\noindent
From fig.~\ref{fig:ps_area}, we note that outside the black hole  there exist at most two PS's or none at all (corresponding to the case where all PS's either hide inside the black hole or the solutions are not real), for both neutral and charged black holes.  Note that the range and values in fig.~\ref{fig:ps_area} are in one-to-one correspondence with the respective plots in 
fig.~\ref{fig:bh_area} and by comparing the two, one arrives at the following conclusions:

\begin{itemize}
\item 
The neutral black holes possessing: two PS's, are located in the $( 1,0)$ region; one PS, are located in the $(1,0), (1,1)$, and $(1,2)$ regions; and zero PSs are located in the $(1,0)$ and $(3,0)$ regions. 

\item
While, the charged black holes possessing: two PSs are located in the $( 2,0)$ region; one PS are located in the $(2,0), (2,1)$, and $(2,2)$ regions; and zero PS's are located in the $(2,0)$ and $(4,0)$ regions.

\end{itemize}
\vskip 0.2cm 
\noindent
Further, the effects of the parameters, $\alpha$ and $\beta$, on the size of the black hole and  (stable and unstable) photon spheres can be seen from the fig.~\ref{fig:alpha, beta on rh} and fig.~\ref{fig:alpha, beta on rps}. For both neutral and charged black holes, the black hole size and also the unstable photon sphere size decrease as the parameter $\alpha$  (or $\beta$) increases. We note the following. While, the stable  photon sphere size decreases as $\alpha$  increases; as $\beta$ increases, its size can decrease or increase depending on the choice of $\alpha$.
\begin{figure}[t!]	
	{\centering

		\subfloat[]{\includegraphics[width=2.5in]{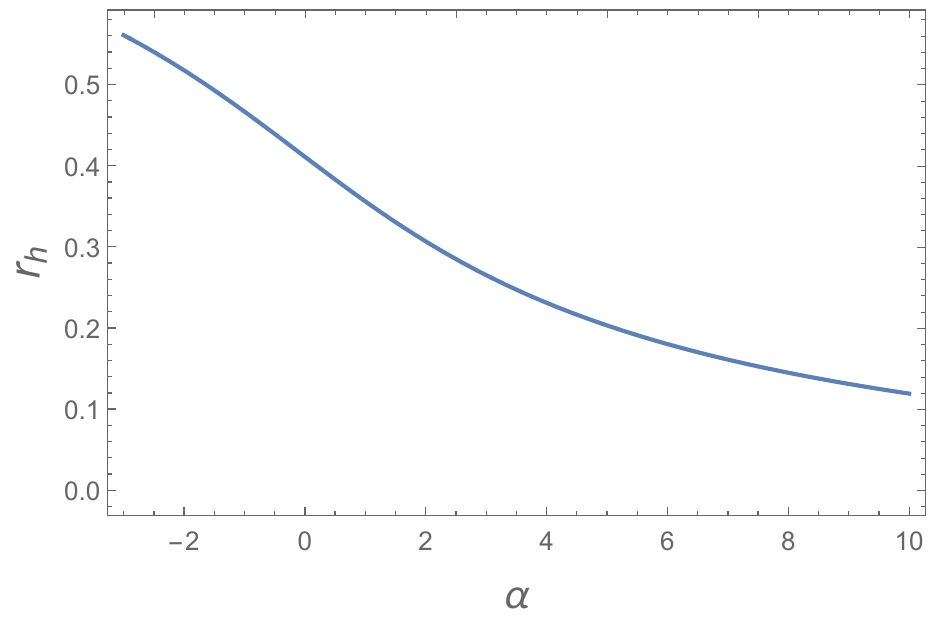}}\hspace{1cm}
		\subfloat[]{\includegraphics[width=2.5in]{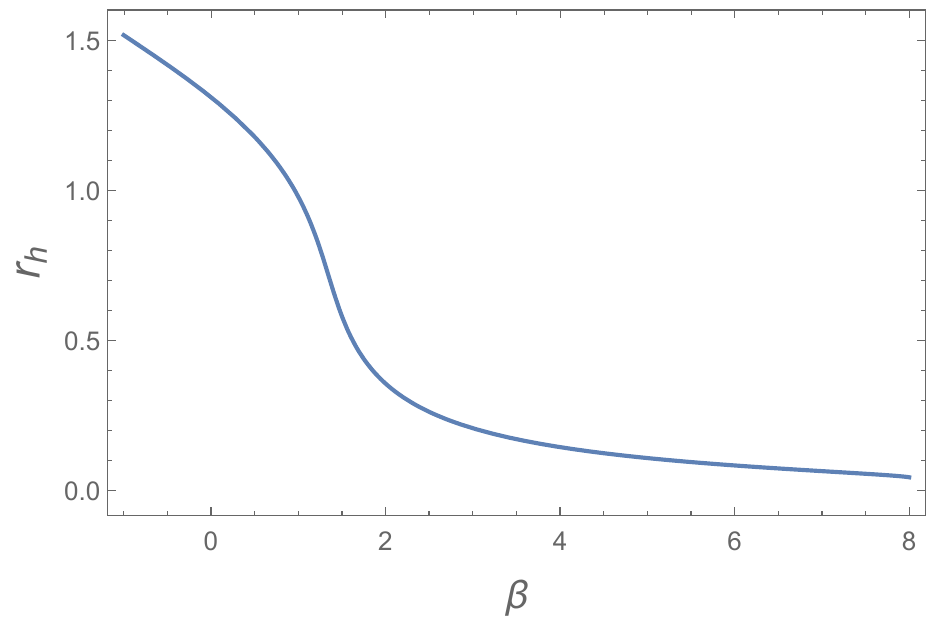}}
		\caption{\footnotesize The black hole size $(r_h)$ vs   (a) the parameter $\alpha$ (Here, we set $Q=0.2$ and $\beta=2$).   (b) the parameter $\beta$ (Here, we set $Q=0.2$ and $\alpha=1$). Similar behaviors can be seen for neutral black hole case as well. \label{fig:alpha, beta on rh}}
	}	
\end{figure}
\begin{figure}[t!]	
	{\centering

		\subfloat[]{\includegraphics[width=2in]{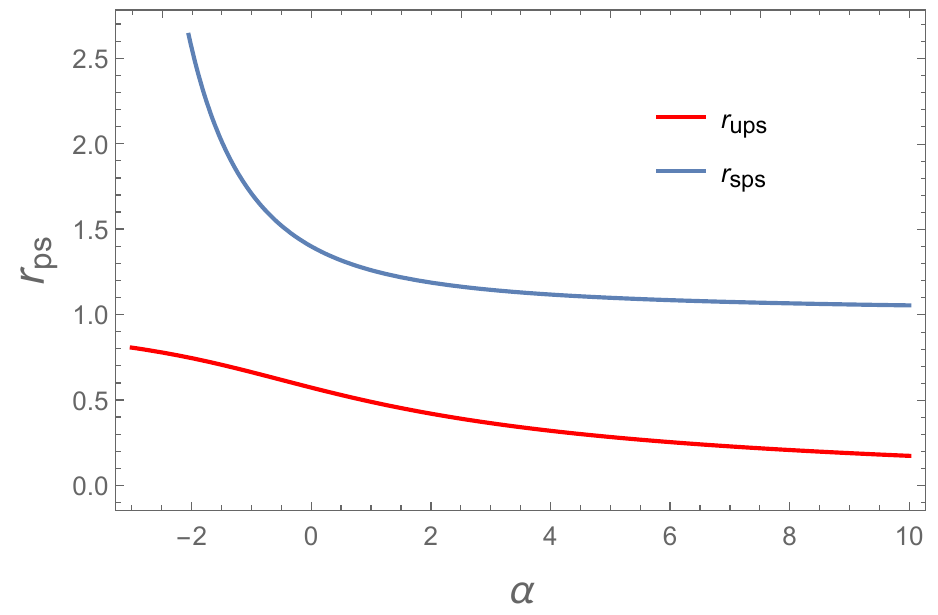}}\hspace{0.15cm}
		\subfloat[]{\includegraphics[width=2in]{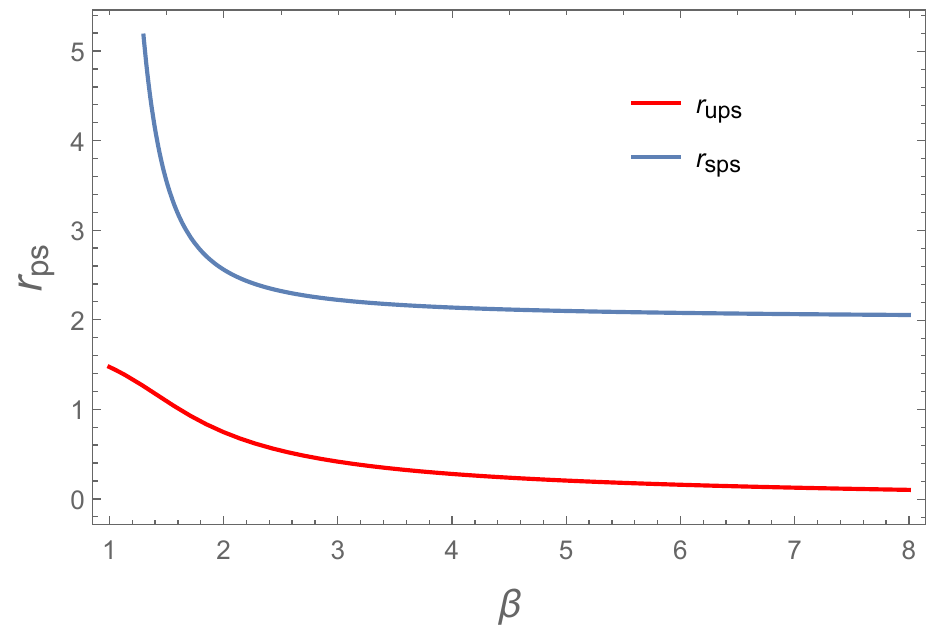}}\hspace{0.15cm}	
		\subfloat[]{\includegraphics[width=2in]{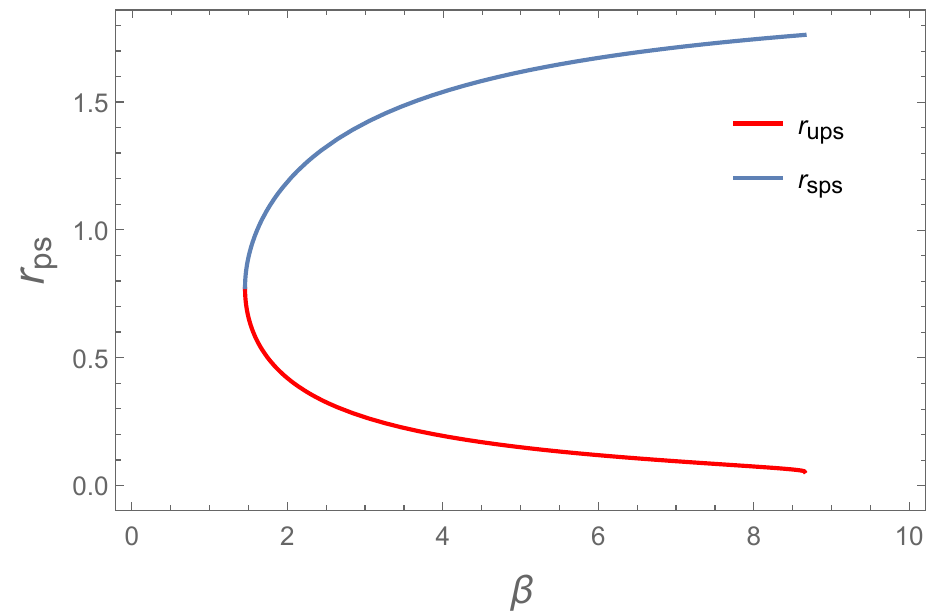}}
		\caption{\footnotesize The photon sphere size $(r_{ps})$  vs   (a) the parameter $\alpha$  (Here, we set $Q=0.2$ and $\beta=2$).  (b) the parameter $\beta$     (Here, we set $Q=0.2$ and $\alpha=-2$). (c) the parameter $\beta$  (Here, we set $Q=0.2$ and $\alpha=2$). Similar behaviors can be seen for neutral black hole case as well. \label{fig:alpha, beta on rps}}
	}	
\end{figure}


\section{Conclusions}  \label{5}
In this paper, we investigated the circular geodesics of massless test particles in the 4-dimensional static and spherically symmetric black holes in dRGT massive gravity theory, where the charged (neutral) black holes can admit at most four (three) event horizons in certain parameter space $(\alpha, \beta)$ of the theory.

\vskip 0.2cm 
\noindent
We found that, similar to the black holes in Einstein's theory of gravity~\cite{Wei:2020rbh},  there exists one unstable  photon sphere  outside the outer horizon for both neutral and charged black holes. In addition, it is possible to have a pair of unstable and stable photon spheres in both neutral and charged black hole backgrounds for specific values of the massive gravity parameters $(\alpha, \beta)$, which do not occur in the standard general theory of gravity~\cite{Wei:2020rbh}. We also observed that certain black holes in this massive gravity theory do not allow any photon spheres to exist, as all the PS's hide inside the horizon of the black hole or they are not real. 
\vskip 0.2cm 
\noindent
To get a better understanding of the appearance of novel PS's in the massive gravity theory, we
followed Duan's topological current $\phi$-mapping theory, and computed the topological charges associated with the photon spheres. We found that, the black holes, having one unstable photon sphere, possess topological charge $Q_t = -1$. This suggests that these black holes have one standard PS and fall in the same topological class of static and stationary black holes of Einstein gravity~\cite{Wei:2020rbh,Cunha:2020azh}. For the case where black holes have two PS's (one unstable (UPS) and the other stable (SPS)),  the topological charges are $ -1$ for UPS, and $+1$ for SPS, with total topological charge being zero. This suggests that these black holes have one exotic PS (having $Q_t = +1$) along with one standard PS, and these new PS's fall in a different topological class.
We should note here that this is a novel topological classification for black holes with even number of PS's. Certain horizonless objects such as the first studied example of ultra compact objects (UCOs) ~\cite{Cunha:2017qtt,Cunha:2020azh} and naked singularities (NSs)~\cite{Wei:2020rbh}, also fall in the same topological class with total topological charge zero, though there is an important difference. For the black holes in massive gravity studied here, the inner PS is unstable, and the outer PS is stable, whereas, this order is reversed
for the horizonless compact objects studied in~\cite{Cunha:2017qtt,Cunha:2020azh,Wei:2020rbh}. 
Further, we also find that the black holes with no photon spheres  always possess  $Q_t = 0$. There  are only a handful of black holes known for which there exist two~\cite{Afshar:2024dhf} or no PS's ~\cite{Junior:2021dyw,Junior:2021svb,Wang:2021ara}. The distinctions between black holes and other compact objects found here through the topological classification of PS's, might provide interesting markers to distinguish geometries with and without horizons in general  and modified theories of gravity.
\vskip 0.2cm 
\noindent
Next, we interpreted the reason for the black holes in massive gravity possessing total topological charge $Q_t= 0$, in contrast to $Q_t = -1$ for black holes in standard Einstein gravity~\cite{Wei:2020rbh}, as arising from the difference in how the effective potential $H (r, \theta)$ falls off at radial infinity, in both cases. 
The black holes in general relativity (with different asymptotic geometries i.e., flat, AdS, or dS), all seem to have their potential function $H (r,\theta)$ behaving identically in the large radius limit, and this matches with their total topological charge being $Q_t = -1$. However, for the black holes in massive gravity theory studied here, the asymptotic behaviour of the potential function 
$H (r, \theta)$ is quite different for the case of even and odd number of PS's cases, categorising them in different topological classes. Thus, we conclude that the asymptotic behavior of the $H (r, \theta) $ plays a major role in deciding the total topological charge of the black holes PS's, rather than the lapse function of the geometry.  
\vskip 0.2 cm \noindent 
Next, for  both the neutral and charged black holes in massive gravity, we studied the landscape of photon spheres in the parameter space of $(\alpha, \beta)$,  and identified the regions (using figs.~\ref{fig:bh_area} and~\ref{fig:ps_area}) where there exist one, two or zero PS's. Our results in these cases are summarised in the table~\ref{table:summary}. 
\begin{table}[t!]\caption{Photon sphere landscape}\label{table:summary} 
\centering{		
\begin{tabular}{|m{2.6 cm} |m{8.9 cm} |m{3.3 cm}|} 
	\hline \hline 
	Black holes in massive gravity   & \hskip 3cm  \text{Horizon structure}  \vskip 0cm ($\#$ of black hole horizons, $\#$ of cosmological horizons) & $\#$ of photon spheres that exist  outside the black hole  \\ \hline  
 \multirow{4}{2.4cm} {Neutral} 	 &  \hskip 4cm $(1,0)$  &  Two, One, Zero \\ \cline{2-3} 
	                             & \hskip 4cm $(1,1)$ & One \\ \cline{2-3}
                                 & \hskip 4cm $(1,2)$ & One \\ \cline{2-3}
	                             & \hskip 4cm $(3,0)$ & Zero \\ \hline \hline
\multirow{4}{2.4cm} {Charged $(Q= 0.2, 0.3)$} 	 &  \hskip 4cm $(2,0)$  &   Two, One, Zero \\ \cline{2-3} 
	                             & \hskip 4cm $(2,1)$ & One \\ \cline{2-3}
	                             & \hskip 4cm $(2,2)$ & One \\ \cline{2-3}
	                             & \hskip 4cm $(4,0)$ & Zero \\ \hline \hline
\end{tabular} }
\end{table}
\vskip 0.2cm 
\noindent
We noted that for both neutral and charged black holes, the black hole size and also the unstable photon sphere size decreases as the parameters $\alpha$  or $\beta$ increase. For the stable  photon sphere we find the following.  Its size decreases as $\alpha$  increases. As $\beta$ increases, its size can decrease or increase depending on the choice of $\alpha$. The effects of the parameters, $\alpha$ and $\beta$, on the size of the black hole and  (stable and unstable) photon spheres are summarised in the fig.~\ref{fig:alpha, beta on rh} and fig.~\ref{fig:alpha, beta on rps}. It might be interesting to study the effects of other parameters, such as charge and graviton mass, on the size of the black holes and photon spheres. We leave these issues for future with a view to finding better approaches to address the numerical complexity of the system.
\vskip 0.2cm 
\noindent
It would be interesting to study the topological properties of PS's for naked singularities and rotating black holes in massive gravity~\cite{Hendi:2022qgi}, as would the cases with different asymptotic geometries, which might through new information, as compared to standard Einstein gravity. It is observed in~\cite{,Junior:2021dyw,Junior:2021svb} that the black holes  that have no PS's, can yield panoramic shadows, seen (almost) all around the equator of the observer’s sky. It would be interesting to check this panoramic  shadow formation   for the black holes in massive gravity having no PS's, as well. 
Due to the existence of one stable PS for certain black holes in massive gravity, one may check for the occurrence of echoes of quasinormal modes~\cite{Huang:2021qwe}. These might be interesting areas to explore in future.
\vskip 0.2cm 
\noindent
When the photon sphere outside the event horizon is unstable (as is the case for the black holes in Einstein gravity), instability of the light ring underlies the sharp shadow boundary and there is the well known correspondence between eikonal quasinormal modes (QNMs) and photon-sphere parameters~\cite{Cardoso:2008bp,Stefanov:2010xz,Dolan:2009nk}. A notable feature of our analysis is the emergence, in certain regions of the massive-gravity parameter space $(\alpha,\beta)$, of an exterior stable photon sphere. 
The existence of multiple photon spheres (with distinct stability properties) has been discussed in others contexts  and certain modified gravity models~\cite{Cardoso:2014sna,Keir:2014oka,Cunha:2017qtt,Junior:2021svb,Cunha:2020azh,Cunha:2022gde,Wei:2020rbh}. Stable light rings are associated with nontrivial phase-space structure and long-lived perturbative configurations and are known to arise in ultra-compact, horizonless configurations such as boson stars and gravastar-like models~\cite{Cardoso:2014sna,Cunha:2017qtt}. In such systems, stable trapping of null geodesics has been linked to slow decay of perturbations and potential nonlinear instabilities~\cite{Keir:2014oka}. The presence of a stable photon sphere has implications for high-frequency perturbations. In general relativity, the eikonal QNM correspondence relies on the instability of the photon orbit, with the damping rate proportional to the Lyapunov exponent of that orbit~\cite{Cardoso:2008bp}. When a stable orbit is present, the Lyapunov exponent becomes imaginary, signaling oscillatory trapping rather than exponential divergence. Similar departures from the standard light-ring/QNM picture have been discussed in spacetimes with multiple light rings or modified effective potentials~\cite{Konoplya:2011qq,Cunha:2017qtt,Shaikh:2018kfv}. Our results therefore indicate that the stable-PS regime represents a qualitative shift in the null-geodesic phase-space structure induced by the graviton mass.

\vskip 0.2cm 
\noindent
From an observational standpoint, unstable photon spheres determine the edge of the black-hole shadow. Observations by the Event Horizon Telescope are consistent with general relativistic expectations, but the coexistence of stable and unstable photon spheres could introduce subdominant lensing structures or brightness features associated with near-trapped null trajectories~\cite{Konoplya:2011qq,Cunha:2017qtt,Shaikh:2018kfv}. In the gravitational-wave sector, increasingly precise ringdown measurements by detectors such as LIGO and Virgo Collaboration probe the QNM spectrum. Thus, deviations from the standard eikonal correspondence may eventually serve as indirect constraints on modified gravity scenarios.
Next generation interferometers may achieve sufficient accuracy to probe small deviations in damping rates or identify additional long-lived components in the late-time signal. A detailed perturbative analysis of quasinormal modes and ray tracing simulations in the allowed parameter region would be necessary to quantify the size of potential observational effects and assess their compatibility with existing and forthcoming data.
In the present case, the emergence of a stable PS in a distinct topological sector provides a diagnostic probe of how graviton-mass effects modify high-frequency propagation, shadow structure, and potentially ringdown physics, and thus merits further theoretical and observational investigation. These are all interesting avenues for future work. 

\section*{Acknowledgements}
The work of C.B. is supported by ARG-MATRICS grant no. ANRF/ARGM/2025/002280/MTR.
We thank Professors Bin Chen, Sudipta Mukherji and Yungui Gong for helpful discussions and encouragement. We also thank the anonymous referees for useful comments which improved the manuscript. 
\bibliographystyle{apsrev4-1}
\bibliography{pss_massive}

\begin{thebibliography}{161}%
\makeatletter
\providecommand \@ifxundefined [1]{%
 \@ifx{#1\undefined}
}%
\providecommand \@ifnum [1]{%
 \ifnum #1\expandafter \@firstoftwo
 \else \expandafter \@secondoftwo
 \fi
}%
\providecommand \@ifx [1]{%
 \ifx #1\expandafter \@firstoftwo
 \else \expandafter \@secondoftwo
 \fi
}%
\providecommand \natexlab [1]{#1}%
\providecommand \enquote  [1]{``#1''}%
\providecommand \bibnamefont  [1]{#1}%
\providecommand \bibfnamefont [1]{#1}%
\providecommand \citenamefont [1]{#1}%
\providecommand \href@noop [0]{\@secondoftwo}%
\providecommand \href [0]{\begingroup \@sanitize@url \@href}%
\providecommand \@href[1]{\@@startlink{#1}\@@href}%
\providecommand \@@href[1]{\endgroup#1\@@endlink}%
\providecommand \@sanitize@url [0]{\catcode `\\12\catcode `\$12\catcode
  `\&12\catcode `\#12\catcode `\^12\catcode `\_12\catcode `\%12\relax}%
\providecommand \@@startlink[1]{}%
\providecommand \@@endlink[0]{}%
\providecommand \url  [0]{\begingroup\@sanitize@url \@url }%
\providecommand \@url [1]{\endgroup\@href {#1}{\urlprefix }}%
\providecommand \urlprefix  [0]{URL }%
\providecommand \Eprint [0]{\href }%
\providecommand \doibase [0]{http://dx.doi.org/}%
\providecommand \selectlanguage [0]{\@gobble}%
\providecommand \bibinfo  [0]{\@secondoftwo}%
\providecommand \bibfield  [0]{\@secondoftwo}%
\providecommand \translation [1]{[#1]}%
\providecommand \BibitemOpen [0]{}%
\providecommand \bibitemStop [0]{}%
\providecommand \bibitemNoStop [0]{.\EOS\space}%
\providecommand \EOS [0]{\spacefactor3000\relax}%
\providecommand \BibitemShut  [1]{\csname bibitem#1\endcsname}%
\let\auto@bib@innerbib\@empty
\bibitem [{\citenamefont {Wei}(2020)}]{Wei:2020rbh}%
  \BibitemOpen
  \bibfield  {author} {\bibinfo {author} {\bibfnamefont {S.-W.}\ \bibnamefont
  {Wei}},\ }\href {\doibase 10.1103/PhysRevD.102.064039} {\bibfield  {journal}
  {\bibinfo  {journal} {Phys. Rev. D}\ }\textbf {\bibinfo {volume} {102}},\
  \bibinfo {pages} {064039} (\bibinfo {year} {2020})},\ \Eprint
  {http://arxiv.org/abs/2006.02112} {arXiv:2006.02112 [gr-qc]} \BibitemShut
  {NoStop}%
\bibitem [{\citenamefont {Cunha}\ \emph {et~al.}(2017)\citenamefont {Cunha},
  \citenamefont {Berti},\ and\ \citenamefont {Herdeiro}}]{Cunha:2017qtt}%
  \BibitemOpen
  \bibfield  {author} {\bibinfo {author} {\bibfnamefont {P.~V.~P.}\
  \bibnamefont {Cunha}}, \bibinfo {author} {\bibfnamefont {E.}~\bibnamefont
  {Berti}}, \ and\ \bibinfo {author} {\bibfnamefont {C.~A.~R.}\ \bibnamefont
  {Herdeiro}},\ }\href {\doibase 10.1103/PhysRevLett.119.251102} {\bibfield
  {journal} {\bibinfo  {journal} {Phys. Rev. Lett.}\ }\textbf {\bibinfo
  {volume} {119}},\ \bibinfo {pages} {251102} (\bibinfo {year} {2017})},\
  \Eprint {http://arxiv.org/abs/1708.04211} {arXiv:1708.04211 [gr-qc]}
  \BibitemShut {NoStop}%
\bibitem [{\citenamefont {Cunha}\ and\ \citenamefont
  {Herdeiro}(2020)}]{Cunha:2020azh}%
  \BibitemOpen
  \bibfield  {author} {\bibinfo {author} {\bibfnamefont {P.~V.~P.}\
  \bibnamefont {Cunha}}\ and\ \bibinfo {author} {\bibfnamefont {C.~A.~R.}\
  \bibnamefont {Herdeiro}},\ }\href {\doibase 10.1103/PhysRevLett.124.181101}
  {\bibfield  {journal} {\bibinfo  {journal} {Phys. Rev. Lett.}\ }\textbf
  {\bibinfo {volume} {124}},\ \bibinfo {pages} {181101} (\bibinfo {year}
  {2020})},\ \Eprint {http://arxiv.org/abs/2003.06445} {arXiv:2003.06445
  [gr-qc]} \BibitemShut {NoStop}%
\bibitem [{\citenamefont {Abbott}\ and\ \citenamefont {et.
  al.}(2016)}]{Abbott}%
  \BibitemOpen
  \bibfield  {author} {\bibinfo {author} {\bibfnamefont {B.~P.}\ \bibnamefont
  {Abbott}}\ and\ \bibinfo {author} {\bibnamefont {et. al.}} (\bibinfo
  {collaboration} {LIGO Scientific Collaboration and Virgo Collaboration}),\
  }\href {\doibase 10.1103/PhysRevLett.116.061102} {\bibfield  {journal}
  {\bibinfo  {journal} {Phys. Rev. Lett.}\ }\textbf {\bibinfo {volume} {116}},\
  \bibinfo {pages} {061102} (\bibinfo {year} {2016})}\BibitemShut {NoStop}%
\bibitem [{\citenamefont {Akiyama}\ \emph
  {et~al.}(2019{\natexlab{a}})\citenamefont {Akiyama} \emph
  {et~al.}}]{EventHorizonTelescope:2019dse}%
  \BibitemOpen
  \bibfield  {author} {\bibinfo {author} {\bibfnamefont {K.}~\bibnamefont
  {Akiyama}} \emph {et~al.} (\bibinfo {collaboration} {Event Horizon
  Telescope}),\ }\href {\doibase 10.3847/2041-8213/ab0ec7} {\bibfield
  {journal} {\bibinfo  {journal} {Astrophys. J. Lett.}\ }\textbf {\bibinfo
  {volume} {875}},\ \bibinfo {pages} {L1} (\bibinfo {year}
  {2019}{\natexlab{a}})},\ \Eprint {http://arxiv.org/abs/1906.11238}
  {arXiv:1906.11238 [astro-ph.GA]} \BibitemShut {NoStop}%
\bibitem [{\citenamefont {Akiyama}\ \emph
  {et~al.}(2019{\natexlab{b}})\citenamefont {Akiyama} \emph
  {et~al.}}]{EventHorizonTelescope:2019pgp}%
  \BibitemOpen
  \bibfield  {author} {\bibinfo {author} {\bibfnamefont {K.}~\bibnamefont
  {Akiyama}} \emph {et~al.} (\bibinfo {collaboration} {Event Horizon
  Telescope}),\ }\href {\doibase 10.3847/2041-8213/ab0f43} {\bibfield
  {journal} {\bibinfo  {journal} {Astrophys. J. Lett.}\ }\textbf {\bibinfo
  {volume} {875}},\ \bibinfo {pages} {L5} (\bibinfo {year}
  {2019}{\natexlab{b}})},\ \Eprint {http://arxiv.org/abs/1906.11242}
  {arXiv:1906.11242 [astro-ph.GA]} \BibitemShut {NoStop}%
\bibitem [{\citenamefont {Akiyama}\ \emph
  {et~al.}(2019{\natexlab{c}})\citenamefont {Akiyama} \emph
  {et~al.}}]{EventHorizonTelescope:2019ggy}%
  \BibitemOpen
  \bibfield  {author} {\bibinfo {author} {\bibfnamefont {K.}~\bibnamefont
  {Akiyama}} \emph {et~al.} (\bibinfo {collaboration} {Event Horizon
  Telescope}),\ }\href {\doibase 10.3847/2041-8213/ab1141} {\bibfield
  {journal} {\bibinfo  {journal} {Astrophys. J. Lett.}\ }\textbf {\bibinfo
  {volume} {875}},\ \bibinfo {pages} {L6} (\bibinfo {year}
  {2019}{\natexlab{c}})},\ \Eprint {http://arxiv.org/abs/1906.11243}
  {arXiv:1906.11243 [astro-ph.GA]} \BibitemShut {NoStop}%
\bibitem [{\citenamefont {Abbott}\ \emph {et~al.}(2016)\citenamefont {Abbott}
  \emph {et~al.}}]{LIGOScientific:2016aoc}%
  \BibitemOpen
  \bibfield  {author} {\bibinfo {author} {\bibfnamefont {B.~P.}\ \bibnamefont
  {Abbott}} \emph {et~al.} (\bibinfo {collaboration} {LIGO Scientific,
  Virgo}),\ }\href {\doibase 10.1103/PhysRevLett.116.061102} {\bibfield
  {journal} {\bibinfo  {journal} {Phys. Rev. Lett.}\ }\textbf {\bibinfo
  {volume} {116}},\ \bibinfo {pages} {061102} (\bibinfo {year} {2016})},\
  \Eprint {http://arxiv.org/abs/1602.03837} {arXiv:1602.03837 [gr-qc]}
  \BibitemShut {NoStop}%
\bibitem [{\citenamefont {Akiyama}\ \emph {et~al.}(2022)\citenamefont {Akiyama}
  \emph {et~al.}}]{EventHorizonTelescope:2022wkp}%
  \BibitemOpen
  \bibfield  {author} {\bibinfo {author} {\bibfnamefont {K.}~\bibnamefont
  {Akiyama}} \emph {et~al.} (\bibinfo {collaboration} {Event Horizon
  Telescope}),\ }\href {\doibase 10.3847/2041-8213/ac6674} {\bibfield
  {journal} {\bibinfo  {journal} {Astrophys. J. Lett.}\ }\textbf {\bibinfo
  {volume} {930}},\ \bibinfo {pages} {L12} (\bibinfo {year} {2022})},\ \Eprint
  {http://arxiv.org/abs/2311.08680} {arXiv:2311.08680 [astro-ph.HE]}
  \BibitemShut {NoStop}%
\bibitem [{\citenamefont {Grandclement}\ \emph {et~al.}(2014)\citenamefont
  {Grandclement}, \citenamefont {Som\'e},\ and\ \citenamefont
  {Gourgoulhon}}]{Grandclement:2014msa}%
  \BibitemOpen
  \bibfield  {author} {\bibinfo {author} {\bibfnamefont {P.}~\bibnamefont
  {Grandclement}}, \bibinfo {author} {\bibfnamefont {C.}~\bibnamefont
  {Som\'e}}, \ and\ \bibinfo {author} {\bibfnamefont {E.}~\bibnamefont
  {Gourgoulhon}},\ }\href {\doibase 10.1103/PhysRevD.90.024068} {\bibfield
  {journal} {\bibinfo  {journal} {Phys. Rev. D}\ }\textbf {\bibinfo {volume}
  {90}},\ \bibinfo {pages} {024068} (\bibinfo {year} {2014})},\ \Eprint
  {http://arxiv.org/abs/1405.4837} {arXiv:1405.4837 [gr-qc]} \BibitemShut
  {NoStop}%
\bibitem [{\citenamefont {Grould}\ \emph {et~al.}(2017)\citenamefont {Grould},
  \citenamefont {Meliani}, \citenamefont {Vincent}, \citenamefont
  {Grandcl\'ement},\ and\ \citenamefont {Gourgoulhon}}]{Grould:2017rzz}%
  \BibitemOpen
  \bibfield  {author} {\bibinfo {author} {\bibfnamefont {M.}~\bibnamefont
  {Grould}}, \bibinfo {author} {\bibfnamefont {Z.}~\bibnamefont {Meliani}},
  \bibinfo {author} {\bibfnamefont {F.~H.}\ \bibnamefont {Vincent}}, \bibinfo
  {author} {\bibfnamefont {P.}~\bibnamefont {Grandcl\'ement}}, \ and\ \bibinfo
  {author} {\bibfnamefont {E.}~\bibnamefont {Gourgoulhon}},\ }\href {\doibase
  10.1088/1361-6382/aa8d39} {\bibfield  {journal} {\bibinfo  {journal} {Class.
  Quant. Grav.}\ }\textbf {\bibinfo {volume} {34}},\ \bibinfo {pages} {215007}
  (\bibinfo {year} {2017})},\ \Eprint {http://arxiv.org/abs/1709.05938}
  {arXiv:1709.05938 [astro-ph.HE]} \BibitemShut {NoStop}%
\bibitem [{\citenamefont {Teodoro}\ \emph
  {et~al.}(2021{\natexlab{a}})\citenamefont {Teodoro}, \citenamefont
  {Collodel},\ and\ \citenamefont {Kunz}}]{Teodoro:2020kok}%
  \BibitemOpen
  \bibfield  {author} {\bibinfo {author} {\bibfnamefont {M.~C.}\ \bibnamefont
  {Teodoro}}, \bibinfo {author} {\bibfnamefont {L.~G.}\ \bibnamefont
  {Collodel}}, \ and\ \bibinfo {author} {\bibfnamefont {J.}~\bibnamefont
  {Kunz}},\ }\href {\doibase 10.1088/1475-7516/2021/03/063} {\bibfield
  {journal} {\bibinfo  {journal} {JCAP}\ }\textbf {\bibinfo {volume} {03}},\
  \bibinfo {pages} {063} (\bibinfo {year} {2021}{\natexlab{a}})},\ \Eprint
  {http://arxiv.org/abs/2011.10288} {arXiv:2011.10288 [gr-qc]} \BibitemShut
  {NoStop}%
\bibitem [{\citenamefont {Teodoro}\ \emph
  {et~al.}(2021{\natexlab{b}})\citenamefont {Teodoro}, \citenamefont
  {Collodel}, \citenamefont {Doneva}, \citenamefont {Kunz}, \citenamefont
  {Nedkova},\ and\ \citenamefont {Yazadjiev}}]{Teodoro:2021ezj}%
  \BibitemOpen
  \bibfield  {author} {\bibinfo {author} {\bibfnamefont {M.~C.}\ \bibnamefont
  {Teodoro}}, \bibinfo {author} {\bibfnamefont {L.~G.}\ \bibnamefont
  {Collodel}}, \bibinfo {author} {\bibfnamefont {D.}~\bibnamefont {Doneva}},
  \bibinfo {author} {\bibfnamefont {J.}~\bibnamefont {Kunz}}, \bibinfo {author}
  {\bibfnamefont {P.}~\bibnamefont {Nedkova}}, \ and\ \bibinfo {author}
  {\bibfnamefont {S.}~\bibnamefont {Yazadjiev}},\ }\href {\doibase
  10.1103/PhysRevD.104.124047} {\bibfield  {journal} {\bibinfo  {journal}
  {Phys. Rev. D}\ }\textbf {\bibinfo {volume} {104}},\ \bibinfo {pages}
  {124047} (\bibinfo {year} {2021}{\natexlab{b}})},\ \Eprint
  {http://arxiv.org/abs/2108.08640} {arXiv:2108.08640 [gr-qc]} \BibitemShut
  {NoStop}%
\bibitem [{\citenamefont {Gibbons}\ and\ \citenamefont
  {Herdeiro}(1999)}]{Gibbons:1999uv}%
  \BibitemOpen
  \bibfield  {author} {\bibinfo {author} {\bibfnamefont {G.~W.}\ \bibnamefont
  {Gibbons}}\ and\ \bibinfo {author} {\bibfnamefont {C.~A.~R.}\ \bibnamefont
  {Herdeiro}},\ }\href {\doibase 10.1088/0264-9381/16/11/311} {\bibfield
  {journal} {\bibinfo  {journal} {Class. Quant. Grav.}\ }\textbf {\bibinfo
  {volume} {16}},\ \bibinfo {pages} {3619} (\bibinfo {year} {1999})},\ \Eprint
  {http://arxiv.org/abs/hep-th/9906098} {arXiv:hep-th/9906098} \BibitemShut
  {NoStop}%
\bibitem [{\citenamefont {Herdeiro}(2000)}]{Herdeiro:2000ap}%
  \BibitemOpen
  \bibfield  {author} {\bibinfo {author} {\bibfnamefont {C.~A.~R.}\
  \bibnamefont {Herdeiro}},\ }\href {\doibase 10.1016/S0550-3213(00)00335-7}
  {\bibfield  {journal} {\bibinfo  {journal} {Nucl. Phys. B}\ }\textbf
  {\bibinfo {volume} {582}},\ \bibinfo {pages} {363} (\bibinfo {year}
  {2000})},\ \Eprint {http://arxiv.org/abs/hep-th/0003063}
  {arXiv:hep-th/0003063} \BibitemShut {NoStop}%
\bibitem [{\citenamefont {Diemer}\ and\ \citenamefont
  {Kunz}(2014)}]{Diemer:2013fza}%
  \BibitemOpen
  \bibfield  {author} {\bibinfo {author} {\bibfnamefont {V.}~\bibnamefont
  {Diemer}}\ and\ \bibinfo {author} {\bibfnamefont {J.}~\bibnamefont {Kunz}},\
  }\href {\doibase 10.1103/PhysRevD.89.084001} {\bibfield  {journal} {\bibinfo
  {journal} {Phys. Rev. D}\ }\textbf {\bibinfo {volume} {89}},\ \bibinfo
  {pages} {084001} (\bibinfo {year} {2014})},\ \Eprint
  {http://arxiv.org/abs/1312.6540} {arXiv:1312.6540 [gr-qc]} \BibitemShut
  {NoStop}%
\bibitem [{\citenamefont {Delgado}\ \emph {et~al.}(2022)\citenamefont
  {Delgado}, \citenamefont {Herdeiro},\ and\ \citenamefont
  {Radu}}]{Delgado:2021jxd}%
  \BibitemOpen
  \bibfield  {author} {\bibinfo {author} {\bibfnamefont {J.~F.~M.}\
  \bibnamefont {Delgado}}, \bibinfo {author} {\bibfnamefont {C.~A.~R.}\
  \bibnamefont {Herdeiro}}, \ and\ \bibinfo {author} {\bibfnamefont
  {E.}~\bibnamefont {Radu}},\ }\href {\doibase 10.1103/PhysRevD.105.064026}
  {\bibfield  {journal} {\bibinfo  {journal} {Phys. Rev. D}\ }\textbf {\bibinfo
  {volume} {105}},\ \bibinfo {pages} {064026} (\bibinfo {year} {2022})},\
  \Eprint {http://arxiv.org/abs/2107.03404} {arXiv:2107.03404 [gr-qc]}
  \BibitemShut {NoStop}%
\bibitem [{\citenamefont {Bambi}\ \emph {et~al.}(2019)\citenamefont {Bambi},
  \citenamefont {Freese}, \citenamefont {Vagnozzi},\ and\ \citenamefont
  {Visinelli}}]{Bambi:2019tjh}%
  \BibitemOpen
  \bibfield  {author} {\bibinfo {author} {\bibfnamefont {C.}~\bibnamefont
  {Bambi}}, \bibinfo {author} {\bibfnamefont {K.}~\bibnamefont {Freese}},
  \bibinfo {author} {\bibfnamefont {S.}~\bibnamefont {Vagnozzi}}, \ and\
  \bibinfo {author} {\bibfnamefont {L.}~\bibnamefont {Visinelli}},\ }\href
  {\doibase 10.1103/PhysRevD.100.044057} {\bibfield  {journal} {\bibinfo
  {journal} {Phys. Rev. D}\ }\textbf {\bibinfo {volume} {100}},\ \bibinfo
  {pages} {044057} (\bibinfo {year} {2019})},\ \Eprint
  {http://arxiv.org/abs/1904.12983} {arXiv:1904.12983 [gr-qc]} \BibitemShut
  {NoStop}%
\bibitem [{\citenamefont {Vagnozzi}\ \emph {et~al.}(2023)\citenamefont
  {Vagnozzi} \emph {et~al.}}]{Vagnozzi:2022moj}%
  \BibitemOpen
  \bibfield  {author} {\bibinfo {author} {\bibfnamefont {S.}~\bibnamefont
  {Vagnozzi}} \emph {et~al.},\ }\href {\doibase 10.1088/1361-6382/acd97b}
  {\bibfield  {journal} {\bibinfo  {journal} {Class. Quant. Grav.}\ }\textbf
  {\bibinfo {volume} {40}},\ \bibinfo {pages} {165007} (\bibinfo {year}
  {2023})},\ \Eprint {http://arxiv.org/abs/2205.07787} {arXiv:2205.07787
  [gr-qc]} \BibitemShut {NoStop}%
\bibitem [{\citenamefont {Khodadi}\ \emph {et~al.}(2024)\citenamefont
  {Khodadi}, \citenamefont {Vagnozzi},\ and\ \citenamefont
  {Firouzjaee}}]{Khodadi:2024ubi}%
  \BibitemOpen
  \bibfield  {author} {\bibinfo {author} {\bibfnamefont {M.}~\bibnamefont
  {Khodadi}}, \bibinfo {author} {\bibfnamefont {S.}~\bibnamefont {Vagnozzi}}, \
  and\ \bibinfo {author} {\bibfnamefont {J.~T.}\ \bibnamefont {Firouzjaee}},\
  }\href {\doibase 10.1038/s41598-024-78264-y} {\bibfield  {journal} {\bibinfo
  {journal} {Sci. Rep.}\ }\textbf {\bibinfo {volume} {14}},\ \bibinfo {pages}
  {26932} (\bibinfo {year} {2024})},\ \Eprint {http://arxiv.org/abs/2408.03241}
  {arXiv:2408.03241 [gr-qc]} \BibitemShut {NoStop}%
\bibitem [{\citenamefont {Stuchlik}\ and\ \citenamefont
  {Hledik}(1999)}]{Stuchlik:1999qk}%
  \BibitemOpen
  \bibfield  {author} {\bibinfo {author} {\bibfnamefont {Z.}~\bibnamefont
  {Stuchlik}}\ and\ \bibinfo {author} {\bibfnamefont {S.}~\bibnamefont
  {Hledik}},\ }\href {\doibase 10.1103/PhysRevD.60.044006} {\bibfield
  {journal} {\bibinfo  {journal} {Phys. Rev. D}\ }\textbf {\bibinfo {volume}
  {60}},\ \bibinfo {pages} {044006} (\bibinfo {year} {1999})}\BibitemShut
  {NoStop}%
\bibitem [{\citenamefont {Claudel}\ \emph {et~al.}(2001)\citenamefont
  {Claudel}, \citenamefont {Virbhadra},\ and\ \citenamefont
  {Ellis}}]{Claudel:2000yi}%
  \BibitemOpen
  \bibfield  {author} {\bibinfo {author} {\bibfnamefont {C.-M.}\ \bibnamefont
  {Claudel}}, \bibinfo {author} {\bibfnamefont {K.~S.}\ \bibnamefont
  {Virbhadra}}, \ and\ \bibinfo {author} {\bibfnamefont {G.~F.~R.}\
  \bibnamefont {Ellis}},\ }\href {\doibase 10.1063/1.1308507} {\bibfield
  {journal} {\bibinfo  {journal} {J. Math. Phys.}\ }\textbf {\bibinfo {volume}
  {42}},\ \bibinfo {pages} {818} (\bibinfo {year} {2001})},\ \Eprint
  {http://arxiv.org/abs/gr-qc/0005050} {arXiv:gr-qc/0005050} \BibitemShut
  {NoStop}%
\bibitem [{\citenamefont {Virbhadra}\ and\ \citenamefont
  {Ellis}(2000)}]{Virbhadra:1999nm}%
  \BibitemOpen
  \bibfield  {author} {\bibinfo {author} {\bibfnamefont {K.~S.}\ \bibnamefont
  {Virbhadra}}\ and\ \bibinfo {author} {\bibfnamefont {G.~F.~R.}\ \bibnamefont
  {Ellis}},\ }\href {\doibase 10.1103/PhysRevD.62.084003} {\bibfield  {journal}
  {\bibinfo  {journal} {Phys. Rev. D}\ }\textbf {\bibinfo {volume} {62}},\
  \bibinfo {pages} {084003} (\bibinfo {year} {2000})},\ \Eprint
  {http://arxiv.org/abs/astro-ph/9904193} {arXiv:astro-ph/9904193} \BibitemShut
  {NoStop}%
\bibitem [{\citenamefont {Virbhadra}\ and\ \citenamefont
  {Ellis}(2002)}]{Virbhadra:2002ju}%
  \BibitemOpen
  \bibfield  {author} {\bibinfo {author} {\bibfnamefont {K.~S.}\ \bibnamefont
  {Virbhadra}}\ and\ \bibinfo {author} {\bibfnamefont {G.~F.~R.}\ \bibnamefont
  {Ellis}},\ }\href {\doibase 10.1103/PhysRevD.65.103004} {\bibfield  {journal}
  {\bibinfo  {journal} {Phys. Rev. D}\ }\textbf {\bibinfo {volume} {65}},\
  \bibinfo {pages} {103004} (\bibinfo {year} {2002})}\BibitemShut {NoStop}%
\bibitem [{\citenamefont {Adler}\ and\ \citenamefont
  {Virbhadra}(2022)}]{Adler:2022qtb}%
  \BibitemOpen
  \bibfield  {author} {\bibinfo {author} {\bibfnamefont {S.~L.}\ \bibnamefont
  {Adler}}\ and\ \bibinfo {author} {\bibfnamefont {K.~S.}\ \bibnamefont
  {Virbhadra}},\ }\href {\doibase 10.1007/s10714-022-02976-7} {\bibfield
  {journal} {\bibinfo  {journal} {Gen. Rel. Grav.}\ }\textbf {\bibinfo {volume}
  {54}},\ \bibinfo {pages} {93} (\bibinfo {year} {2022})},\ \Eprint
  {http://arxiv.org/abs/2205.04628} {arXiv:2205.04628 [gr-qc]} \BibitemShut
  {NoStop}%
\bibitem [{\citenamefont {Levin}\ and\ \citenamefont
  {Perez-Giz}(2008)}]{Levin:2008mq}%
  \BibitemOpen
  \bibfield  {author} {\bibinfo {author} {\bibfnamefont {J.}~\bibnamefont
  {Levin}}\ and\ \bibinfo {author} {\bibfnamefont {G.}~\bibnamefont
  {Perez-Giz}},\ }\href {\doibase 10.1103/PhysRevD.77.103005} {\bibfield
  {journal} {\bibinfo  {journal} {Phys. Rev. D}\ }\textbf {\bibinfo {volume}
  {77}},\ \bibinfo {pages} {103005} (\bibinfo {year} {2008})},\ \Eprint
  {http://arxiv.org/abs/0802.0459} {arXiv:0802.0459 [gr-qc]} \BibitemShut
  {NoStop}%
\bibitem [{\citenamefont {Pugliese}\ \emph {et~al.}(2011)\citenamefont
  {Pugliese}, \citenamefont {Quevedo},\ and\ \citenamefont
  {Ruffini}}]{Pugliese:2010ps}%
  \BibitemOpen
  \bibfield  {author} {\bibinfo {author} {\bibfnamefont {D.}~\bibnamefont
  {Pugliese}}, \bibinfo {author} {\bibfnamefont {H.}~\bibnamefont {Quevedo}}, \
  and\ \bibinfo {author} {\bibfnamefont {R.}~\bibnamefont {Ruffini}},\ }\href
  {\doibase 10.1103/PhysRevD.83.024021} {\bibfield  {journal} {\bibinfo
  {journal} {Phys. Rev. D}\ }\textbf {\bibinfo {volume} {83}},\ \bibinfo
  {pages} {024021} (\bibinfo {year} {2011})},\ \Eprint
  {http://arxiv.org/abs/1012.5411} {arXiv:1012.5411 [astro-ph.HE]} \BibitemShut
  {NoStop}%
\bibitem [{\citenamefont {Hackmann}\ \emph {et~al.}(2010)\citenamefont
  {Hackmann}, \citenamefont {Lammerzahl}, \citenamefont {Kagramanova},\ and\
  \citenamefont {Kunz}}]{Hackmann:2010zz}%
  \BibitemOpen
  \bibfield  {author} {\bibinfo {author} {\bibfnamefont {E.}~\bibnamefont
  {Hackmann}}, \bibinfo {author} {\bibfnamefont {C.}~\bibnamefont
  {Lammerzahl}}, \bibinfo {author} {\bibfnamefont {V.}~\bibnamefont
  {Kagramanova}}, \ and\ \bibinfo {author} {\bibfnamefont {J.}~\bibnamefont
  {Kunz}},\ }\href {\doibase 10.1103/PhysRevD.81.044020} {\bibfield  {journal}
  {\bibinfo  {journal} {Phys. Rev. D}\ }\textbf {\bibinfo {volume} {81}},\
  \bibinfo {pages} {044020} (\bibinfo {year} {2010})},\ \Eprint
  {http://arxiv.org/abs/1009.6117} {arXiv:1009.6117 [gr-qc]} \BibitemShut
  {NoStop}%
\bibitem [{\citenamefont {Villanueva}\ \emph {et~al.}(2013)\citenamefont
  {Villanueva}, \citenamefont {Saavedra}, \citenamefont {Olivares},\ and\
  \citenamefont {Cruz}}]{Villanueva:2013zta}%
  \BibitemOpen
  \bibfield  {author} {\bibinfo {author} {\bibfnamefont {J.~R.}\ \bibnamefont
  {Villanueva}}, \bibinfo {author} {\bibfnamefont {J.}~\bibnamefont
  {Saavedra}}, \bibinfo {author} {\bibfnamefont {M.}~\bibnamefont {Olivares}},
  \ and\ \bibinfo {author} {\bibfnamefont {N.}~\bibnamefont {Cruz}},\ }\href
  {\doibase 10.1007/s10509-012-1333-x} {\bibfield  {journal} {\bibinfo
  {journal} {Astrophys. Space Sci.}\ }\textbf {\bibinfo {volume} {344}},\
  \bibinfo {pages} {437} (\bibinfo {year} {2013})}\BibitemShut {NoStop}%
\bibitem [{\citenamefont {Wei}\ and\ \citenamefont {Liu}(2018)}]{Wei:2017mwc}%
  \BibitemOpen
  \bibfield  {author} {\bibinfo {author} {\bibfnamefont {S.-W.}\ \bibnamefont
  {Wei}}\ and\ \bibinfo {author} {\bibfnamefont {Y.-X.}\ \bibnamefont {Liu}},\
  }\href {\doibase 10.1103/PhysRevD.97.104027} {\bibfield  {journal} {\bibinfo
  {journal} {Phys. Rev. D}\ }\textbf {\bibinfo {volume} {97}},\ \bibinfo
  {pages} {104027} (\bibinfo {year} {2018})},\ \Eprint
  {http://arxiv.org/abs/1711.01522} {arXiv:1711.01522 [gr-qc]} \BibitemShut
  {NoStop}%
\bibitem [{\citenamefont {Chandrasekhar}\ and\ \citenamefont
  {Mohapatra}(2019)}]{Chandrasekhar:2018sjg}%
  \BibitemOpen
  \bibfield  {author} {\bibinfo {author} {\bibfnamefont {B.}~\bibnamefont
  {Chandrasekhar}}\ and\ \bibinfo {author} {\bibfnamefont {S.}~\bibnamefont
  {Mohapatra}},\ }\href {\doibase 10.1016/j.physletb.2019.02.042} {\bibfield
  {journal} {\bibinfo  {journal} {Phys. Lett. B}\ }\textbf {\bibinfo {volume}
  {791}},\ \bibinfo {pages} {367} (\bibinfo {year} {2019})},\ \Eprint
  {http://arxiv.org/abs/1805.05088} {arXiv:1805.05088 [hep-th]} \BibitemShut
  {NoStop}%
\bibitem [{\citenamefont {Leh\'ebel}\ and\ \citenamefont
  {Cardoso}(2022)}]{Lehebel:2022yyz}%
  \BibitemOpen
  \bibfield  {author} {\bibinfo {author} {\bibfnamefont {A.}~\bibnamefont
  {Leh\'ebel}}\ and\ \bibinfo {author} {\bibfnamefont {V.}~\bibnamefont
  {Cardoso}},\ }\href {\doibase 10.1103/PhysRevD.105.064014} {\bibfield
  {journal} {\bibinfo  {journal} {Phys. Rev. D}\ }\textbf {\bibinfo {volume}
  {105}},\ \bibinfo {pages} {064014} (\bibinfo {year} {2022})},\ \Eprint
  {http://arxiv.org/abs/2202.08850} {arXiv:2202.08850 [gr-qc]} \BibitemShut
  {NoStop}%
\bibitem [{\citenamefont {Bozza}(2002)}]{Bozza:2002zj}%
  \BibitemOpen
  \bibfield  {author} {\bibinfo {author} {\bibfnamefont {V.}~\bibnamefont
  {Bozza}},\ }\href {\doibase 10.1103/PhysRevD.66.103001} {\bibfield  {journal}
  {\bibinfo  {journal} {Phys. Rev. D}\ }\textbf {\bibinfo {volume} {66}},\
  \bibinfo {pages} {103001} (\bibinfo {year} {2002})},\ \Eprint
  {http://arxiv.org/abs/gr-qc/0208075} {arXiv:gr-qc/0208075} \BibitemShut
  {NoStop}%
\bibitem [{\citenamefont {Cardoso}\ \emph {et~al.}(2009)\citenamefont
  {Cardoso}, \citenamefont {Miranda}, \citenamefont {Berti}, \citenamefont
  {Witek},\ and\ \citenamefont {Zanchin}}]{Cardoso:2008bp}%
  \BibitemOpen
  \bibfield  {author} {\bibinfo {author} {\bibfnamefont {V.}~\bibnamefont
  {Cardoso}}, \bibinfo {author} {\bibfnamefont {A.~S.}\ \bibnamefont
  {Miranda}}, \bibinfo {author} {\bibfnamefont {E.}~\bibnamefont {Berti}},
  \bibinfo {author} {\bibfnamefont {H.}~\bibnamefont {Witek}}, \ and\ \bibinfo
  {author} {\bibfnamefont {V.~T.}\ \bibnamefont {Zanchin}},\ }\href {\doibase
  10.1103/PhysRevD.79.064016} {\bibfield  {journal} {\bibinfo  {journal} {Phys.
  Rev. D}\ }\textbf {\bibinfo {volume} {79}},\ \bibinfo {pages} {064016}
  (\bibinfo {year} {2009})},\ \Eprint {http://arxiv.org/abs/0812.1806}
  {arXiv:0812.1806 [hep-th]} \BibitemShut {NoStop}%
\bibitem [{\citenamefont {Hioki}\ and\ \citenamefont
  {Maeda}(2009)}]{Hioki:2009na}%
  \BibitemOpen
  \bibfield  {author} {\bibinfo {author} {\bibfnamefont {K.}~\bibnamefont
  {Hioki}}\ and\ \bibinfo {author} {\bibfnamefont {K.-i.}\ \bibnamefont
  {Maeda}},\ }\href {\doibase 10.1103/PhysRevD.80.024042} {\bibfield  {journal}
  {\bibinfo  {journal} {Phys. Rev. D}\ }\textbf {\bibinfo {volume} {80}},\
  \bibinfo {pages} {024042} (\bibinfo {year} {2009})},\ \Eprint
  {http://arxiv.org/abs/0904.3575} {arXiv:0904.3575 [astro-ph.HE]} \BibitemShut
  {NoStop}%
\bibitem [{\citenamefont {Collodel}\ \emph {et~al.}(2018)\citenamefont
  {Collodel}, \citenamefont {Kleihaus},\ and\ \citenamefont
  {Kunz}}]{Collodel:2017end}%
  \BibitemOpen
  \bibfield  {author} {\bibinfo {author} {\bibfnamefont {L.~G.}\ \bibnamefont
  {Collodel}}, \bibinfo {author} {\bibfnamefont {B.}~\bibnamefont {Kleihaus}},
  \ and\ \bibinfo {author} {\bibfnamefont {J.}~\bibnamefont {Kunz}},\ }\href
  {\doibase 10.1103/PhysRevLett.120.201103} {\bibfield  {journal} {\bibinfo
  {journal} {Phys. Rev. Lett.}\ }\textbf {\bibinfo {volume} {120}},\ \bibinfo
  {pages} {201103} (\bibinfo {year} {2018})},\ \Eprint
  {http://arxiv.org/abs/1711.05191} {arXiv:1711.05191 [gr-qc]} \BibitemShut
  {NoStop}%
\bibitem [{\citenamefont {Ye}\ and\ \citenamefont {Wei}(2023)}]{Ye:2023gmk}%
  \BibitemOpen
  \bibfield  {author} {\bibinfo {author} {\bibfnamefont {X.}~\bibnamefont
  {Ye}}\ and\ \bibinfo {author} {\bibfnamefont {S.-W.}\ \bibnamefont {Wei}},\
  }\href {\doibase 10.1088/1475-7516/2023/07/049} {\bibfield  {journal}
  {\bibinfo  {journal} {JCAP}\ }\textbf {\bibinfo {volume} {07}},\ \bibinfo
  {pages} {049} (\bibinfo {year} {2023})},\ \Eprint
  {http://arxiv.org/abs/2301.04786} {arXiv:2301.04786 [gr-qc]} \BibitemShut
  {NoStop}%
\bibitem [{\citenamefont {Cunha}\ \emph {et~al.}(2023)\citenamefont {Cunha},
  \citenamefont {Herdeiro}, \citenamefont {Radu},\ and\ \citenamefont
  {Sanchis-Gual}}]{Cunha:2022gde}%
  \BibitemOpen
  \bibfield  {author} {\bibinfo {author} {\bibfnamefont {P.~V.~P.}\
  \bibnamefont {Cunha}}, \bibinfo {author} {\bibfnamefont {C.}~\bibnamefont
  {Herdeiro}}, \bibinfo {author} {\bibfnamefont {E.}~\bibnamefont {Radu}}, \
  and\ \bibinfo {author} {\bibfnamefont {N.}~\bibnamefont {Sanchis-Gual}},\
  }\href {\doibase 10.1103/PhysRevLett.130.061401} {\bibfield  {journal}
  {\bibinfo  {journal} {Phys. Rev. Lett.}\ }\textbf {\bibinfo {volume} {130}},\
  \bibinfo {pages} {061401} (\bibinfo {year} {2023})},\ \Eprint
  {http://arxiv.org/abs/2207.13713} {arXiv:2207.13713 [gr-qc]} \BibitemShut
  {NoStop}%
\bibitem [{\citenamefont {Wei}\ and\ \citenamefont {Liu}(2023)}]{Wei:2022mzv}%
  \BibitemOpen
  \bibfield  {author} {\bibinfo {author} {\bibfnamefont {S.-W.}\ \bibnamefont
  {Wei}}\ and\ \bibinfo {author} {\bibfnamefont {Y.-X.}\ \bibnamefont {Liu}},\
  }\href {\doibase 10.1103/PhysRevD.107.064006} {\bibfield  {journal} {\bibinfo
   {journal} {Phys. Rev. D}\ }\textbf {\bibinfo {volume} {107}},\ \bibinfo
  {pages} {064006} (\bibinfo {year} {2023})},\ \Eprint
  {http://arxiv.org/abs/2207.08397} {arXiv:2207.08397 [gr-qc]} \BibitemShut
  {NoStop}%
\bibitem [{\citenamefont {Ghosh}\ and\ \citenamefont
  {Sarkar}(2021)}]{Ghosh:2021txu}%
  \BibitemOpen
  \bibfield  {author} {\bibinfo {author} {\bibfnamefont {R.}~\bibnamefont
  {Ghosh}}\ and\ \bibinfo {author} {\bibfnamefont {S.}~\bibnamefont {Sarkar}},\
  }\href {\doibase 10.1103/PhysRevD.104.044019} {\bibfield  {journal} {\bibinfo
   {journal} {Phys. Rev. D}\ }\textbf {\bibinfo {volume} {104}},\ \bibinfo
  {pages} {044019} (\bibinfo {year} {2021})},\ \Eprint
  {http://arxiv.org/abs/2107.07370} {arXiv:2107.07370 [gr-qc]} \BibitemShut
  {NoStop}%
\bibitem [{\citenamefont {Ghosh}\ \emph {et~al.}(2023)\citenamefont {Ghosh},
  \citenamefont {Sk},\ and\ \citenamefont {Sarkar}}]{Ghosh:2023kge}%
  \BibitemOpen
  \bibfield  {author} {\bibinfo {author} {\bibfnamefont {R.}~\bibnamefont
  {Ghosh}}, \bibinfo {author} {\bibfnamefont {S.}~\bibnamefont {Sk}}, \ and\
  \bibinfo {author} {\bibfnamefont {S.}~\bibnamefont {Sarkar}},\ }\href
  {\doibase 10.1103/PhysRevD.108.L041501} {\bibfield  {journal} {\bibinfo
  {journal} {Phys. Rev. D}\ }\textbf {\bibinfo {volume} {108}},\ \bibinfo
  {pages} {L041501} (\bibinfo {year} {2023})},\ \Eprint
  {http://arxiv.org/abs/2306.14193} {arXiv:2306.14193 [gr-qc]} \BibitemShut
  {NoStop}%
\bibitem [{\citenamefont {Glampedakis}\ \emph {et~al.}(2017)\citenamefont
  {Glampedakis}, \citenamefont {Pappas}, \citenamefont {Silva},\ and\
  \citenamefont {Berti}}]{Glampedakis:2017dvb}%
  \BibitemOpen
  \bibfield  {author} {\bibinfo {author} {\bibfnamefont {K.}~\bibnamefont
  {Glampedakis}}, \bibinfo {author} {\bibfnamefont {G.}~\bibnamefont {Pappas}},
  \bibinfo {author} {\bibfnamefont {H.~O.}\ \bibnamefont {Silva}}, \ and\
  \bibinfo {author} {\bibfnamefont {E.}~\bibnamefont {Berti}},\ }\href
  {\doibase 10.1103/PhysRevD.96.064054} {\bibfield  {journal} {\bibinfo
  {journal} {Phys. Rev. D}\ }\textbf {\bibinfo {volume} {96}},\ \bibinfo
  {pages} {064054} (\bibinfo {year} {2017})},\ \Eprint
  {http://arxiv.org/abs/1706.07658} {arXiv:1706.07658 [gr-qc]} \BibitemShut
  {NoStop}%
\bibitem [{\citenamefont {Berti}\ and\ \citenamefont
  {Kokkotas}(2005)}]{Berti:2005eb}%
  \BibitemOpen
  \bibfield  {author} {\bibinfo {author} {\bibfnamefont {E.}~\bibnamefont
  {Berti}}\ and\ \bibinfo {author} {\bibfnamefont {K.~D.}\ \bibnamefont
  {Kokkotas}},\ }\href {\doibase 10.1103/PhysRevD.71.124008} {\bibfield
  {journal} {\bibinfo  {journal} {Phys. Rev. D}\ }\textbf {\bibinfo {volume}
  {71}},\ \bibinfo {pages} {124008} (\bibinfo {year} {2005})},\ \Eprint
  {http://arxiv.org/abs/gr-qc/0502065} {arXiv:gr-qc/0502065} \BibitemShut
  {NoStop}%
\bibitem [{\citenamefont {Cardoso}\ \emph {et~al.}(2016)\citenamefont
  {Cardoso}, \citenamefont {Franzin},\ and\ \citenamefont
  {Pani}}]{Cardoso:2016rao}%
  \BibitemOpen
  \bibfield  {author} {\bibinfo {author} {\bibfnamefont {V.}~\bibnamefont
  {Cardoso}}, \bibinfo {author} {\bibfnamefont {E.}~\bibnamefont {Franzin}}, \
  and\ \bibinfo {author} {\bibfnamefont {P.}~\bibnamefont {Pani}},\ }\href
  {\doibase 10.1103/PhysRevLett.116.171101} {\bibfield  {journal} {\bibinfo
  {journal} {Phys. Rev. Lett.}\ }\textbf {\bibinfo {volume} {116}},\ \bibinfo
  {pages} {171101} (\bibinfo {year} {2016})},\ \bibinfo {note} {[Erratum:
  Phys.Rev.Lett. 117, 089902 (2016)]},\ \Eprint
  {http://arxiv.org/abs/1602.07309} {arXiv:1602.07309 [gr-qc]} \BibitemShut
  {NoStop}%
\bibitem [{\citenamefont {Decanini}\ and\ \citenamefont
  {Folacci}(2010)}]{Decanini:2009mu}%
  \BibitemOpen
  \bibfield  {author} {\bibinfo {author} {\bibfnamefont {Y.}~\bibnamefont
  {Decanini}}\ and\ \bibinfo {author} {\bibfnamefont {A.}~\bibnamefont
  {Folacci}},\ }\href {\doibase 10.1103/PhysRevD.81.024031} {\bibfield
  {journal} {\bibinfo  {journal} {Phys. Rev. D}\ }\textbf {\bibinfo {volume}
  {81}},\ \bibinfo {pages} {024031} (\bibinfo {year} {2010})},\ \Eprint
  {http://arxiv.org/abs/0906.2601} {arXiv:0906.2601 [gr-qc]} \BibitemShut
  {NoStop}%
\bibitem [{\citenamefont {Stefanov}\ \emph {et~al.}(2010)\citenamefont
  {Stefanov}, \citenamefont {Yazadjiev},\ and\ \citenamefont
  {Gyulchev}}]{Stefanov:2010xz}%
  \BibitemOpen
  \bibfield  {author} {\bibinfo {author} {\bibfnamefont {I.~Z.}\ \bibnamefont
  {Stefanov}}, \bibinfo {author} {\bibfnamefont {S.~S.}\ \bibnamefont
  {Yazadjiev}}, \ and\ \bibinfo {author} {\bibfnamefont {G.~G.}\ \bibnamefont
  {Gyulchev}},\ }\href {\doibase 10.1103/PhysRevLett.104.251103} {\bibfield
  {journal} {\bibinfo  {journal} {Phys. Rev. Lett.}\ }\textbf {\bibinfo
  {volume} {104}},\ \bibinfo {pages} {251103} (\bibinfo {year} {2010})},\
  \Eprint {http://arxiv.org/abs/1003.1609} {arXiv:1003.1609 [gr-qc]}
  \BibitemShut {NoStop}%
\bibitem [{\citenamefont {Cunha}\ and\ \citenamefont
  {Herdeiro}(2018)}]{Cunha:2018acu}%
  \BibitemOpen
  \bibfield  {author} {\bibinfo {author} {\bibfnamefont {P.~V.~P.}\
  \bibnamefont {Cunha}}\ and\ \bibinfo {author} {\bibfnamefont {C.~A.~R.}\
  \bibnamefont {Herdeiro}},\ }\href {\doibase 10.1007/s10714-018-2361-9}
  {\bibfield  {journal} {\bibinfo  {journal} {Gen. Rel. Grav.}\ }\textbf
  {\bibinfo {volume} {50}},\ \bibinfo {pages} {42} (\bibinfo {year} {2018})},\
  \Eprint {http://arxiv.org/abs/1801.00860} {arXiv:1801.00860 [gr-qc]}
  \BibitemShut {NoStop}%
\bibitem [{\citenamefont {Hod}(2018{\natexlab{a}})}]{Hod:2018kql}%
  \BibitemOpen
  \bibfield  {author} {\bibinfo {author} {\bibfnamefont {S.}~\bibnamefont
  {Hod}},\ }\href {\doibase 10.1140/epjc/s10052-018-5905-y} {\bibfield
  {journal} {\bibinfo  {journal} {Eur. Phys. J. C}\ }\textbf {\bibinfo {volume}
  {78}},\ \bibinfo {pages} {417} (\bibinfo {year} {2018}{\natexlab{a}})},\
  \Eprint {http://arxiv.org/abs/1811.04948} {arXiv:1811.04948 [gr-qc]}
  \BibitemShut {NoStop}%
\bibitem [{\citenamefont {Guo}\ \emph {et~al.}(2023)\citenamefont {Guo},
  \citenamefont {Lu}, \citenamefont {Wang}, \citenamefont {Wu},\ and\
  \citenamefont {Yang}}]{Guo:2022ghl}%
  \BibitemOpen
  \bibfield  {author} {\bibinfo {author} {\bibfnamefont {G.}~\bibnamefont
  {Guo}}, \bibinfo {author} {\bibfnamefont {Y.}~\bibnamefont {Lu}}, \bibinfo
  {author} {\bibfnamefont {P.}~\bibnamefont {Wang}}, \bibinfo {author}
  {\bibfnamefont {H.}~\bibnamefont {Wu}}, \ and\ \bibinfo {author}
  {\bibfnamefont {H.}~\bibnamefont {Yang}},\ }\href {\doibase
  10.1103/PhysRevD.107.124037} {\bibfield  {journal} {\bibinfo  {journal}
  {Phys. Rev. D}\ }\textbf {\bibinfo {volume} {107}},\ \bibinfo {pages}
  {124037} (\bibinfo {year} {2023})},\ \Eprint
  {http://arxiv.org/abs/2212.12901} {arXiv:2212.12901 [gr-qc]} \BibitemShut
  {NoStop}%
\bibitem [{\citenamefont {Wu}\ and\ \citenamefont {Wei}(2023)}]{Wu:2023eml}%
  \BibitemOpen
  \bibfield  {author} {\bibinfo {author} {\bibfnamefont {S.-P.}\ \bibnamefont
  {Wu}}\ and\ \bibinfo {author} {\bibfnamefont {S.-W.}\ \bibnamefont {Wei}},\
  }\href {\doibase 10.1103/PhysRevD.108.104041} {\bibfield  {journal} {\bibinfo
   {journal} {Phys. Rev. D}\ }\textbf {\bibinfo {volume} {108}},\ \bibinfo
  {pages} {104041} (\bibinfo {year} {2023})},\ \Eprint
  {http://arxiv.org/abs/2307.14003} {arXiv:2307.14003 [gr-qc]} \BibitemShut
  {NoStop}%
\bibitem [{\citenamefont {Sadeghi}\ and\ \citenamefont
  {Afshar}(2024)}]{Sadeghi:2024itx}%
  \BibitemOpen
  \bibfield  {author} {\bibinfo {author} {\bibfnamefont {J.}~\bibnamefont
  {Sadeghi}}\ and\ \bibinfo {author} {\bibfnamefont {M.~A.~S.}\ \bibnamefont
  {Afshar}},\ }\href {\doibase 10.1016/j.astropartphys.2024.102994} {\bibfield
  {journal} {\bibinfo  {journal} {Astropart. Phys.}\ }\textbf {\bibinfo
  {volume} {162}},\ \bibinfo {pages} {102994} (\bibinfo {year} {2024})},\
  \Eprint {http://arxiv.org/abs/2405.06568} {arXiv:2405.06568 [gr-qc]}
  \BibitemShut {NoStop}%
\bibitem [{\citenamefont {Afshar}\ and\ \citenamefont
  {Sadeghi}(2025{\natexlab{a}})}]{Afshar:2024bgi}%
  \BibitemOpen
  \bibfield  {author} {\bibinfo {author} {\bibfnamefont {M.~A.~S.}\
  \bibnamefont {Afshar}}\ and\ \bibinfo {author} {\bibfnamefont
  {J.}~\bibnamefont {Sadeghi}},\ }\href {\doibase 10.1088/1674-1137/ada127}
  {\bibfield  {journal} {\bibinfo  {journal} {Chin. Phys. C}\ }\textbf
  {\bibinfo {volume} {49}},\ \bibinfo {pages} {035107} (\bibinfo {year}
  {2025}{\natexlab{a}})},\ \Eprint {http://arxiv.org/abs/2405.18798}
  {arXiv:2405.18798 [gr-qc]} \BibitemShut {NoStop}%
\bibitem [{\citenamefont {Liu}\ \emph {et~al.}(2024)\citenamefont {Liu},
  \citenamefont {Wu},\ and\ \citenamefont {Wang}}]{Liu:2024soc}%
  \BibitemOpen
  \bibfield  {author} {\bibinfo {author} {\bibfnamefont {W.}~\bibnamefont
  {Liu}}, \bibinfo {author} {\bibfnamefont {D.}~\bibnamefont {Wu}}, \ and\
  \bibinfo {author} {\bibfnamefont {J.}~\bibnamefont {Wang}},\ }\href {\doibase
  10.1016/j.physletb.2024.139052} {\bibfield  {journal} {\bibinfo  {journal}
  {Phys. Lett. B}\ }\textbf {\bibinfo {volume} {858}},\ \bibinfo {pages}
  {139052} (\bibinfo {year} {2024})},\ \Eprint
  {http://arxiv.org/abs/2408.05569} {arXiv:2408.05569 [gr-qc]} \BibitemShut
  {NoStop}%
\bibitem [{\citenamefont {Song}\ \emph {et~al.}(2025)\citenamefont {Song},
  \citenamefont {Li}, \citenamefont {Cen}, \citenamefont {Diao}, \citenamefont
  {Zhao},\ and\ \citenamefont {Shi}}]{Song:2025vhw}%
  \BibitemOpen
  \bibfield  {author} {\bibinfo {author} {\bibfnamefont {Y.}~\bibnamefont
  {Song}}, \bibinfo {author} {\bibfnamefont {J.}~\bibnamefont {Li}}, \bibinfo
  {author} {\bibfnamefont {Y.}~\bibnamefont {Cen}}, \bibinfo {author}
  {\bibfnamefont {K.}~\bibnamefont {Diao}}, \bibinfo {author} {\bibfnamefont
  {X.}~\bibnamefont {Zhao}}, \ and\ \bibinfo {author} {\bibfnamefont
  {S.}~\bibnamefont {Shi}},\ }\href@noop {} {\  (\bibinfo {year} {2025})},\
  \Eprint {http://arxiv.org/abs/2504.05061} {arXiv:2504.05061 [gr-qc]}
  \BibitemShut {NoStop}%
\bibitem [{\citenamefont {Sadeghi}\ \emph {et~al.}(2024)\citenamefont
  {Sadeghi}, \citenamefont {Afshar}, \citenamefont {Noori~Gashti},\ and\
  \citenamefont {Alipour}}]{Sadeghi:2023dxy}%
  \BibitemOpen
  \bibfield  {author} {\bibinfo {author} {\bibfnamefont {J.}~\bibnamefont
  {Sadeghi}}, \bibinfo {author} {\bibfnamefont {M.~A.~S.}\ \bibnamefont
  {Afshar}}, \bibinfo {author} {\bibfnamefont {S.}~\bibnamefont
  {Noori~Gashti}}, \ and\ \bibinfo {author} {\bibfnamefont {M.~R.}\
  \bibnamefont {Alipour}},\ }\href {\doibase
  10.1016/j.astropartphys.2023.102920} {\bibfield  {journal} {\bibinfo
  {journal} {Astropart. Phys.}\ }\textbf {\bibinfo {volume} {156}},\ \bibinfo
  {pages} {102920} (\bibinfo {year} {2024})},\ \Eprint
  {http://arxiv.org/abs/2307.12873} {arXiv:2307.12873 [gr-qc]} \BibitemShut
  {NoStop}%
\bibitem [{\citenamefont {Noori~Gashti}\ \emph {et~al.}(2025)\citenamefont
  {Noori~Gashti}, \citenamefont {Afshar}, \citenamefont {Alipour},
  \citenamefont {Sakall{\i}}, \citenamefont {Pourhassan},\ and\ \citenamefont
  {Sadeghi}}]{NooriGashti:2025rwl}%
  \BibitemOpen
  \bibfield  {author} {\bibinfo {author} {\bibfnamefont {S.}~\bibnamefont
  {Noori~Gashti}}, \bibinfo {author} {\bibfnamefont {M.~A.~S.}\ \bibnamefont
  {Afshar}}, \bibinfo {author} {\bibfnamefont {M.~R.}\ \bibnamefont {Alipour}},
  \bibinfo {author} {\bibfnamefont {I.}~\bibnamefont {Sakall{\i}}}, \bibinfo
  {author} {\bibfnamefont {B.}~\bibnamefont {Pourhassan}}, \ and\ \bibinfo
  {author} {\bibfnamefont {J.}~\bibnamefont {Sadeghi}},\ }\href@noop {} {\
  (\bibinfo {year} {2025})},\ \Eprint {http://arxiv.org/abs/2504.11939}
  {arXiv:2504.11939 [hep-th]} \BibitemShut {NoStop}%
\bibitem [{\citenamefont {Hod}(2018{\natexlab{b}})}]{Hod:2017zpi}%
  \BibitemOpen
  \bibfield  {author} {\bibinfo {author} {\bibfnamefont {S.}~\bibnamefont
  {Hod}},\ }\href {\doibase 10.1016/j.physletb.2017.11.021} {\bibfield
  {journal} {\bibinfo  {journal} {Phys. Lett. B}\ }\textbf {\bibinfo {volume}
  {776}},\ \bibinfo {pages} {1} (\bibinfo {year} {2018}{\natexlab{b}})},\
  \Eprint {http://arxiv.org/abs/1710.00836} {arXiv:1710.00836 [gr-qc]}
  \BibitemShut {NoStop}%
\bibitem [{\citenamefont {Hod}(2013)}]{Hod:2013jhd}%
  \BibitemOpen
  \bibfield  {author} {\bibinfo {author} {\bibfnamefont {S.}~\bibnamefont
  {Hod}},\ }\href {\doibase 10.1016/j.physletb.2013.10.047} {\bibfield
  {journal} {\bibinfo  {journal} {Phys. Lett. B}\ }\textbf {\bibinfo {volume}
  {727}},\ \bibinfo {pages} {345} (\bibinfo {year} {2013})},\ \Eprint
  {http://arxiv.org/abs/1701.06587} {arXiv:1701.06587 [gr-qc]} \BibitemShut
  {NoStop}%
\bibitem [{\citenamefont {Abbott}\ and\ \citenamefont {et.
  al.}(2017)}]{LIGO2017}%
  \BibitemOpen
  \bibfield  {author} {\bibinfo {author} {\bibfnamefont {B.~P.}\ \bibnamefont
  {Abbott}}\ and\ \bibinfo {author} {\bibnamefont {et. al.}} (\bibinfo
  {collaboration} {LIGO Scientific and Virgo Collaboration}),\ }\href {\doibase
  10.1103/PhysRevLett.118.221101} {\bibfield  {journal} {\bibinfo  {journal}
  {Phys. Rev. Lett.}\ }\textbf {\bibinfo {volume} {118}},\ \bibinfo {pages}
  {221101} (\bibinfo {year} {2017})}\BibitemShut {NoStop}%
\bibitem [{\citenamefont {de~Rham}(2014{\natexlab{a}})}]{deRham2014Review}%
  \BibitemOpen
  \bibfield  {author} {\bibinfo {author} {\bibfnamefont {C.}~\bibnamefont
  {de~Rham}},\ }\href {\doibase 10.12942/lrr-2014-7} {\bibfield  {journal}
  {\bibinfo  {journal} {Living Rev. Rel.}\ }\textbf {\bibinfo {volume} {17}},\
  \bibinfo {pages} {7} (\bibinfo {year} {2014}{\natexlab{a}})},\ \Eprint
  {http://arxiv.org/abs/1401.4173} {arXiv:1401.4173 [hep-th]} \BibitemShut
  {NoStop}%
\bibitem [{\citenamefont {Dvali}\ \emph
  {et~al.}(2000{\natexlab{a}})\citenamefont {Dvali}, \citenamefont
  {Gabadadze},\ and\ \citenamefont {Porrati}}]{MassiveIb}%
  \BibitemOpen
  \bibfield  {author} {\bibinfo {author} {\bibfnamefont {G.}~\bibnamefont
  {Dvali}}, \bibinfo {author} {\bibfnamefont {G.}~\bibnamefont {Gabadadze}}, \
  and\ \bibinfo {author} {\bibfnamefont {M.}~\bibnamefont {Porrati}},\
  }\href@noop {} {\bibfield  {journal} {\bibinfo  {journal} {Physics Letters
  B}\ }\textbf {\bibinfo {volume} {484}},\ \bibinfo {pages} {112} (\bibinfo
  {year} {2000}{\natexlab{a}})}\BibitemShut {NoStop}%
\bibitem [{\citenamefont {Dvali}\ \emph
  {et~al.}(2000{\natexlab{b}})\citenamefont {Dvali}, \citenamefont
  {Gabadadze},\ and\ \citenamefont {Porrati}}]{MassiveIc}%
  \BibitemOpen
  \bibfield  {author} {\bibinfo {author} {\bibfnamefont {G.}~\bibnamefont
  {Dvali}}, \bibinfo {author} {\bibfnamefont {G.}~\bibnamefont {Gabadadze}}, \
  and\ \bibinfo {author} {\bibfnamefont {M.}~\bibnamefont {Porrati}},\
  }\href@noop {} {\bibfield  {journal} {\bibinfo  {journal} {Physics Letters
  B}\ }\textbf {\bibinfo {volume} {485}},\ \bibinfo {pages} {208} (\bibinfo
  {year} {2000}{\natexlab{b}})}\BibitemShut {NoStop}%
\bibitem [{\citenamefont {Fierz}\ and\ \citenamefont
  {Pauli}(1939)}]{Fierz1939}%
  \BibitemOpen
  \bibfield  {author} {\bibinfo {author} {\bibfnamefont {M.}~\bibnamefont
  {Fierz}}\ and\ \bibinfo {author} {\bibfnamefont {W.~E.}\ \bibnamefont
  {Pauli}},\ }\href@noop {} {\bibfield  {journal} {\bibinfo  {journal}
  {Proceedings of the Royal Society of London. Series A. Mathematical and
  Physical Sciences}\ }\textbf {\bibinfo {volume} {173}},\ \bibinfo {pages}
  {211} (\bibinfo {year} {1939})}\BibitemShut {NoStop}%
\bibitem [{\citenamefont {Boulware}\ and\ \citenamefont
  {Deser}(1972)}]{BDghost}%
  \BibitemOpen
  \bibfield  {author} {\bibinfo {author} {\bibfnamefont {D.~G.}\ \bibnamefont
  {Boulware}}\ and\ \bibinfo {author} {\bibfnamefont {S.}~\bibnamefont
  {Deser}},\ }\href {\doibase 10.1103/PhysRevD.6.3368} {\bibfield  {journal}
  {\bibinfo  {journal} {Phys. Rev. D}\ }\textbf {\bibinfo {volume} {6}},\
  \bibinfo {pages} {3368} (\bibinfo {year} {1972})}\BibitemShut {NoStop}%
\bibitem [{\citenamefont {Bergshoeff}\ \emph {et~al.}(2009)\citenamefont
  {Bergshoeff}, \citenamefont {Hohm},\ and\ \citenamefont
  {Townsend}}]{Newmasssive}%
  \BibitemOpen
  \bibfield  {author} {\bibinfo {author} {\bibfnamefont {E.~A.}\ \bibnamefont
  {Bergshoeff}}, \bibinfo {author} {\bibfnamefont {O.}~\bibnamefont {Hohm}}, \
  and\ \bibinfo {author} {\bibfnamefont {P.~K.}\ \bibnamefont {Townsend}},\
  }\href@noop {} {\bibfield  {journal} {\bibinfo  {journal} {Physical Review
  Letters}\ }\textbf {\bibinfo {volume} {102}},\ \bibinfo {pages} {201301}
  (\bibinfo {year} {2009})}\BibitemShut {NoStop}%
\bibitem [{\citenamefont {de~Rham}\ \emph
  {et~al.}(2011{\natexlab{a}})\citenamefont {de~Rham}, \citenamefont
  {Gabadadze},\ and\ \citenamefont {Tolley}}]{dRGTI}%
  \BibitemOpen
  \bibfield  {author} {\bibinfo {author} {\bibfnamefont {C.}~\bibnamefont
  {de~Rham}}, \bibinfo {author} {\bibfnamefont {G.}~\bibnamefont {Gabadadze}},
  \ and\ \bibinfo {author} {\bibfnamefont {A.~J.}\ \bibnamefont {Tolley}},\
  }\href {\doibase 10.1103/PhysRevLett.106.231101} {\bibfield  {journal}
  {\bibinfo  {journal} {Phys. Rev. Lett.}\ }\textbf {\bibinfo {volume} {106}},\
  \bibinfo {pages} {231101} (\bibinfo {year} {2011}{\natexlab{a}})}\BibitemShut
  {NoStop}%
\bibitem [{\citenamefont {de~Rham}\ \emph {et~al.}(2012)\citenamefont
  {de~Rham}, \citenamefont {Gabadadze},\ and\ \citenamefont {Tolley}}]{dRGTII}%
  \BibitemOpen
  \bibfield  {author} {\bibinfo {author} {\bibfnamefont {C.}~\bibnamefont
  {de~Rham}}, \bibinfo {author} {\bibfnamefont {G.}~\bibnamefont {Gabadadze}},
  \ and\ \bibinfo {author} {\bibfnamefont {A.~J.}\ \bibnamefont {Tolley}},\
  }\href@noop {} {\bibfield  {journal} {\bibinfo  {journal} {Physics Letters
  B}\ }\textbf {\bibinfo {volume} {711}},\ \bibinfo {pages} {190} (\bibinfo
  {year} {2012})}\BibitemShut {NoStop}%
\bibitem [{\citenamefont {Myung}\ \emph {et~al.}(2011)\citenamefont {Myung},
  \citenamefont {Kim}, \citenamefont {Moon},\ and\ \citenamefont
  {Park}}]{NewM1}%
  \BibitemOpen
  \bibfield  {author} {\bibinfo {author} {\bibfnamefont {Y.~S.}\ \bibnamefont
  {Myung}}, \bibinfo {author} {\bibfnamefont {Y.-W.}\ \bibnamefont {Kim}},
  \bibinfo {author} {\bibfnamefont {T.}~\bibnamefont {Moon}}, \ and\ \bibinfo
  {author} {\bibfnamefont {Y.-J.}\ \bibnamefont {Park}},\ }\href {\doibase
  10.1103/PhysRevD.84.024044} {\bibfield  {journal} {\bibinfo  {journal} {Phys.
  Rev. D}\ }\textbf {\bibinfo {volume} {84}},\ \bibinfo {pages} {024044}
  (\bibinfo {year} {2011})}\BibitemShut {NoStop}%
\bibitem [{\citenamefont {Bergshoeff}\ \emph {et~al.}(2011)\citenamefont
  {Bergshoeff}, \citenamefont {Hohm}, \citenamefont {Rosseel},\ and\
  \citenamefont {Townsend}}]{NewM2}%
  \BibitemOpen
  \bibfield  {author} {\bibinfo {author} {\bibfnamefont {E.~A.}\ \bibnamefont
  {Bergshoeff}}, \bibinfo {author} {\bibfnamefont {O.}~\bibnamefont {Hohm}},
  \bibinfo {author} {\bibfnamefont {J.}~\bibnamefont {Rosseel}}, \ and\
  \bibinfo {author} {\bibfnamefont {P.~K.}\ \bibnamefont {Townsend}},\ }\href
  {\doibase 10.1103/PhysRevD.83.104038} {\bibfield  {journal} {\bibinfo
  {journal} {Phys. Rev. D}\ }\textbf {\bibinfo {volume} {83}},\ \bibinfo
  {pages} {104038} (\bibinfo {year} {2011})}\BibitemShut {NoStop}%
\bibitem [{\citenamefont {Kim}\ \emph {et~al.}(2013)\citenamefont {Kim},
  \citenamefont {Kulkarni},\ and\ \citenamefont {Yi}}]{NewM3}%
  \BibitemOpen
  \bibfield  {author} {\bibinfo {author} {\bibfnamefont {W.}~\bibnamefont
  {Kim}}, \bibinfo {author} {\bibfnamefont {S.}~\bibnamefont {Kulkarni}}, \
  and\ \bibinfo {author} {\bibfnamefont {S.-H.}\ \bibnamefont {Yi}},\
  }\href@noop {} {\bibfield  {journal} {\bibinfo  {journal} {Journal of High
  Energy Physics}\ }\textbf {\bibinfo {volume} {2013}},\ \bibinfo {pages} {41}
  (\bibinfo {year} {2013})}\BibitemShut {NoStop}%
\bibitem [{\citenamefont {Ay\'on-Beato}\ \emph {et~al.}(2014)\citenamefont
  {Ay\'on-Beato}, \citenamefont {Hassa\"{\i}ne},\ and\ \citenamefont
  {Ju\'arez-Aubry}}]{NewM4}%
  \BibitemOpen
  \bibfield  {author} {\bibinfo {author} {\bibfnamefont {E.}~\bibnamefont
  {Ay\'on-Beato}}, \bibinfo {author} {\bibfnamefont {M.}~\bibnamefont
  {Hassa\"{\i}ne}}, \ and\ \bibinfo {author} {\bibfnamefont {M.~M.}\
  \bibnamefont {Ju\'arez-Aubry}},\ }\href {\doibase 10.1103/PhysRevD.90.044026}
  {\bibfield  {journal} {\bibinfo  {journal} {Phys. Rev. D}\ }\textbf {\bibinfo
  {volume} {90}},\ \bibinfo {pages} {044026} (\bibinfo {year}
  {2014})}\BibitemShut {NoStop}%
\bibitem [{\citenamefont {Myung}(2015)}]{NewM5}%
  \BibitemOpen
  \bibfield  {author} {\bibinfo {author} {\bibfnamefont {Y.~S.}\ \bibnamefont
  {Myung}},\ }\href@noop {} {\bibfield  {journal} {\bibinfo  {journal}
  {Advances in High Energy Physics}\ }\textbf {\bibinfo {volume} {2015}}
  (\bibinfo {year} {2015})}\BibitemShut {NoStop}%
\bibitem [{\citenamefont {Hassan}\ and\ \citenamefont {Rosen}(2012)}]{HassanI}%
  \BibitemOpen
  \bibfield  {author} {\bibinfo {author} {\bibfnamefont {S.~F.}\ \bibnamefont
  {Hassan}}\ and\ \bibinfo {author} {\bibfnamefont {R.~A.}\ \bibnamefont
  {Rosen}},\ }\href {\doibase 10.1103/PhysRevLett.108.041101} {\bibfield
  {journal} {\bibinfo  {journal} {Phys. Rev. Lett.}\ }\textbf {\bibinfo
  {volume} {108}},\ \bibinfo {pages} {041101} (\bibinfo {year}
  {2012})}\BibitemShut {NoStop}%
\bibitem [{\citenamefont {Hassan}\ \emph {et~al.}(2012)\citenamefont {Hassan},
  \citenamefont {Rosen},\ and\ \citenamefont {Schmidt-May}}]{HassanII}%
  \BibitemOpen
  \bibfield  {author} {\bibinfo {author} {\bibfnamefont {S.~F.}\ \bibnamefont
  {Hassan}}, \bibinfo {author} {\bibfnamefont {R.~A.}\ \bibnamefont {Rosen}}, \
  and\ \bibinfo {author} {\bibfnamefont {A.}~\bibnamefont {Schmidt-May}},\
  }\href@noop {} {\bibfield  {journal} {\bibinfo  {journal} {Journal of High
  Energy Physics}\ }\textbf {\bibinfo {volume} {2012}},\ \bibinfo {pages} {26}
  (\bibinfo {year} {2012})}\BibitemShut {NoStop}%
\bibitem [{\citenamefont {Cai}\ \emph {et~al.}(2013)\citenamefont {Cai},
  \citenamefont {Easson}, \citenamefont {Gao},\ and\ \citenamefont
  {Saridakis}}]{BHMassiveI}%
  \BibitemOpen
  \bibfield  {author} {\bibinfo {author} {\bibfnamefont {Y.-F.}\ \bibnamefont
  {Cai}}, \bibinfo {author} {\bibfnamefont {D.~A.}\ \bibnamefont {Easson}},
  \bibinfo {author} {\bibfnamefont {C.}~\bibnamefont {Gao}}, \ and\ \bibinfo
  {author} {\bibfnamefont {E.~N.}\ \bibnamefont {Saridakis}},\ }\href {\doibase
  10.1103/PhysRevD.87.064001} {\bibfield  {journal} {\bibinfo  {journal} {Phys.
  Rev. D}\ }\textbf {\bibinfo {volume} {87}},\ \bibinfo {pages} {064001}
  (\bibinfo {year} {2013})}\BibitemShut {NoStop}%
\bibitem [{\citenamefont {Kodama}\ and\ \citenamefont
  {Arraut}(2014)}]{BHMassiveII}%
  \BibitemOpen
  \bibfield  {author} {\bibinfo {author} {\bibfnamefont {H.}~\bibnamefont
  {Kodama}}\ and\ \bibinfo {author} {\bibfnamefont {I.}~\bibnamefont
  {Arraut}},\ }\href@noop {} {\bibfield  {journal} {\bibinfo  {journal}
  {Progress of Theoretical and Experimental Physics}\ }\textbf {\bibinfo
  {volume} {2014}} (\bibinfo {year} {2014})}\BibitemShut {NoStop}%
\bibitem [{\citenamefont {Zou}\ \emph {et~al.}(2017)\citenamefont {Zou},
  \citenamefont {Yue},\ and\ \citenamefont {Zhang}}]{BHMassiveIII}%
  \BibitemOpen
  \bibfield  {author} {\bibinfo {author} {\bibfnamefont {D.-C.}\ \bibnamefont
  {Zou}}, \bibinfo {author} {\bibfnamefont {R.}~\bibnamefont {Yue}}, \ and\
  \bibinfo {author} {\bibfnamefont {M.}~\bibnamefont {Zhang}},\ }\href@noop {}
  {\bibfield  {journal} {\bibinfo  {journal} {The European Physical Journal C}\
  }\textbf {\bibinfo {volume} {77}},\ \bibinfo {pages} {256} (\bibinfo {year}
  {2017})}\BibitemShut {NoStop}%
\bibitem [{\citenamefont {Tannukij}\ \emph {et~al.}(2017)\citenamefont
  {Tannukij}, \citenamefont {Wongjun},\ and\ \citenamefont
  {Ghosh}}]{BHMassiveIV}%
  \BibitemOpen
  \bibfield  {author} {\bibinfo {author} {\bibfnamefont {L.}~\bibnamefont
  {Tannukij}}, \bibinfo {author} {\bibfnamefont {P.}~\bibnamefont {Wongjun}}, \
  and\ \bibinfo {author} {\bibfnamefont {S.~G.}\ \bibnamefont {Ghosh}},\ }\href
  {\doibase 10.1140/epjc/s10052-017-5426-0} {\bibfield  {journal} {\bibinfo
  {journal} {Eur. Phys. J. C}\ }\textbf {\bibinfo {volume} {77}},\ \bibinfo
  {pages} {846} (\bibinfo {year} {2017})},\ \Eprint
  {http://arxiv.org/abs/1701.05332} {arXiv:1701.05332 [gr-qc]} \BibitemShut
  {NoStop}%
\bibitem [{\citenamefont {Katsuragawa}\ \emph {et~al.}(2016)\citenamefont
  {Katsuragawa}, \citenamefont {Nojiri}, \citenamefont {Odintsov},\ and\
  \citenamefont {Yamazaki}}]{Katsuragawa}%
  \BibitemOpen
  \bibfield  {author} {\bibinfo {author} {\bibfnamefont {T.}~\bibnamefont
  {Katsuragawa}}, \bibinfo {author} {\bibfnamefont {S.}~\bibnamefont {Nojiri}},
  \bibinfo {author} {\bibfnamefont {S.~D.}\ \bibnamefont {Odintsov}}, \ and\
  \bibinfo {author} {\bibfnamefont {M.}~\bibnamefont {Yamazaki}},\ }\href
  {\doibase 10.1103/PhysRevD.93.124013} {\bibfield  {journal} {\bibinfo
  {journal} {Phys. Rev. D}\ }\textbf {\bibinfo {volume} {93}},\ \bibinfo
  {pages} {124013} (\bibinfo {year} {2016})}\BibitemShut {NoStop}%
\bibitem [{\citenamefont {Saridakis}(2013)}]{Saridakis}%
  \BibitemOpen
  \bibfield  {author} {\bibinfo {author} {\bibfnamefont {E.~N.}\ \bibnamefont
  {Saridakis}},\ }\href@noop {} {\bibfield  {journal} {\bibinfo  {journal}
  {Classical and Quantum Gravity}\ }\textbf {\bibinfo {volume} {30}},\ \bibinfo
  {pages} {075003} (\bibinfo {year} {2013})}\BibitemShut {NoStop}%
\bibitem [{\citenamefont {Cai}\ \emph {et~al.}(2012)\citenamefont {Cai},
  \citenamefont {Gao},\ and\ \citenamefont {Saridakis}}]{YFCai}%
  \BibitemOpen
  \bibfield  {author} {\bibinfo {author} {\bibfnamefont {Y.-F.}\ \bibnamefont
  {Cai}}, \bibinfo {author} {\bibfnamefont {C.}~\bibnamefont {Gao}}, \ and\
  \bibinfo {author} {\bibfnamefont {E.~N.}\ \bibnamefont {Saridakis}},\ }\href
  {\doibase 10.1088/1475-7516/2012/10/048} {\bibfield  {journal} {\bibinfo
  {journal} {JCAP}\ }\textbf {\bibinfo {volume} {10}},\ \bibinfo {pages} {048}
  (\bibinfo {year} {2012})},\ \Eprint {http://arxiv.org/abs/1207.3786}
  {arXiv:1207.3786 [astro-ph.CO]} \BibitemShut {NoStop}%
\bibitem [{\citenamefont {Leon}\ \emph {et~al.}(2013)\citenamefont {Leon},
  \citenamefont {Saavedra},\ and\ \citenamefont {Saridakis}}]{Leon}%
  \BibitemOpen
  \bibfield  {author} {\bibinfo {author} {\bibfnamefont {G.}~\bibnamefont
  {Leon}}, \bibinfo {author} {\bibfnamefont {J.}~\bibnamefont {Saavedra}}, \
  and\ \bibinfo {author} {\bibfnamefont {E.~N.}\ \bibnamefont {Saridakis}},\
  }\href@noop {} {\bibfield  {journal} {\bibinfo  {journal} {Classical and
  Quantum Gravity}\ }\textbf {\bibinfo {volume} {30}},\ \bibinfo {pages}
  {135001} (\bibinfo {year} {2013})}\BibitemShut {NoStop}%
\bibitem [{\citenamefont {Hinterbichler}\ \emph {et~al.}(2013)\citenamefont
  {Hinterbichler}, \citenamefont {Stokes},\ and\ \citenamefont
  {Trodden}}]{Hinterbichler}%
  \BibitemOpen
  \bibfield  {author} {\bibinfo {author} {\bibfnamefont {K.}~\bibnamefont
  {Hinterbichler}}, \bibinfo {author} {\bibfnamefont {J.}~\bibnamefont
  {Stokes}}, \ and\ \bibinfo {author} {\bibfnamefont {M.}~\bibnamefont
  {Trodden}},\ }\href {\doibase https://doi.org/10.1016/j.physletb.2013.07.009}
  {\bibfield  {journal} {\bibinfo  {journal} {Physics Letters B}\ }\textbf
  {\bibinfo {volume} {725}},\ \bibinfo {pages} {1 } (\bibinfo {year}
  {2013})}\BibitemShut {NoStop}%
\bibitem [{\citenamefont {Fasiello}\ and\ \citenamefont
  {Tolley}(2013)}]{Fasiello}%
  \BibitemOpen
  \bibfield  {author} {\bibinfo {author} {\bibfnamefont {M.}~\bibnamefont
  {Fasiello}}\ and\ \bibinfo {author} {\bibfnamefont {A.~J.}\ \bibnamefont
  {Tolley}},\ }\href@noop {} {\bibfield  {journal} {\bibinfo  {journal}
  {Journal of Cosmology and Astroparticle Physics}\ }\textbf {\bibinfo {volume}
  {2013}},\ \bibinfo {pages} {002} (\bibinfo {year} {2013})}\BibitemShut
  {NoStop}%
\bibitem [{\citenamefont {Bamba}\ \emph {et~al.}(2014)\citenamefont {Bamba},
  \citenamefont {Hossain}, \citenamefont {Myrzakulov}, \citenamefont {Nojiri},\
  and\ \citenamefont {Sami}}]{Bamba}%
  \BibitemOpen
  \bibfield  {author} {\bibinfo {author} {\bibfnamefont {K.}~\bibnamefont
  {Bamba}}, \bibinfo {author} {\bibfnamefont {M.~W.}\ \bibnamefont {Hossain}},
  \bibinfo {author} {\bibfnamefont {R.}~\bibnamefont {Myrzakulov}}, \bibinfo
  {author} {\bibfnamefont {S.}~\bibnamefont {Nojiri}}, \ and\ \bibinfo {author}
  {\bibfnamefont {M.}~\bibnamefont {Sami}},\ }\href {\doibase
  10.1103/PhysRevD.89.083518} {\bibfield  {journal} {\bibinfo  {journal} {Phys.
  Rev. D}\ }\textbf {\bibinfo {volume} {89}},\ \bibinfo {pages} {083518}
  (\bibinfo {year} {2014})}\BibitemShut {NoStop}%
\bibitem [{\citenamefont {Vegh}(2013)}]{Vegh}%
  \BibitemOpen
  \bibfield  {author} {\bibinfo {author} {\bibfnamefont {D.}~\bibnamefont
  {Vegh}},\ }\href@noop {} {\  (\bibinfo {year} {2013})},\ \Eprint
  {http://arxiv.org/abs/1301.0537} {arXiv:1301.0537 [hep-th]} \BibitemShut
  {NoStop}%
\bibitem [{\citenamefont {Hinterbichler}(2012)}]{Hinterbichler:2011tt}%
  \BibitemOpen
  \bibfield  {author} {\bibinfo {author} {\bibfnamefont {K.}~\bibnamefont
  {Hinterbichler}},\ }\href {\doibase 10.1103/RevModPhys.84.671} {\bibfield
  {journal} {\bibinfo  {journal} {Rev. Mod. Phys.}\ }\textbf {\bibinfo {volume}
  {84}},\ \bibinfo {pages} {671} (\bibinfo {year} {2012})},\ \Eprint
  {http://arxiv.org/abs/1105.3735} {arXiv:1105.3735 [hep-th]} \BibitemShut
  {NoStop}%
\bibitem [{\citenamefont {de~Rham}(2014{\natexlab{b}})}]{deRham:2014zqa}%
  \BibitemOpen
  \bibfield  {author} {\bibinfo {author} {\bibfnamefont {C.}~\bibnamefont
  {de~Rham}},\ }\href {\doibase 10.12942/lrr-2014-7} {\bibfield  {journal}
  {\bibinfo  {journal} {Living Rev. Rel.}\ }\textbf {\bibinfo {volume} {17}},\
  \bibinfo {pages} {7} (\bibinfo {year} {2014}{\natexlab{b}})},\ \Eprint
  {http://arxiv.org/abs/1401.4173} {arXiv:1401.4173 [hep-th]} \BibitemShut
  {NoStop}%
\bibitem [{\citenamefont {G{\"u}mr{\"u}k{\c{c}}{\"u}o{\u{g}}lu}\ \emph
  {et~al.}(2011)\citenamefont {G{\"u}mr{\"u}k{\c{c}}{\"u}o{\u{g}}lu},
  \citenamefont {Lin},\ and\ \citenamefont {Mukohyama}}]{Gumrukcuoglu}%
  \BibitemOpen
  \bibfield  {author} {\bibinfo {author} {\bibfnamefont {A.~E.}\ \bibnamefont
  {G{\"u}mr{\"u}k{\c{c}}{\"u}o{\u{g}}lu}}, \bibinfo {author} {\bibfnamefont
  {C.}~\bibnamefont {Lin}}, \ and\ \bibinfo {author} {\bibfnamefont
  {S.}~\bibnamefont {Mukohyama}},\ }\href@noop {} {\bibfield  {journal}
  {\bibinfo  {journal} {Journal of Cosmology and Astroparticle Physics}\
  }\textbf {\bibinfo {volume} {2011}},\ \bibinfo {pages} {030} (\bibinfo {year}
  {2011})}\BibitemShut {NoStop}%
\bibitem [{\citenamefont {Gratia}\ \emph {et~al.}(2012)\citenamefont {Gratia},
  \citenamefont {Hu},\ and\ \citenamefont {Wyman}}]{Gratia}%
  \BibitemOpen
  \bibfield  {author} {\bibinfo {author} {\bibfnamefont {P.}~\bibnamefont
  {Gratia}}, \bibinfo {author} {\bibfnamefont {W.}~\bibnamefont {Hu}}, \ and\
  \bibinfo {author} {\bibfnamefont {M.}~\bibnamefont {Wyman}},\ }\href
  {\doibase 10.1103/PhysRevD.86.061504} {\bibfield  {journal} {\bibinfo
  {journal} {Phys. Rev. D}\ }\textbf {\bibinfo {volume} {86}},\ \bibinfo
  {pages} {061504} (\bibinfo {year} {2012})}\BibitemShut {NoStop}%
\bibitem [{\citenamefont {Kobayashi}\ \emph {et~al.}(2012)\citenamefont
  {Kobayashi}, \citenamefont {Siino}, \citenamefont {Yamaguchi},\ and\
  \citenamefont {Yoshida}}]{Kobayash}%
  \BibitemOpen
  \bibfield  {author} {\bibinfo {author} {\bibfnamefont {T.}~\bibnamefont
  {Kobayashi}}, \bibinfo {author} {\bibfnamefont {M.}~\bibnamefont {Siino}},
  \bibinfo {author} {\bibfnamefont {M.}~\bibnamefont {Yamaguchi}}, \ and\
  \bibinfo {author} {\bibfnamefont {D.}~\bibnamefont {Yoshida}},\ }\href
  {\doibase 10.1103/PhysRevD.86.061505} {\bibfield  {journal} {\bibinfo
  {journal} {Phys. Rev. D}\ }\textbf {\bibinfo {volume} {86}},\ \bibinfo
  {pages} {061505} (\bibinfo {year} {2012})}\BibitemShut {NoStop}%
\bibitem [{\citenamefont {Deffayet}(2001)}]{DeffayetI}%
  \BibitemOpen
  \bibfield  {author} {\bibinfo {author} {\bibfnamefont {C.}~\bibnamefont
  {Deffayet}},\ }\href {\doibase 10.1016/S0370-2693(01)00160-5} {\bibfield
  {journal} {\bibinfo  {journal} {Phys. Lett. B}\ }\textbf {\bibinfo {volume}
  {502}},\ \bibinfo {pages} {199} (\bibinfo {year} {2001})},\ \Eprint
  {http://arxiv.org/abs/hep-th/0010186} {arXiv:hep-th/0010186} \BibitemShut
  {NoStop}%
\bibitem [{\citenamefont {Deffayet}\ \emph {et~al.}(2002)\citenamefont
  {Deffayet}, \citenamefont {Dvali},\ and\ \citenamefont
  {Gabadadze}}]{DeffayetII}%
  \BibitemOpen
  \bibfield  {author} {\bibinfo {author} {\bibfnamefont {C.}~\bibnamefont
  {Deffayet}}, \bibinfo {author} {\bibfnamefont {G.}~\bibnamefont {Dvali}}, \
  and\ \bibinfo {author} {\bibfnamefont {G.}~\bibnamefont {Gabadadze}},\ }\href
  {\doibase 10.1103/PhysRevD.65.044023} {\bibfield  {journal} {\bibinfo
  {journal} {Phys. Rev. D}\ }\textbf {\bibinfo {volume} {65}},\ \bibinfo
  {pages} {044023} (\bibinfo {year} {2002})}\BibitemShut {NoStop}%
\bibitem [{\citenamefont {Dvali}\ \emph {et~al.}(2003)\citenamefont {Dvali},
  \citenamefont {Gabadadze},\ and\ \citenamefont {Shifman}}]{DvaliI}%
  \BibitemOpen
  \bibfield  {author} {\bibinfo {author} {\bibfnamefont {G.}~\bibnamefont
  {Dvali}}, \bibinfo {author} {\bibfnamefont {G.}~\bibnamefont {Gabadadze}}, \
  and\ \bibinfo {author} {\bibfnamefont {M.}~\bibnamefont {Shifman}},\ }\href
  {\doibase 10.1103/PhysRevD.67.044020} {\bibfield  {journal} {\bibinfo
  {journal} {Phys. Rev. D}\ }\textbf {\bibinfo {volume} {67}},\ \bibinfo
  {pages} {044020} (\bibinfo {year} {2003})}\BibitemShut {NoStop}%
\bibitem [{\citenamefont {Dvali}\ \emph {et~al.}(2007)\citenamefont {Dvali},
  \citenamefont {Hofmann},\ and\ \citenamefont {Khoury}}]{DvaliII}%
  \BibitemOpen
  \bibfield  {author} {\bibinfo {author} {\bibfnamefont {G.}~\bibnamefont
  {Dvali}}, \bibinfo {author} {\bibfnamefont {S.}~\bibnamefont {Hofmann}}, \
  and\ \bibinfo {author} {\bibfnamefont {J.}~\bibnamefont {Khoury}},\ }\href
  {\doibase 10.1103/PhysRevD.76.084006} {\bibfield  {journal} {\bibinfo
  {journal} {Phys. Rev. D}\ }\textbf {\bibinfo {volume} {76}},\ \bibinfo
  {pages} {084006} (\bibinfo {year} {2007})}\BibitemShut {NoStop}%
\bibitem [{\citenamefont {Will}(2014)}]{Will}%
  \BibitemOpen
  \bibfield  {author} {\bibinfo {author} {\bibfnamefont {C.~M.}\ \bibnamefont
  {Will}},\ }\href@noop {} {\bibfield  {journal} {\bibinfo  {journal} {Living
  reviews in relativity}\ }\textbf {\bibinfo {volume} {17}},\ \bibinfo {pages}
  {4} (\bibinfo {year} {2014})}\BibitemShut {NoStop}%
\bibitem [{\citenamefont {Mohseni}(2011)}]{Mohseni}%
  \BibitemOpen
  \bibfield  {author} {\bibinfo {author} {\bibfnamefont {M.}~\bibnamefont
  {Mohseni}},\ }\href {\doibase 10.1103/PhysRevD.84.064026} {\bibfield
  {journal} {\bibinfo  {journal} {Phys. Rev. D}\ }\textbf {\bibinfo {volume}
  {84}},\ \bibinfo {pages} {064026} (\bibinfo {year} {2011})}\BibitemShut
  {NoStop}%
\bibitem [{\citenamefont {Gumrukcuoglu}\ \emph {et~al.}(2012)\citenamefont
  {Gumrukcuoglu}, \citenamefont {Kuroyanagi}, \citenamefont {Lin},
  \citenamefont {Mukohyama},\ and\ \citenamefont {Tanahashi}}]{GumrukcuogluII}%
  \BibitemOpen
  \bibfield  {author} {\bibinfo {author} {\bibfnamefont {A.}~\bibnamefont
  {Gumrukcuoglu}}, \bibinfo {author} {\bibfnamefont {S.}~\bibnamefont
  {Kuroyanagi}}, \bibinfo {author} {\bibfnamefont {C.}~\bibnamefont {Lin}},
  \bibinfo {author} {\bibfnamefont {S.}~\bibnamefont {Mukohyama}}, \ and\
  \bibinfo {author} {\bibfnamefont {N.}~\bibnamefont {Tanahashi}},\ }\href
  {\doibase 10.1088/0264-9381/29/23/235026} {\bibfield  {journal} {\bibinfo
  {journal} {Class. Quant. Grav.}\ }\textbf {\bibinfo {volume} {29}},\ \bibinfo
  {pages} {235026} (\bibinfo {year} {2012})},\ \Eprint
  {http://arxiv.org/abs/1208.5975} {arXiv:1208.5975 [hep-th]} \BibitemShut
  {NoStop}%
\bibitem [{\citenamefont {Hendi}\ \emph
  {et~al.}(2017{\natexlab{a}})\citenamefont {Hendi}, \citenamefont {Bordbar},
  \citenamefont {Eslam~Panah},\ and\ \citenamefont {Panahiyan}}]{NeutronMass}%
  \BibitemOpen
  \bibfield  {author} {\bibinfo {author} {\bibfnamefont {S.}~\bibnamefont
  {Hendi}}, \bibinfo {author} {\bibfnamefont {G.}~\bibnamefont {Bordbar}},
  \bibinfo {author} {\bibfnamefont {B.}~\bibnamefont {Eslam~Panah}}, \ and\
  \bibinfo {author} {\bibfnamefont {S.}~\bibnamefont {Panahiyan}},\ }\href
  {\doibase 10.1088/1475-7516/2017/07/004} {\bibfield  {journal} {\bibinfo
  {journal} {JCAP}\ }\textbf {\bibinfo {volume} {07}},\ \bibinfo {pages} {004}
  (\bibinfo {year} {2017}{\natexlab{a}})},\ \Eprint
  {http://arxiv.org/abs/1701.01039} {arXiv:1701.01039 [gr-qc]} \BibitemShut
  {NoStop}%
\bibitem [{\citenamefont {Rhoades}\ and\ \citenamefont
  {Ruffini}(1974)}]{Ruffini}%
  \BibitemOpen
  \bibfield  {author} {\bibinfo {author} {\bibfnamefont {C.~E.}\ \bibnamefont
  {Rhoades}}\ and\ \bibinfo {author} {\bibfnamefont {R.}~\bibnamefont
  {Ruffini}},\ }\href {\doibase 10.1103/PhysRevLett.32.324} {\bibfield
  {journal} {\bibinfo  {journal} {Phys. Rev. Lett.}\ }\textbf {\bibinfo
  {volume} {32}},\ \bibinfo {pages} {324} (\bibinfo {year} {1974})}\BibitemShut
  {NoStop}%
\bibitem [{\citenamefont {Eslam~Panah}\ and\ \citenamefont
  {Liu}(2019)}]{EslamPanah:2018evk}%
  \BibitemOpen
  \bibfield  {author} {\bibinfo {author} {\bibfnamefont {B.}~\bibnamefont
  {Eslam~Panah}}\ and\ \bibinfo {author} {\bibfnamefont {H.~L.}\ \bibnamefont
  {Liu}},\ }\href {\doibase 10.1103/PhysRevD.99.104074} {\bibfield  {journal}
  {\bibinfo  {journal} {Phys. Rev. D}\ }\textbf {\bibinfo {volume} {99}},\
  \bibinfo {pages} {104074} (\bibinfo {year} {2019})},\ \Eprint
  {http://arxiv.org/abs/1805.10650} {arXiv:1805.10650 [gr-qc]} \BibitemShut
  {NoStop}%
\bibitem [{\citenamefont {Babichev}\ \emph {et~al.}(2016)\citenamefont
  {Babichev}, \citenamefont {Marzola}, \citenamefont {Raidal}, \citenamefont
  {Schmidt-May}, \citenamefont {Urban}, \citenamefont {Veermäe},\ and\
  \citenamefont {von Strauss}}]{Schmidt-May2016DarkMatter}%
  \BibitemOpen
  \bibfield  {author} {\bibinfo {author} {\bibfnamefont {E.}~\bibnamefont
  {Babichev}}, \bibinfo {author} {\bibfnamefont {L.}~\bibnamefont {Marzola}},
  \bibinfo {author} {\bibfnamefont {M.}~\bibnamefont {Raidal}}, \bibinfo
  {author} {\bibfnamefont {A.}~\bibnamefont {Schmidt-May}}, \bibinfo {author}
  {\bibfnamefont {F.}~\bibnamefont {Urban}}, \bibinfo {author} {\bibfnamefont
  {H.}~\bibnamefont {Veermäe}}, \ and\ \bibinfo {author} {\bibfnamefont
  {M.}~\bibnamefont {von Strauss}},\ }\href {\doibase
  10.1088/1475-7516/2016/09/016} {\bibfield  {journal} {\bibinfo  {journal}
  {JCAP}\ }\textbf {\bibinfo {volume} {09}},\ \bibinfo {pages} {016} (\bibinfo
  {year} {2016})},\ \Eprint {http://arxiv.org/abs/1607.03497} {arXiv:1607.03497
  [hep-th]} \BibitemShut {NoStop}%
\bibitem [{\citenamefont {Akrami}\ \emph {et~al.}(2013)\citenamefont {Akrami},
  \citenamefont {Koivisto},\ and\ \citenamefont
  {Sandstad}}]{MassiveCosmology2013}%
  \BibitemOpen
  \bibfield  {author} {\bibinfo {author} {\bibfnamefont {Y.}~\bibnamefont
  {Akrami}}, \bibinfo {author} {\bibfnamefont {T.~S.}\ \bibnamefont
  {Koivisto}}, \ and\ \bibinfo {author} {\bibfnamefont {M.}~\bibnamefont
  {Sandstad}},\ }\href@noop {} {\bibfield  {journal} {\bibinfo  {journal}
  {Journal of High Energy Physics}\ }\textbf {\bibinfo {volume} {2013}},\
  \bibinfo {pages} {99} (\bibinfo {year} {2013})}\BibitemShut {NoStop}%
\bibitem [{\citenamefont {Akrami}\ \emph {et~al.}(2015)\citenamefont {Akrami},
  \citenamefont {Hassan}, \citenamefont {K{\"o}nnig}, \citenamefont
  {Schmidt-May},\ and\ \citenamefont {Solomon}}]{MassiveCosmology2015}%
  \BibitemOpen
  \bibfield  {author} {\bibinfo {author} {\bibfnamefont {Y.}~\bibnamefont
  {Akrami}}, \bibinfo {author} {\bibfnamefont {S.~F.}\ \bibnamefont {Hassan}},
  \bibinfo {author} {\bibfnamefont {F.}~\bibnamefont {K{\"o}nnig}}, \bibinfo
  {author} {\bibfnamefont {A.}~\bibnamefont {Schmidt-May}}, \ and\ \bibinfo
  {author} {\bibfnamefont {A.~R.}\ \bibnamefont {Solomon}},\ }\href@noop {}
  {\bibfield  {journal} {\bibinfo  {journal} {Physics Letters B}\ }\textbf
  {\bibinfo {volume} {748}},\ \bibinfo {pages} {37} (\bibinfo {year}
  {2015})}\BibitemShut {NoStop}%
\bibitem [{\citenamefont {Bachas}\ and\ \citenamefont
  {Lavdas}(2018)}]{MGinString2018}%
  \BibitemOpen
  \bibfield  {author} {\bibinfo {author} {\bibfnamefont {C.}~\bibnamefont
  {Bachas}}\ and\ \bibinfo {author} {\bibfnamefont {I.}~\bibnamefont
  {Lavdas}},\ }\href@noop {} {\bibfield  {journal} {\bibinfo  {journal}
  {Journal of High Energy Physics}\ }\textbf {\bibinfo {volume} {2018}},\
  \bibinfo {pages} {3} (\bibinfo {year} {2018})}\BibitemShut {NoStop}%
\bibitem [{\citenamefont {Geng}\ and\ \citenamefont
  {Karch}(2020)}]{Geng:2020qvw}%
  \BibitemOpen
  \bibfield  {author} {\bibinfo {author} {\bibfnamefont {H.}~\bibnamefont
  {Geng}}\ and\ \bibinfo {author} {\bibfnamefont {A.}~\bibnamefont {Karch}},\
  }\href {\doibase 10.1007/JHEP09(2020)121} {\bibfield  {journal} {\bibinfo
  {journal} {JHEP}\ }\textbf {\bibinfo {volume} {09}},\ \bibinfo {pages} {121}
  (\bibinfo {year} {2020})},\ \Eprint {http://arxiv.org/abs/2006.02438}
  {arXiv:2006.02438 [hep-th]} \BibitemShut {NoStop}%
\bibitem [{\citenamefont {Geng}\ \emph {et~al.}(2021)\citenamefont {Geng},
  \citenamefont {Karch}, \citenamefont {Perez-Pardavila}, \citenamefont {Raju},
  \citenamefont {Randall}, \citenamefont {Riojas},\ and\ \citenamefont
  {Shashi}}]{Geng:2020fxl}%
  \BibitemOpen
  \bibfield  {author} {\bibinfo {author} {\bibfnamefont {H.}~\bibnamefont
  {Geng}}, \bibinfo {author} {\bibfnamefont {A.}~\bibnamefont {Karch}},
  \bibinfo {author} {\bibfnamefont {C.}~\bibnamefont {Perez-Pardavila}},
  \bibinfo {author} {\bibfnamefont {S.}~\bibnamefont {Raju}}, \bibinfo {author}
  {\bibfnamefont {L.}~\bibnamefont {Randall}}, \bibinfo {author} {\bibfnamefont
  {M.}~\bibnamefont {Riojas}}, \ and\ \bibinfo {author} {\bibfnamefont
  {S.}~\bibnamefont {Shashi}},\ }\href {\doibase 10.21468/SciPostPhys.10.5.103}
  {\bibfield  {journal} {\bibinfo  {journal} {SciPost Phys.}\ }\textbf
  {\bibinfo {volume} {10}},\ \bibinfo {pages} {103} (\bibinfo {year} {2021})},\
  \Eprint {http://arxiv.org/abs/2012.04671} {arXiv:2012.04671 [hep-th]}
  \BibitemShut {NoStop}%
\bibitem [{\citenamefont {Ahmed}\ \emph {et~al.}(2023)\citenamefont {Ahmed},
  \citenamefont {Cong}, \citenamefont {Kubiz\v{n}\'ak}, \citenamefont {Mann},\
  and\ \citenamefont {Visser}}]{Ahmed:2023snm}%
  \BibitemOpen
  \bibfield  {author} {\bibinfo {author} {\bibfnamefont {M.~B.}\ \bibnamefont
  {Ahmed}}, \bibinfo {author} {\bibfnamefont {W.}~\bibnamefont {Cong}},
  \bibinfo {author} {\bibfnamefont {D.}~\bibnamefont {Kubiz\v{n}\'ak}},
  \bibinfo {author} {\bibfnamefont {R.~B.}\ \bibnamefont {Mann}}, \ and\
  \bibinfo {author} {\bibfnamefont {M.~R.}\ \bibnamefont {Visser}},\ }\href
  {\doibase 10.1103/PhysRevLett.130.181401} {\bibfield  {journal} {\bibinfo
  {journal} {Phys. Rev. Lett.}\ }\textbf {\bibinfo {volume} {130}},\ \bibinfo
  {pages} {181401} (\bibinfo {year} {2023})},\ \Eprint
  {http://arxiv.org/abs/2302.08163} {arXiv:2302.08163 [hep-th]} \BibitemShut
  {NoStop}%
\bibitem [{\citenamefont {Mann}(2025)}]{Mann:2025xrb}%
  \BibitemOpen
  \bibfield  {author} {\bibinfo {author} {\bibfnamefont {R.~B.}\ \bibnamefont
  {Mann}},\ }\href {\doibase 10.1142/S0218271825420015} {\bibfield  {journal}
  {\bibinfo  {journal} {Int. J. Mod. Phys. D}\ }\textbf {\bibinfo {volume}
  {34}},\ \bibinfo {pages} {2542001} (\bibinfo {year} {2025})},\ \Eprint
  {http://arxiv.org/abs/2508.01830} {arXiv:2508.01830 [gr-qc]} \BibitemShut
  {NoStop}%
\bibitem [{\citenamefont {Cai}\ \emph {et~al.}(2015)\citenamefont {Cai},
  \citenamefont {Hu}, \citenamefont {Pan},\ and\ \citenamefont
  {Zhang}}]{Cai2015}%
  \BibitemOpen
  \bibfield  {author} {\bibinfo {author} {\bibfnamefont {R.-G.}\ \bibnamefont
  {Cai}}, \bibinfo {author} {\bibfnamefont {Y.-P.}\ \bibnamefont {Hu}},
  \bibinfo {author} {\bibfnamefont {Q.-Y.}\ \bibnamefont {Pan}}, \ and\
  \bibinfo {author} {\bibfnamefont {Y.-L.}\ \bibnamefont {Zhang}},\ }\href
  {\doibase 10.1103/PhysRevD.91.024032} {\bibfield  {journal} {\bibinfo
  {journal} {Phys. Rev. D}\ }\textbf {\bibinfo {volume} {91}},\ \bibinfo
  {pages} {024032} (\bibinfo {year} {2015})}\BibitemShut {NoStop}%
\bibitem [{\citenamefont {Hendi}\ \emph
  {et~al.}(2017{\natexlab{b}})\citenamefont {Hendi}, \citenamefont {Panah},\
  and\ \citenamefont {Panahiyan}}]{PVMassV}%
  \BibitemOpen
  \bibfield  {author} {\bibinfo {author} {\bibfnamefont {S.}~\bibnamefont
  {Hendi}}, \bibinfo {author} {\bibfnamefont {B.~E.}\ \bibnamefont {Panah}}, \
  and\ \bibinfo {author} {\bibfnamefont {S.}~\bibnamefont {Panahiyan}},\
  }\href@noop {} {\bibfield  {journal} {\bibinfo  {journal} {Physics Letters
  B}\ }\textbf {\bibinfo {volume} {769}},\ \bibinfo {pages} {191} (\bibinfo
  {year} {2017}{\natexlab{b}})}\BibitemShut {NoStop}%
\bibitem [{\citenamefont {Hendi}\ \emph
  {et~al.}(2017{\natexlab{c}})\citenamefont {Hendi}, \citenamefont {Mann},
  \citenamefont {Panahiyan},\ and\ \citenamefont {Eslam~Panah}}]{PVMassIV}%
  \BibitemOpen
  \bibfield  {author} {\bibinfo {author} {\bibfnamefont {S.~H.}\ \bibnamefont
  {Hendi}}, \bibinfo {author} {\bibfnamefont {R.~B.}\ \bibnamefont {Mann}},
  \bibinfo {author} {\bibfnamefont {S.}~\bibnamefont {Panahiyan}}, \ and\
  \bibinfo {author} {\bibfnamefont {B.}~\bibnamefont {Eslam~Panah}},\ }\href
  {\doibase 10.1103/PhysRevD.95.021501} {\bibfield  {journal} {\bibinfo
  {journal} {Phys. Rev. D}\ }\textbf {\bibinfo {volume} {95}},\ \bibinfo
  {pages} {021501} (\bibinfo {year} {2017}{\natexlab{c}})}\BibitemShut
  {NoStop}%
\bibitem [{\citenamefont {Alberte}\ and\ \citenamefont
  {Khmelnitsky}(2015)}]{Alberte}%
  \BibitemOpen
  \bibfield  {author} {\bibinfo {author} {\bibfnamefont {L.}~\bibnamefont
  {Alberte}}\ and\ \bibinfo {author} {\bibfnamefont {A.}~\bibnamefont
  {Khmelnitsky}},\ }\href {\doibase 10.1103/PhysRevD.91.046006} {\bibfield
  {journal} {\bibinfo  {journal} {Phys. Rev. D}\ }\textbf {\bibinfo {volume}
  {91}},\ \bibinfo {pages} {046006} (\bibinfo {year} {2015})}\BibitemShut
  {NoStop}%
\bibitem [{\citenamefont {Zhou}\ \emph {et~al.}(2015)\citenamefont {Zhou},
  \citenamefont {Wu},\ and\ \citenamefont {Ling}}]{Zhou}%
  \BibitemOpen
  \bibfield  {author} {\bibinfo {author} {\bibfnamefont {Z.}~\bibnamefont
  {Zhou}}, \bibinfo {author} {\bibfnamefont {J.-P.}\ \bibnamefont {Wu}}, \ and\
  \bibinfo {author} {\bibfnamefont {Y.}~\bibnamefont {Ling}},\ }\href@noop {}
  {\bibfield  {journal} {\bibinfo  {journal} {Journal of High Energy Physics}\
  }\textbf {\bibinfo {volume} {2015}},\ \bibinfo {pages} {67} (\bibinfo {year}
  {2015})}\BibitemShut {NoStop}%
\bibitem [{\citenamefont {Dehyadegari}\ \emph {et~al.}(2017)\citenamefont
  {Dehyadegari}, \citenamefont {Kord~Zangeneh},\ and\ \citenamefont
  {Sheykhi}}]{Dehyadegari}%
  \BibitemOpen
  \bibfield  {author} {\bibinfo {author} {\bibfnamefont {A.}~\bibnamefont
  {Dehyadegari}}, \bibinfo {author} {\bibfnamefont {M.}~\bibnamefont
  {Kord~Zangeneh}}, \ and\ \bibinfo {author} {\bibfnamefont {A.}~\bibnamefont
  {Sheykhi}},\ }\href {\doibase 10.1016/j.physletb.2017.08.029} {\bibfield
  {journal} {\bibinfo  {journal} {Phys. Lett. B}\ }\textbf {\bibinfo {volume}
  {773}},\ \bibinfo {pages} {344} (\bibinfo {year} {2017})},\ \Eprint
  {http://arxiv.org/abs/1703.00975} {arXiv:1703.00975 [hep-th]} \BibitemShut
  {NoStop}%
\bibitem [{\citenamefont {Hendi}\ \emph
  {et~al.}(2017{\natexlab{d}})\citenamefont {Hendi}, \citenamefont
  {Eslam~Panah}, \citenamefont {Panahiyan},\ and\ \citenamefont
  {Momennia}}]{Magmass}%
  \BibitemOpen
  \bibfield  {author} {\bibinfo {author} {\bibfnamefont {S.}~\bibnamefont
  {Hendi}}, \bibinfo {author} {\bibfnamefont {B.}~\bibnamefont {Eslam~Panah}},
  \bibinfo {author} {\bibfnamefont {S.}~\bibnamefont {Panahiyan}}, \ and\
  \bibinfo {author} {\bibfnamefont {M.}~\bibnamefont {Momennia}},\ }\href
  {\doibase 10.1016/j.physletb.2017.10.053} {\bibfield  {journal} {\bibinfo
  {journal} {Phys. Lett. B}\ }\textbf {\bibinfo {volume} {775}},\ \bibinfo
  {pages} {251} (\bibinfo {year} {2017}{\natexlab{d}})},\ \Eprint
  {http://arxiv.org/abs/1704.00996} {arXiv:1704.00996 [gr-qc]} \BibitemShut
  {NoStop}%
\bibitem [{\citenamefont {Dehghani}\ and\ \citenamefont
  {Hendi}(2020)}]{Dehghani:2019thq}%
  \BibitemOpen
  \bibfield  {author} {\bibinfo {author} {\bibfnamefont {A.}~\bibnamefont
  {Dehghani}}\ and\ \bibinfo {author} {\bibfnamefont {S.~H.}\ \bibnamefont
  {Hendi}},\ }\href {\doibase 10.1088/1361-6382/ab5eb4} {\bibfield  {journal}
  {\bibinfo  {journal} {Class. Quant. Grav.}\ }\textbf {\bibinfo {volume}
  {37}},\ \bibinfo {pages} {024001} (\bibinfo {year} {2020})},\ \Eprint
  {http://arxiv.org/abs/1909.00956} {arXiv:1909.00956 [hep-th]} \BibitemShut
  {NoStop}%
\bibitem [{\citenamefont {Hendi}\ \emph
  {et~al.}(2016{\natexlab{a}})\citenamefont {Hendi}, \citenamefont {Panahiyan},
  \citenamefont {Eslam~Panah},\ and\ \citenamefont {Momennia}}]{Hendi:2015eca}%
  \BibitemOpen
  \bibfield  {author} {\bibinfo {author} {\bibfnamefont {S.~H.}\ \bibnamefont
  {Hendi}}, \bibinfo {author} {\bibfnamefont {S.}~\bibnamefont {Panahiyan}},
  \bibinfo {author} {\bibfnamefont {B.}~\bibnamefont {Eslam~Panah}}, \ and\
  \bibinfo {author} {\bibfnamefont {M.}~\bibnamefont {Momennia}},\ }\href
  {\doibase 10.1002/andp.201600180} {\bibfield  {journal} {\bibinfo  {journal}
  {Annalen Phys.}\ }\textbf {\bibinfo {volume} {528}},\ \bibinfo {pages} {819}
  (\bibinfo {year} {2016}{\natexlab{a}})},\ \Eprint
  {http://arxiv.org/abs/1506.07262} {arXiv:1506.07262 [hep-th]} \BibitemShut
  {NoStop}%
\bibitem [{\citenamefont {Akbarieh}\ \emph {et~al.}(2021)\citenamefont
  {Akbarieh}, \citenamefont {Kazempour},\ and\ \citenamefont
  {Shao}}]{Akbarieh:2021vhv}%
  \BibitemOpen
  \bibfield  {author} {\bibinfo {author} {\bibfnamefont {A.~R.}\ \bibnamefont
  {Akbarieh}}, \bibinfo {author} {\bibfnamefont {S.}~\bibnamefont {Kazempour}},
  \ and\ \bibinfo {author} {\bibfnamefont {L.}~\bibnamefont {Shao}},\ }\href
  {\doibase 10.1103/PhysRevD.103.123518} {\bibfield  {journal} {\bibinfo
  {journal} {Phys. Rev. D}\ }\textbf {\bibinfo {volume} {103}},\ \bibinfo
  {pages} {123518} (\bibinfo {year} {2021})},\ \Eprint
  {http://arxiv.org/abs/2105.03744} {arXiv:2105.03744 [gr-qc]} \BibitemShut
  {NoStop}%
\bibitem [{\citenamefont {Chen}\ \emph {et~al.}(2023)\citenamefont {Chen},
  \citenamefont {He},\ and\ \citenamefont {Tao}}]{Chen:2023ddv}%
  \BibitemOpen
  \bibfield  {author} {\bibinfo {author} {\bibfnamefont {D.}~\bibnamefont
  {Chen}}, \bibinfo {author} {\bibfnamefont {Y.}~\bibnamefont {He}}, \ and\
  \bibinfo {author} {\bibfnamefont {J.}~\bibnamefont {Tao}},\ }\href {\doibase
  10.1140/epjc/s10052-023-11983-0} {\bibfield  {journal} {\bibinfo  {journal}
  {Eur. Phys. J. C}\ }\textbf {\bibinfo {volume} {83}},\ \bibinfo {pages} {872}
  (\bibinfo {year} {2023})},\ \Eprint {http://arxiv.org/abs/2306.13286}
  {arXiv:2306.13286 [gr-qc]} \BibitemShut {NoStop}%
\bibitem [{\citenamefont {Yerra}\ and\ \citenamefont
  {Bhamidipati}(2021{\natexlab{a}})}]{Yerra:2020tzg}%
  \BibitemOpen
  \bibfield  {author} {\bibinfo {author} {\bibfnamefont {P.~K.}\ \bibnamefont
  {Yerra}}\ and\ \bibinfo {author} {\bibfnamefont {C.}~\bibnamefont
  {Bhamidipati}},\ }\href {\doibase 10.1016/j.physletb.2021.136450} {\bibfield
  {journal} {\bibinfo  {journal} {Phys. Lett. B}\ }\textbf {\bibinfo {volume}
  {819}},\ \bibinfo {pages} {136450} (\bibinfo {year} {2021}{\natexlab{a}})},\
  \Eprint {http://arxiv.org/abs/2007.11515} {arXiv:2007.11515 [hep-th]}
  \BibitemShut {NoStop}%
\bibitem [{\citenamefont {Yerra}\ and\ \citenamefont
  {Bhamidipati}(2020)}]{Yerra:2020oph}%
  \BibitemOpen
  \bibfield  {author} {\bibinfo {author} {\bibfnamefont {P.~K.}\ \bibnamefont
  {Yerra}}\ and\ \bibinfo {author} {\bibfnamefont {C.}~\bibnamefont
  {Bhamidipati}},\ }\href {\doibase 10.1142/S0217751X20501201} {\bibfield
  {journal} {\bibinfo  {journal} {Int. J. Mod. Phys. A}\ }\textbf {\bibinfo
  {volume} {35}},\ \bibinfo {pages} {2050120} (\bibinfo {year} {2020})},\
  \Eprint {http://arxiv.org/abs/2006.07775} {arXiv:2006.07775 [hep-th]}
  \BibitemShut {NoStop}%
\bibitem [{\citenamefont {Yerra}\ and\ \citenamefont
  {Bhamidipati}(2021{\natexlab{b}})}]{Yerra:2021hnh}%
  \BibitemOpen
  \bibfield  {author} {\bibinfo {author} {\bibfnamefont {P.~K.}\ \bibnamefont
  {Yerra}}\ and\ \bibinfo {author} {\bibfnamefont {C.}~\bibnamefont
  {Bhamidipati}},\ }\href {\doibase 10.1103/PhysRevD.104.104049} {\bibfield
  {journal} {\bibinfo  {journal} {Phys. Rev. D}\ }\textbf {\bibinfo {volume}
  {104}},\ \bibinfo {pages} {104049} (\bibinfo {year} {2021}{\natexlab{b}})},\
  \Eprint {http://arxiv.org/abs/2107.04504} {arXiv:2107.04504 [gr-qc]}
  \BibitemShut {NoStop}%
\bibitem [{\citenamefont {H\"og\r{a}s}\ and\ \citenamefont
  {M\"ortsell}(2021)}]{Hogas:2021saw}%
  \BibitemOpen
  \bibfield  {author} {\bibinfo {author} {\bibfnamefont {M.}~\bibnamefont
  {H\"og\r{a}s}}\ and\ \bibinfo {author} {\bibfnamefont {E.}~\bibnamefont
  {M\"ortsell}},\ }\href@noop {} {\  (\bibinfo {year} {2021})},\ \Eprint
  {http://arxiv.org/abs/2106.09030} {arXiv:2106.09030 [astro-ph.CO]}
  \BibitemShut {NoStop}%
\bibitem [{\citenamefont {Caravano}\ \emph {et~al.}(2021)\citenamefont
  {Caravano}, \citenamefont {L\"uben},\ and\ \citenamefont
  {Weller}}]{Caravano:2021aum}%
  \BibitemOpen
  \bibfield  {author} {\bibinfo {author} {\bibfnamefont {A.}~\bibnamefont
  {Caravano}}, \bibinfo {author} {\bibfnamefont {M.}~\bibnamefont {L\"uben}}, \
  and\ \bibinfo {author} {\bibfnamefont {J.}~\bibnamefont {Weller}},\
  }\href@noop {} {\  (\bibinfo {year} {2021})},\ \Eprint
  {http://arxiv.org/abs/2101.08791} {arXiv:2101.08791 [gr-qc]} \BibitemShut
  {NoStop}%
\bibitem [{\citenamefont {Chabab}\ \emph {et~al.}(2019)\citenamefont {Chabab},
  \citenamefont {El~Moumni}, \citenamefont {Iraoui},\ and\ \citenamefont
  {Masmar}}]{Chabab:2019mlu}%
  \BibitemOpen
  \bibfield  {author} {\bibinfo {author} {\bibfnamefont {M.}~\bibnamefont
  {Chabab}}, \bibinfo {author} {\bibfnamefont {H.}~\bibnamefont {El~Moumni}},
  \bibinfo {author} {\bibfnamefont {S.}~\bibnamefont {Iraoui}}, \ and\ \bibinfo
  {author} {\bibfnamefont {K.}~\bibnamefont {Masmar}},\ }\href {\doibase
  10.1140/epjc/s10052-019-6850-0} {\bibfield  {journal} {\bibinfo  {journal}
  {Eur. Phys. J. C}\ }\textbf {\bibinfo {volume} {79}},\ \bibinfo {pages} {342}
  (\bibinfo {year} {2019})},\ \Eprint {http://arxiv.org/abs/1904.03532}
  {arXiv:1904.03532 [hep-th]} \BibitemShut {NoStop}%
\bibitem [{\citenamefont {Wu}\ \emph {et~al.}(2020)\citenamefont {Wu},
  \citenamefont {Wang}, \citenamefont {Xu},\ and\ \citenamefont
  {Yang}}]{Wu:2020fij}%
  \BibitemOpen
  \bibfield  {author} {\bibinfo {author} {\bibfnamefont {B.}~\bibnamefont
  {Wu}}, \bibinfo {author} {\bibfnamefont {C.}~\bibnamefont {Wang}}, \bibinfo
  {author} {\bibfnamefont {Z.-M.}\ \bibnamefont {Xu}}, \ and\ \bibinfo {author}
  {\bibfnamefont {W.-L.}\ \bibnamefont {Yang}},\ }\href@noop {} {\  (\bibinfo
  {year} {2020})},\ \Eprint {http://arxiv.org/abs/2006.09021} {arXiv:2006.09021
  [gr-qc]} \BibitemShut {NoStop}%
\bibitem [{\citenamefont {de~Rham}\ \emph
  {et~al.}(2011{\natexlab{b}})\citenamefont {de~Rham}, \citenamefont
  {Gabadadze},\ and\ \citenamefont {Tolley}}]{deRham:2010kj}%
  \BibitemOpen
  \bibfield  {author} {\bibinfo {author} {\bibfnamefont {C.}~\bibnamefont
  {de~Rham}}, \bibinfo {author} {\bibfnamefont {G.}~\bibnamefont {Gabadadze}},
  \ and\ \bibinfo {author} {\bibfnamefont {A.~J.}\ \bibnamefont {Tolley}},\
  }\href {\doibase 10.1103/PhysRevLett.106.231101} {\bibfield  {journal}
  {\bibinfo  {journal} {Phys. Rev. Lett.}\ }\textbf {\bibinfo {volume} {106}},\
  \bibinfo {pages} {231101} (\bibinfo {year} {2011}{\natexlab{b}})},\ \Eprint
  {http://arxiv.org/abs/1011.1232} {arXiv:1011.1232 [hep-th]} \BibitemShut
  {NoStop}%
\bibitem [{\citenamefont {Ghosh}\ \emph {et~al.}(2016)\citenamefont {Ghosh},
  \citenamefont {Tannukij},\ and\ \citenamefont {Wongjun}}]{Ghosh:2015cva}%
  \BibitemOpen
  \bibfield  {author} {\bibinfo {author} {\bibfnamefont {S.~G.}\ \bibnamefont
  {Ghosh}}, \bibinfo {author} {\bibfnamefont {L.}~\bibnamefont {Tannukij}}, \
  and\ \bibinfo {author} {\bibfnamefont {P.}~\bibnamefont {Wongjun}},\ }\href
  {\doibase 10.1140/epjc/s10052-016-3943-x} {\bibfield  {journal} {\bibinfo
  {journal} {Eur. Phys. J. C}\ }\textbf {\bibinfo {volume} {76}},\ \bibinfo
  {pages} {119} (\bibinfo {year} {2016})},\ \Eprint
  {http://arxiv.org/abs/1506.07119} {arXiv:1506.07119 [gr-qc]} \BibitemShut
  {NoStop}%
\bibitem [{\citenamefont {Hendi}\ \emph {et~al.}(2023)\citenamefont {Hendi},
  \citenamefont {Jafarzade},\ and\ \citenamefont
  {Eslam~Panah}}]{Hendi:2022qgi}%
  \BibitemOpen
  \bibfield  {author} {\bibinfo {author} {\bibfnamefont {S.~H.}\ \bibnamefont
  {Hendi}}, \bibinfo {author} {\bibfnamefont {K.}~\bibnamefont {Jafarzade}}, \
  and\ \bibinfo {author} {\bibfnamefont {B.}~\bibnamefont {Eslam~Panah}},\
  }\href {\doibase 10.1088/1475-7516/2023/02/022} {\bibfield  {journal}
  {\bibinfo  {journal} {JCAP}\ }\textbf {\bibinfo {volume} {02}},\ \bibinfo
  {pages} {022} (\bibinfo {year} {2023})},\ \Eprint
  {http://arxiv.org/abs/2206.05132} {arXiv:2206.05132 [gr-qc]} \BibitemShut
  {NoStop}%
\bibitem [{\citenamefont {Zhang}\ and\ \citenamefont {Li}(2016)}]{HZhang}%
  \BibitemOpen
  \bibfield  {author} {\bibinfo {author} {\bibfnamefont {H.}~\bibnamefont
  {Zhang}}\ and\ \bibinfo {author} {\bibfnamefont {X.-Z.}\ \bibnamefont {Li}},\
  }\href {\doibase 10.1103/PhysRevD.93.124039} {\bibfield  {journal} {\bibinfo
  {journal} {Phys. Rev. D}\ }\textbf {\bibinfo {volume} {93}},\ \bibinfo
  {pages} {124039} (\bibinfo {year} {2016})}\BibitemShut {NoStop}%
\bibitem [{\citenamefont {Xu}\ \emph {et~al.}(2015)\citenamefont {Xu},
  \citenamefont {Cao},\ and\ \citenamefont {Hu}}]{PVMassI}%
  \BibitemOpen
  \bibfield  {author} {\bibinfo {author} {\bibfnamefont {J.}~\bibnamefont
  {Xu}}, \bibinfo {author} {\bibfnamefont {L.-M.}\ \bibnamefont {Cao}}, \ and\
  \bibinfo {author} {\bibfnamefont {Y.-P.}\ \bibnamefont {Hu}},\ }\href
  {\doibase 10.1103/PhysRevD.91.124033} {\bibfield  {journal} {\bibinfo
  {journal} {Phys. Rev. D}\ }\textbf {\bibinfo {volume} {91}},\ \bibinfo
  {pages} {124033} (\bibinfo {year} {2015})}\BibitemShut {NoStop}%
\bibitem [{\citenamefont {Hendi}\ \emph {et~al.}(2015)\citenamefont {Hendi},
  \citenamefont {Panah},\ and\ \citenamefont {Panahiyan}}]{PVMassII}%
  \BibitemOpen
  \bibfield  {author} {\bibinfo {author} {\bibfnamefont {S.~H.}\ \bibnamefont
  {Hendi}}, \bibinfo {author} {\bibfnamefont {B.~E.}\ \bibnamefont {Panah}}, \
  and\ \bibinfo {author} {\bibfnamefont {S.}~\bibnamefont {Panahiyan}},\
  }\href@noop {} {\bibfield  {journal} {\bibinfo  {journal} {Journal of High
  Energy Physics}\ }\textbf {\bibinfo {volume} {2015}},\ \bibinfo {pages} {157}
  (\bibinfo {year} {2015})}\BibitemShut {NoStop}%
\bibitem [{\citenamefont {Hendi}\ \emph
  {et~al.}(2016{\natexlab{b}})\citenamefont {Hendi}, \citenamefont {Panah},\
  and\ \citenamefont {Panahiyan}}]{PVMassIII}%
  \BibitemOpen
  \bibfield  {author} {\bibinfo {author} {\bibfnamefont {S.}~\bibnamefont
  {Hendi}}, \bibinfo {author} {\bibfnamefont {B.~E.}\ \bibnamefont {Panah}}, \
  and\ \bibinfo {author} {\bibfnamefont {S.}~\bibnamefont {Panahiyan}},\
  }\href@noop {} {\bibfield  {journal} {\bibinfo  {journal} {Classical and
  Quantum Gravity}\ }\textbf {\bibinfo {volume} {33}},\ \bibinfo {pages}
  {235007} (\bibinfo {year} {2016}{\natexlab{b}})}\BibitemShut {NoStop}%
\bibitem [{\citenamefont {Davison}(2013)}]{Davison:2013jba}%
  \BibitemOpen
  \bibfield  {author} {\bibinfo {author} {\bibfnamefont {R.~A.}\ \bibnamefont
  {Davison}},\ }\href {\doibase 10.1103/PhysRevD.88.086003} {\bibfield
  {journal} {\bibinfo  {journal} {Phys. Rev. D}\ }\textbf {\bibinfo {volume}
  {88}},\ \bibinfo {pages} {086003} (\bibinfo {year} {2013})},\ \Eprint
  {http://arxiv.org/abs/1306.5792} {arXiv:1306.5792 [hep-th]} \BibitemShut
  {NoStop}%
\bibitem [{\citenamefont {Hartnoll}\ \emph
  {et~al.}(2008{\natexlab{a}})\citenamefont {Hartnoll}, \citenamefont
  {Herzog},\ and\ \citenamefont {Horowitz}}]{Hartnoll:2008vx}%
  \BibitemOpen
  \bibfield  {author} {\bibinfo {author} {\bibfnamefont {S.~A.}\ \bibnamefont
  {Hartnoll}}, \bibinfo {author} {\bibfnamefont {C.~P.}\ \bibnamefont
  {Herzog}}, \ and\ \bibinfo {author} {\bibfnamefont {G.~T.}\ \bibnamefont
  {Horowitz}},\ }\href {\doibase 10.1103/PhysRevLett.101.031601} {\bibfield
  {journal} {\bibinfo  {journal} {Phys. Rev. Lett.}\ }\textbf {\bibinfo
  {volume} {101}},\ \bibinfo {pages} {031601} (\bibinfo {year}
  {2008}{\natexlab{a}})},\ \Eprint {http://arxiv.org/abs/0803.3295}
  {arXiv:0803.3295 [hep-th]} \BibitemShut {NoStop}%
\bibitem [{\citenamefont {Hartnoll}\ \emph
  {et~al.}(2008{\natexlab{b}})\citenamefont {Hartnoll}, \citenamefont
  {Herzog},\ and\ \citenamefont {Horowitz}}]{Hartnoll:2008kx}%
  \BibitemOpen
  \bibfield  {author} {\bibinfo {author} {\bibfnamefont {S.~A.}\ \bibnamefont
  {Hartnoll}}, \bibinfo {author} {\bibfnamefont {C.~P.}\ \bibnamefont
  {Herzog}}, \ and\ \bibinfo {author} {\bibfnamefont {G.~T.}\ \bibnamefont
  {Horowitz}},\ }\href {\doibase 10.1088/1126-6708/2008/12/015} {\bibfield
  {journal} {\bibinfo  {journal} {JHEP}\ }\textbf {\bibinfo {volume} {12}},\
  \bibinfo {pages} {015} (\bibinfo {year} {2008}{\natexlab{b}})},\ \Eprint
  {http://arxiv.org/abs/0810.1563} {arXiv:0810.1563 [hep-th]} \BibitemShut
  {NoStop}%
\bibitem [{\citenamefont {Gregory}\ \emph {et~al.}(2009)\citenamefont
  {Gregory}, \citenamefont {Kanno},\ and\ \citenamefont
  {Soda}}]{Gregory:2009fj}%
  \BibitemOpen
  \bibfield  {author} {\bibinfo {author} {\bibfnamefont {R.}~\bibnamefont
  {Gregory}}, \bibinfo {author} {\bibfnamefont {S.}~\bibnamefont {Kanno}}, \
  and\ \bibinfo {author} {\bibfnamefont {J.}~\bibnamefont {Soda}},\ }\href
  {\doibase 10.1088/1126-6708/2009/10/010} {\bibfield  {journal} {\bibinfo
  {journal} {JHEP}\ }\textbf {\bibinfo {volume} {10}},\ \bibinfo {pages} {010}
  (\bibinfo {year} {2009})},\ \Eprint {http://arxiv.org/abs/0907.3203}
  {arXiv:0907.3203 [hep-th]} \BibitemShut {NoStop}%
\bibitem [{\citenamefont {Barclay}\ \emph {et~al.}(2010)\citenamefont
  {Barclay}, \citenamefont {Gregory}, \citenamefont {Kanno},\ and\
  \citenamefont {Sutcliffe}}]{Barclay:2010up}%
  \BibitemOpen
  \bibfield  {author} {\bibinfo {author} {\bibfnamefont {L.}~\bibnamefont
  {Barclay}}, \bibinfo {author} {\bibfnamefont {R.}~\bibnamefont {Gregory}},
  \bibinfo {author} {\bibfnamefont {S.}~\bibnamefont {Kanno}}, \ and\ \bibinfo
  {author} {\bibfnamefont {P.}~\bibnamefont {Sutcliffe}},\ }\href {\doibase
  10.1007/JHEP12(2010)029} {\bibfield  {journal} {\bibinfo  {journal} {JHEP}\
  }\textbf {\bibinfo {volume} {12}},\ \bibinfo {pages} {029} (\bibinfo {year}
  {2010})},\ \Eprint {http://arxiv.org/abs/1009.1991} {arXiv:1009.1991
  [hep-th]} \BibitemShut {NoStop}%
\bibitem [{\citenamefont {Alberte}\ \emph {et~al.}(2016)\citenamefont
  {Alberte}, \citenamefont {Baggioli}, \citenamefont {Khmelnitsky},\ and\
  \citenamefont {Pujolas}}]{Alberte:2015isw}%
  \BibitemOpen
  \bibfield  {author} {\bibinfo {author} {\bibfnamefont {L.}~\bibnamefont
  {Alberte}}, \bibinfo {author} {\bibfnamefont {M.}~\bibnamefont {Baggioli}},
  \bibinfo {author} {\bibfnamefont {A.}~\bibnamefont {Khmelnitsky}}, \ and\
  \bibinfo {author} {\bibfnamefont {O.}~\bibnamefont {Pujolas}},\ }\href
  {\doibase 10.1007/JHEP02(2016)114} {\bibfield  {journal} {\bibinfo  {journal}
  {JHEP}\ }\textbf {\bibinfo {volume} {02}},\ \bibinfo {pages} {114} (\bibinfo
  {year} {2016})},\ \Eprint {http://arxiv.org/abs/1510.09089} {arXiv:1510.09089
  [hep-th]} \BibitemShut {NoStop}%
\bibitem [{\citenamefont {Alberte}\ \emph {et~al.}(2018)\citenamefont
  {Alberte}, \citenamefont {Ammon}, \citenamefont {Jim\'enez-Alba},
  \citenamefont {Baggioli},\ and\ \citenamefont {Pujol\`as}}]{Alberte:2017oqx}%
  \BibitemOpen
  \bibfield  {author} {\bibinfo {author} {\bibfnamefont {L.}~\bibnamefont
  {Alberte}}, \bibinfo {author} {\bibfnamefont {M.}~\bibnamefont {Ammon}},
  \bibinfo {author} {\bibfnamefont {A.}~\bibnamefont {Jim\'enez-Alba}},
  \bibinfo {author} {\bibfnamefont {M.}~\bibnamefont {Baggioli}}, \ and\
  \bibinfo {author} {\bibfnamefont {O.}~\bibnamefont {Pujol\`as}},\ }\href
  {\doibase 10.1103/PhysRevLett.120.171602} {\bibfield  {journal} {\bibinfo
  {journal} {Phys. Rev. Lett.}\ }\textbf {\bibinfo {volume} {120}},\ \bibinfo
  {pages} {171602} (\bibinfo {year} {2018})},\ \Eprint
  {http://arxiv.org/abs/1711.03100} {arXiv:1711.03100 [hep-th]} \BibitemShut
  {NoStop}%
\bibitem [{\citenamefont {Zhang}\ \emph {et~al.}(2019)\citenamefont {Zhang},
  \citenamefont {Hu},\ and\ \citenamefont {Zhang}}]{Zhang:2019oes}%
  \BibitemOpen
  \bibfield  {author} {\bibinfo {author} {\bibfnamefont {H.}~\bibnamefont
  {Zhang}}, \bibinfo {author} {\bibfnamefont {Y.-p.}\ \bibnamefont {Hu}}, \
  and\ \bibinfo {author} {\bibfnamefont {Y.}~\bibnamefont {Zhang}},\ }\href
  {\doibase 10.1016/j.dark.2018.100257} {\bibfield  {journal} {\bibinfo
  {journal} {Phys. Dark Univ.}\ }\textbf {\bibinfo {volume} {23}},\ \bibinfo
  {pages} {100257} (\bibinfo {year} {2019})},\ \Eprint
  {http://arxiv.org/abs/1901.09331} {arXiv:1901.09331 [gr-qc]} \BibitemShut
  {NoStop}%
\bibitem [{\citenamefont {Blake}\ and\ \citenamefont
  {Tong}(2013)}]{Blake:2013bqa}%
  \BibitemOpen
  \bibfield  {author} {\bibinfo {author} {\bibfnamefont {M.}~\bibnamefont
  {Blake}}\ and\ \bibinfo {author} {\bibfnamefont {D.}~\bibnamefont {Tong}},\
  }\href {\doibase 10.1103/PhysRevD.88.106004} {\bibfield  {journal} {\bibinfo
  {journal} {Phys. Rev. D}\ }\textbf {\bibinfo {volume} {88}},\ \bibinfo
  {pages} {106004} (\bibinfo {year} {2013})},\ \Eprint
  {http://arxiv.org/abs/1308.4970} {arXiv:1308.4970 [hep-th]} \BibitemShut
  {NoStop}%
\bibitem [{\citenamefont {Cao}\ and\ \citenamefont {Peng}(2015)}]{Cao:2015cza}%
  \BibitemOpen
  \bibfield  {author} {\bibinfo {author} {\bibfnamefont {L.-M.}\ \bibnamefont
  {Cao}}\ and\ \bibinfo {author} {\bibfnamefont {Y.}~\bibnamefont {Peng}},\
  }\href {\doibase 10.1103/PhysRevD.92.124052} {\bibfield  {journal} {\bibinfo
  {journal} {Phys. Rev. D}\ }\textbf {\bibinfo {volume} {92}},\ \bibinfo
  {pages} {124052} (\bibinfo {year} {2015})},\ \Eprint
  {http://arxiv.org/abs/1509.08738} {arXiv:1509.08738 [hep-th]} \BibitemShut
  {NoStop}%
\bibitem [{\citenamefont {Baggioli}\ and\ \citenamefont
  {Pujolas}(2015)}]{Baggioli:2014roa}%
  \BibitemOpen
  \bibfield  {author} {\bibinfo {author} {\bibfnamefont {M.}~\bibnamefont
  {Baggioli}}\ and\ \bibinfo {author} {\bibfnamefont {O.}~\bibnamefont
  {Pujolas}},\ }\href {\doibase 10.1103/PhysRevLett.114.251602} {\bibfield
  {journal} {\bibinfo  {journal} {Phys. Rev. Lett.}\ }\textbf {\bibinfo
  {volume} {114}},\ \bibinfo {pages} {251602} (\bibinfo {year} {2015})},\
  \Eprint {http://arxiv.org/abs/1411.1003} {arXiv:1411.1003 [hep-th]}
  \BibitemShut {NoStop}%
\bibitem [{\citenamefont {Gialamas}\ and\ \citenamefont
  {Tamvakis}(2023{\natexlab{a}})}]{Gialamas:2023aim}%
  \BibitemOpen
  \bibfield  {author} {\bibinfo {author} {\bibfnamefont {I.~D.}\ \bibnamefont
  {Gialamas}}\ and\ \bibinfo {author} {\bibfnamefont {K.}~\bibnamefont
  {Tamvakis}},\ }\href {\doibase 10.1103/PhysRevD.107.104012} {\bibfield
  {journal} {\bibinfo  {journal} {Phys. Rev. D}\ }\textbf {\bibinfo {volume}
  {107}},\ \bibinfo {pages} {104012} (\bibinfo {year} {2023}{\natexlab{a}})},\
  \Eprint {http://arxiv.org/abs/2303.11353} {arXiv:2303.11353 [gr-qc]}
  \BibitemShut {NoStop}%
\bibitem [{\citenamefont {Gialamas}\ and\ \citenamefont
  {Tamvakis}(2023{\natexlab{b}})}]{Gialamas:2023lxj}%
  \BibitemOpen
  \bibfield  {author} {\bibinfo {author} {\bibfnamefont {I.~D.}\ \bibnamefont
  {Gialamas}}\ and\ \bibinfo {author} {\bibfnamefont {K.}~\bibnamefont
  {Tamvakis}},\ }\href {\doibase 10.1103/PhysRevD.108.104023} {\bibfield
  {journal} {\bibinfo  {journal} {Phys. Rev. D}\ }\textbf {\bibinfo {volume}
  {108}},\ \bibinfo {pages} {104023} (\bibinfo {year} {2023}{\natexlab{b}})},\
  \Eprint {http://arxiv.org/abs/2307.05673} {arXiv:2307.05673 [gr-qc]}
  \BibitemShut {NoStop}%
\bibitem [{\citenamefont {Gialamas}\ and\ \citenamefont
  {Tamvakis}(2024)}]{Gialamas:2023fly}%
  \BibitemOpen
  \bibfield  {author} {\bibinfo {author} {\bibfnamefont {I.~D.}\ \bibnamefont
  {Gialamas}}\ and\ \bibinfo {author} {\bibfnamefont {K.}~\bibnamefont
  {Tamvakis}},\ }\href {\doibase 10.1088/1475-7516/2024/03/016} {\bibfield
  {journal} {\bibinfo  {journal} {JCAP}\ }\textbf {\bibinfo {volume} {03}},\
  \bibinfo {pages} {016} (\bibinfo {year} {2024})},\ \Eprint
  {http://arxiv.org/abs/2311.14799} {arXiv:2311.14799 [gr-qc]} \BibitemShut
  {NoStop}%
\bibitem [{\citenamefont {Yerra}\ \emph {et~al.}(2025)\citenamefont {Yerra},
  \citenamefont {Mukherji},\ and\ \citenamefont {Bhamidipati}}]{Yerra:2024stj}%
  \BibitemOpen
  \bibfield  {author} {\bibinfo {author} {\bibfnamefont {P.~K.}\ \bibnamefont
  {Yerra}}, \bibinfo {author} {\bibfnamefont {S.}~\bibnamefont {Mukherji}}, \
  and\ \bibinfo {author} {\bibfnamefont {C.}~\bibnamefont {Bhamidipati}},\
  }\href {\doibase 10.1103/lj4b-j3tr} {\bibfield  {journal} {\bibinfo
  {journal} {Phys. Rev. D}\ }\textbf {\bibinfo {volume} {111}},\ \bibinfo
  {pages} {124018} (\bibinfo {year} {2025})},\ \Eprint
  {http://arxiv.org/abs/2411.01261} {arXiv:2411.01261 [gr-qc]} \BibitemShut
  {NoStop}%
\bibitem [{\citenamefont {Duan}(1984)}]{Duan:1984ws}%
  \BibitemOpen
  \bibfield  {author} {\bibinfo {author} {\bibfnamefont {Y.-S.}\ \bibnamefont
  {Duan}},\ }\href@noop {} {\bibfield  {journal} {\bibinfo  {journal}
  {SLAC-PUB-3301}\ } (\bibinfo {year} {1984})}\BibitemShut {NoStop}%
\bibitem [{\citenamefont {Duan}\ and\ \citenamefont {Ge}(1979)}]{Duan:2018rbd}%
  \BibitemOpen
  \bibfield  {author} {\bibinfo {author} {\bibfnamefont {Y.-S.}\ \bibnamefont
  {Duan}}\ and\ \bibinfo {author} {\bibfnamefont {M.-L.}\ \bibnamefont {Ge}},\
  }\href {\doibase 10.1142/9789813237278_0001} {\bibfield  {journal} {\bibinfo
  {journal} {Sci. Sin.}\ }\textbf {\bibinfo {volume} {9}},\ \bibinfo {pages}
  {1072} (\bibinfo {year} {1979})}\BibitemShut {NoStop}%
\bibitem [{\citenamefont {Afshar}\ and\ \citenamefont
  {Sadeghi}(2025{\natexlab{b}})}]{Afshar:2024dhf}%
  \BibitemOpen
  \bibfield  {author} {\bibinfo {author} {\bibfnamefont {M.~A.~S.}\
  \bibnamefont {Afshar}}\ and\ \bibinfo {author} {\bibfnamefont
  {J.}~\bibnamefont {Sadeghi}},\ }\href {\doibase 10.1016/j.aop.2025.169953}
  {\bibfield  {journal} {\bibinfo  {journal} {Annals Phys.}\ }\textbf {\bibinfo
  {volume} {474}},\ \bibinfo {pages} {169953} (\bibinfo {year}
  {2025}{\natexlab{b}})},\ \Eprint {http://arxiv.org/abs/2412.06357}
  {arXiv:2412.06357 [gr-qc]} \BibitemShut {NoStop}%
\bibitem [{\citenamefont {Junior}\ \emph {et~al.}(2021)\citenamefont {Junior},
  \citenamefont {Cunha}, \citenamefont {Herdeiro},\ and\ \citenamefont
  {Crispino}}]{Junior:2021dyw}%
  \BibitemOpen
  \bibfield  {author} {\bibinfo {author} {\bibfnamefont {H.~C. D.~L.}\
  \bibnamefont {Junior}}, \bibinfo {author} {\bibfnamefont {P.~V.~P.}\
  \bibnamefont {Cunha}}, \bibinfo {author} {\bibfnamefont {C.~A.~R.}\
  \bibnamefont {Herdeiro}}, \ and\ \bibinfo {author} {\bibfnamefont {L.~C.~B.}\
  \bibnamefont {Crispino}},\ }\href {\doibase 10.1103/PhysRevD.104.044018}
  {\bibfield  {journal} {\bibinfo  {journal} {Phys. Rev. D}\ }\textbf {\bibinfo
  {volume} {104}},\ \bibinfo {pages} {044018} (\bibinfo {year} {2021})},\
  \Eprint {http://arxiv.org/abs/2104.09577} {arXiv:2104.09577 [gr-qc]}
  \BibitemShut {NoStop}%
\bibitem [{\citenamefont {Junior}\ \emph {et~al.}(2022)\citenamefont {Junior},
  \citenamefont {Yang}, \citenamefont {Crispino}, \citenamefont {Cunha},\ and\
  \citenamefont {Herdeiro}}]{Junior:2021svb}%
  \BibitemOpen
  \bibfield  {author} {\bibinfo {author} {\bibfnamefont {H.~C. D.~L.}\
  \bibnamefont {Junior}}, \bibinfo {author} {\bibfnamefont {J.-Z.}\
  \bibnamefont {Yang}}, \bibinfo {author} {\bibfnamefont {L.~C.~B.}\
  \bibnamefont {Crispino}}, \bibinfo {author} {\bibfnamefont {P.~V.~P.}\
  \bibnamefont {Cunha}}, \ and\ \bibinfo {author} {\bibfnamefont {C.~A.~R.}\
  \bibnamefont {Herdeiro}},\ }\href {\doibase 10.1103/PhysRevD.105.064070}
  {\bibfield  {journal} {\bibinfo  {journal} {Phys. Rev. D}\ }\textbf {\bibinfo
  {volume} {105}},\ \bibinfo {pages} {064070} (\bibinfo {year} {2022})},\
  \Eprint {http://arxiv.org/abs/2112.10802} {arXiv:2112.10802 [gr-qc]}
  \BibitemShut {NoStop}%
\bibitem [{\citenamefont {Wang}\ \emph {et~al.}(2021)\citenamefont {Wang},
  \citenamefont {Chen},\ and\ \citenamefont {Jing}}]{Wang:2021ara}%
  \BibitemOpen
  \bibfield  {author} {\bibinfo {author} {\bibfnamefont {M.}~\bibnamefont
  {Wang}}, \bibinfo {author} {\bibfnamefont {S.}~\bibnamefont {Chen}}, \ and\
  \bibinfo {author} {\bibfnamefont {J.}~\bibnamefont {Jing}},\ }\href {\doibase
  10.1103/PhysRevD.104.084021} {\bibfield  {journal} {\bibinfo  {journal}
  {Phys. Rev. D}\ }\textbf {\bibinfo {volume} {104}},\ \bibinfo {pages}
  {084021} (\bibinfo {year} {2021})},\ \Eprint
  {http://arxiv.org/abs/2104.12304} {arXiv:2104.12304 [gr-qc]} \BibitemShut
  {NoStop}%
\bibitem [{\citenamefont {Huang}\ \emph {et~al.}(2022)\citenamefont {Huang},
  \citenamefont {Ou}, \citenamefont {Lai},\ and\ \citenamefont
  {Lu}}]{Huang:2021qwe}%
  \BibitemOpen
  \bibfield  {author} {\bibinfo {author} {\bibfnamefont {H.}~\bibnamefont
  {Huang}}, \bibinfo {author} {\bibfnamefont {M.-Y.}\ \bibnamefont {Ou}},
  \bibinfo {author} {\bibfnamefont {M.-Y.}\ \bibnamefont {Lai}}, \ and\
  \bibinfo {author} {\bibfnamefont {H.}~\bibnamefont {Lu}},\ }\href {\doibase
  10.1103/PhysRevD.105.104049} {\bibfield  {journal} {\bibinfo  {journal}
  {Phys. Rev. D}\ }\textbf {\bibinfo {volume} {105}},\ \bibinfo {pages}
  {104049} (\bibinfo {year} {2022})},\ \Eprint
  {http://arxiv.org/abs/2112.14780} {arXiv:2112.14780 [hep-th]} \BibitemShut
  {NoStop}%
\bibitem [{\citenamefont {Dolan}\ and\ \citenamefont
  {Ottewill}(2009)}]{Dolan:2009nk}%
  \BibitemOpen
  \bibfield  {author} {\bibinfo {author} {\bibfnamefont {S.~R.}\ \bibnamefont
  {Dolan}}\ and\ \bibinfo {author} {\bibfnamefont {A.~C.}\ \bibnamefont
  {Ottewill}},\ }\href {\doibase 10.1088/0264-9381/26/22/225003} {\bibfield
  {journal} {\bibinfo  {journal} {Class. Quant. Grav.}\ }\textbf {\bibinfo
  {volume} {26}},\ \bibinfo {pages} {225003} (\bibinfo {year} {2009})},\
  \Eprint {http://arxiv.org/abs/0908.0329} {arXiv:0908.0329 [gr-qc]}
  \BibitemShut {NoStop}%
\bibitem [{\citenamefont {Cardoso}\ \emph {et~al.}(2014)\citenamefont
  {Cardoso}, \citenamefont {Crispino}, \citenamefont {Macedo}, \citenamefont
  {Okawa},\ and\ \citenamefont {Pani}}]{Cardoso:2014sna}%
  \BibitemOpen
  \bibfield  {author} {\bibinfo {author} {\bibfnamefont {V.}~\bibnamefont
  {Cardoso}}, \bibinfo {author} {\bibfnamefont {L.~C.~B.}\ \bibnamefont
  {Crispino}}, \bibinfo {author} {\bibfnamefont {C.~F.~B.}\ \bibnamefont
  {Macedo}}, \bibinfo {author} {\bibfnamefont {H.}~\bibnamefont {Okawa}}, \
  and\ \bibinfo {author} {\bibfnamefont {P.}~\bibnamefont {Pani}},\ }\href
  {\doibase 10.1103/PhysRevD.90.044069} {\bibfield  {journal} {\bibinfo
  {journal} {Phys. Rev. D}\ }\textbf {\bibinfo {volume} {90}},\ \bibinfo
  {pages} {044069} (\bibinfo {year} {2014})},\ \Eprint
  {http://arxiv.org/abs/1406.5510} {arXiv:1406.5510 [gr-qc]} \BibitemShut
  {NoStop}%
\bibitem [{\citenamefont {Keir}(2016)}]{Keir:2014oka}%
  \BibitemOpen
  \bibfield  {author} {\bibinfo {author} {\bibfnamefont {J.}~\bibnamefont
  {Keir}},\ }\href {\doibase 10.1088/0264-9381/33/13/135009} {\bibfield
  {journal} {\bibinfo  {journal} {Class. Quant. Grav.}\ }\textbf {\bibinfo
  {volume} {33}},\ \bibinfo {pages} {135009} (\bibinfo {year} {2016})},\
  \Eprint {http://arxiv.org/abs/1404.7036} {arXiv:1404.7036 [gr-qc]}
  \BibitemShut {NoStop}%
\bibitem [{\citenamefont {Konoplya}\ and\ \citenamefont
  {Zhidenko}(2011)}]{Konoplya:2011qq}%
  \BibitemOpen
  \bibfield  {author} {\bibinfo {author} {\bibfnamefont {R.~A.}\ \bibnamefont
  {Konoplya}}\ and\ \bibinfo {author} {\bibfnamefont {A.}~\bibnamefont
  {Zhidenko}},\ }\href {\doibase 10.1103/RevModPhys.83.793} {\bibfield
  {journal} {\bibinfo  {journal} {Rev. Mod. Phys.}\ }\textbf {\bibinfo {volume}
  {83}},\ \bibinfo {pages} {793} (\bibinfo {year} {2011})},\ \Eprint
  {http://arxiv.org/abs/1102.4014} {arXiv:1102.4014 [gr-qc]} \BibitemShut
  {NoStop}%
\bibitem [{\citenamefont {Shaikh}(2018)}]{Shaikh:2018kfv}%
  \BibitemOpen
  \bibfield  {author} {\bibinfo {author} {\bibfnamefont {R.}~\bibnamefont
  {Shaikh}},\ }\href {\doibase 10.1103/PhysRevD.98.024044} {\bibfield
  {journal} {\bibinfo  {journal} {Phys. Rev. D}\ }\textbf {\bibinfo {volume}
  {98}},\ \bibinfo {pages} {024044} (\bibinfo {year} {2018})},\ \Eprint
  {http://arxiv.org/abs/1803.11422} {arXiv:1803.11422 [gr-qc]} \BibitemShut
  {NoStop}%
\end{thebibliography}%
\end{document}